\documentclass[a4paper,11pt]{article}
\usepackage{jcappub}
\usepackage{geometry}
\usepackage[utf8]{inputenc}
\usepackage{graphicx}
\usepackage{booktabs}
\usepackage{tabularx}
\usepackage{textcomp}
\usepackage{comment}
\usepackage{xspace}
\usepackage{amsmath}
\usepackage{amsfonts}
\usepackage{amssymb}
\usepackage{upgreek}
\usepackage[dvipsnames,table]{xcolor}
\usepackage[normalem]{ulem}
\usepackage{csquotes}

\hypersetup{urlcolor=BlueViolet,
	    citecolor=Plum,
	    linkcolor=PineGreen}
\usepackage[official]{eurosym}
\definecolor{lightgray}{rgb}{0.9,0.9,0.9}	    
\definecolor{green}{rgb}{0,0.5,0}
\definecolor{red}{rgb}{1,0,0}
\definecolor{blue}{rgb}{0,0,0.5}

\begin{document}

\title{Fine-grained dark matter substructure and axion haloscopes}

\author[a,b]{Ciaran A. J. O'Hare,}
\author[c]{Giovanni Pierobon,}
\affiliation[a]{Sydney Consortium for Particle Physics and Cosmology,
 School of Physics, The University of Sydney, NSW 2006, Australia }
\affiliation[b]{ARC Centre of Excellence for Dark Matter Particle Physics, School of Physics, The University of Sydney, NSW 2006, Australia}

\affiliation[c]{Sydney Consortium for Particle Physics and Cosmology, School of Physics, The University of New South Wales, NSW 2052, Sydney, Australia}
\emailAdd{ciaran.ohare@sydney.edu.au}
\emailAdd{g.pierobon@unsw.edu.au}

\date{\today}
\smallskip
\abstract{
The velocity distribution of cold dark matter within galaxies is expected to exhibit ultra-fine-grained substructure as a result of the many foldings of an initially smooth phase-space sheet under gravity. The innumerable folds of this sheet in the inner regions of a galactic halo would appear on solar-system scales like an extremely large number of spatially overlapping streams of dark matter particles. Some of these streams may also receive amplified densities if dark matter undergoes enhanced clustering in the early Universe, as is the case for axions in the post-inflationary scenario. This ultra-fine-grained dark matter substructure is usually considered irrelevant and undetectable for most direct detection experiments, but for axion haloscopes, this may not be the case. We develop and explore several plausible models for the degree of fine-grained axion dark matter substructure so as to evaluate its impact on the detectability of axions. We find that not only is this substructure detectable in haloscope experiments, but that it may enhance the detectability of the axion if data is analysed with a high-enough frequency resolution. This conclusion motivates ongoing high-resolution analyses of axion data by haloscope collaborations, which have the potential to reveal evidence of the QCD axion even if their nominal experimental sensitivity does not reach the required level under the Standard Halo Model assumption.
}

\maketitle

\section{Introduction}
The internal structures of cold dark matter (CDM) halos are formed from the gravitational collapse of an initial smooth density field---a process that repeatedly folds the thin phase space sheet until a fully phase-mixed distribution emerges. Despite the subsequent billions of years of collapse, mergers and accretion, the distribution of collisionless dark matter particles inside galaxies is still expected to reflect, albeit in a highly complex manner, this folding process (see e.g.~\cite{Abel:2011ui, Falck:2012ai, Afshordi:2008mx, Zavala:2013bha, Ramachandra:2017vyl, Shandarin:2019jjv, Stucker:2019mjw, Stucker:2021vyx} for more discussion). If this finely-foliated phase space distribution were observed at a single spatial point (as dark matter detectors on Earth are trying to do), the layers would manifest as numerous overlapping streams---dynamically cold groups of particles on common orbits. 

\begin{figure}
    \centering
    \includegraphics[width=0.99\linewidth]{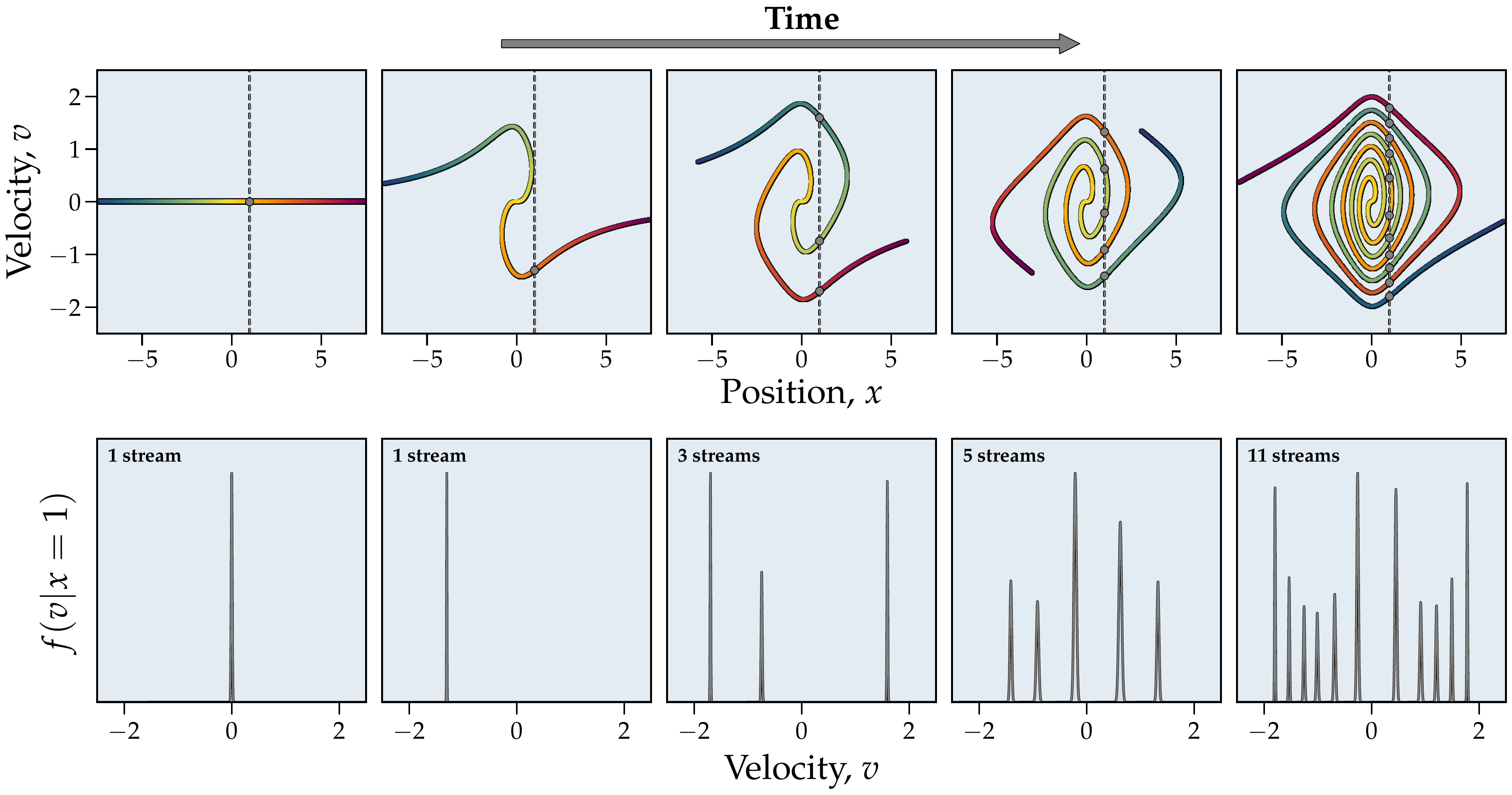}
    \caption{A cartoon of how the process of phase mixing leads to a velocity distribution composed of a large number of streams. The dark-matter distribution starts out as a thin sheet (one-dimensional here because the phase space is two-dimensional) with a thickness set by the initial velocities the dark matter inherits when it is produced. The sheet then folds in on itself due to gravity. After many folds, the velocity distribution of dark matter measured at a single spatial point (bottom row of panels) is made up of the streaming particle motions at a number of discrete velocities set by each overlapping fold. In the vicinity of the Sun in the inner halo of the Milky Way, a vast number of folds would be overlapping, meaning the velocity distribution $f(v)$ would appear almost perfectly smooth despite fundamentally being composed of fine-grained streams.}
    \label{fig:PhaseMixing}
\end{figure}

In Fig.~\ref{fig:PhaseMixing}, we present a cartoon illustration of the process of phase mixing, inspired by similar depictions in e.g.~\cite{Abel:2011ui, Schulz:2013}. Dark matter begins as a thin sheet with a small but finite width set by its initial velocity dispersion. Gravity causes this sheet to fold in on itself, each fold foliating the phase space distribution to an increasingly fine degree. Although the spatial distribution of particles becomes extremely smooth in the process, if the \textit{velocity} distribution of those particles were measured at a single point---i.e.~$f(v|x)$ as shown in the bottom panels---one would see that it is fundamentally composed of as many narrow streams as there are folds of the sheet overlapping at that point. Each stream would carry a velocity dispersion similar to the sheet's initial thickness in phase space. 

The question of how finely foliated is the phase space of the Milky Way in the solar neighbourhood is one that is highly relevant for attempts to directly detect dark matter. One answer was provided by studies such as Refs.~\cite{White:2008as, Vogelsberger:2010gd}, which extrapolated the evolution of the dark matter's phase space distribution beyond the resolution of N-body simulations, revealing an extremely high degree of phase mixing in the inner regions of Milky-Way-like halos (far beyond what can be captured in Fig.~\ref{fig:PhaseMixing}). The dark matter density in the solar neighbourhood, $\rho_{\rm DM} \approx 0.4$~GeV~cm$^{-3}\approx 0.01\,M_\odot~{\rm pc}^{-3}$~\cite{Read:2014qva,deSalas:2020hbh} is expected to be composed of some $10^{14}$ overlapping streams of particles, with around $10^6$ of the densest ones composing half of that density. For experiments looking to detect dark matter on Earth, one concludes from this result that the velocity distribution can be considered to be essentially indistinguishable from one that is exactly smooth. This justifies the use of the so-called Standard Halo Model in direct detection analyses, in which the local dark-matter velocity distribution is described by an isotropic and exactly smooth Gaussian~\cite{Evans:2018bqy}, i.e.~$f(\mathbf{v}) \sim \exp(-\mathbf{v}^2/2\sigma_v^2)$. But while the distribution would \textit{appear} smooth to a dark-matter detector without the ability to precisely reconstruct the exact values of the incoming particles' velocities, the distribution cannot be \textit{exactly} smooth.

These ideas apply to cold dark matter halos in general, and so are the most intriguing for dark matter candidates where the assumptions of cold and collisionless behaviour hold the most strongly. Arguably the ``coldest''~\footnote{By cold, we are referring here to dark matter's free-streaming motion around the time it is produced. For example, a 100 GeV-mass thermal relic dark matter candidate undergoes random free-streaming motions, preventing it from clustering into halos below around the Earth mass~\cite{Green:2003un, Wang:2019ftp, Bechtol:2022koa}. On the other hand, the gravitational collapse of axion structures is only inhibited by its wave-like nature, which sets a cutoff known as the axion Jeans scale, which scales inversely with the axion mass. Axions in the neV-meV mass range that we are interested in here can therefore continue to cluster on scales many orders of magnitude smaller than GeV-scale thermal relic dark matter.} of all dark matter candidates are low-mass bosonic particles, also known as wave-like dark matter, which includes the axion~\cite{Peccei:1977hh, Peccei:1977ur, Weinberg:1977ma, Wilczek:1977pj, Kim:2008hd, Kim:1979if, Shifman:1979if, Dine:1981rt, Zhitnitsky:1980tq}. Our primary motivation in this study is the specific model called the ``QCD axion''~\cite{Peccei:1977hh, Peccei:1977ur}. The QCD axion stands out as perhaps the most promising candidate for dark matter because its existence in Nature is predicted independently. We refer to Ref.~\cite{Kaplan:2025bgy} for a recent reinforcement of the need for the QCD axion in our Universe. See also Refs.~\cite{DiLuzio:2020wdo, ParticleDataGroup:2024cfk, Baryakhtar:2025jwh, OHare:2024nmr, Marsh:2015xka, Adams:2022pbo} for reviews on axions and axion dark matter.

Among the various proposed axion production mechanisms, the most predictive example is the so-called ``post-inflationary'' misalignment scenario, in which the birth of the axion as a pseudo-Nambu-Goldstone boson following a symmetry-breaking phase transition occurs \textit{after} the end of inflation. In this scenario, the initial value of the axion field differs between causally disconnected patches of the Universe at the phase transition. The subsequent evolution of this inhomogeneous field involves non-trivial dynamics due to the emergence of topological defects. In the case of the QCD axion, which acquires its mass around the QCD phase transition, we expect these dynamics to bestow the dark matter distribution with density inhomogeneities on scales set by the horizon size at this epoch. Correctly predicting the abundance of dark matter and the spectrum of its perturbations in this scenario demands challenging multi-scale numerical simulations~\cite{Vaquero:2018tib, Buschmann:2019icd, Buschmann:2021sdq, Gorghetto:2018myk, Gorghetto:2020qws, OHare:2021zrq, Kim:2024wku, Saikawa:2024bta, Benabou:2024msj}.

One generic late-time conclusion of this cosmological scenario that was pointed out many years ago~\cite{Hogan:1988mp, Kolb:1994fi, Kolb:1995bu}, but has now been corroborated more quantitatively thanks to recent simulations~\cite{Vaquero:2018tib, Eggemeier:2019khm, Pierobon:2023ozb}, is that these inhomogeneities seed the formation of dark matter ``miniclusters'' or ``minihalos''. Most of the axion dark matter in this scenario should be bound within miniclusters by around the time of galaxy formation~\cite{Eggemeier:2019khm, Eggemeier:2022hqa}. However, the most massive miniclusters merge into loosely bound minicluster halos, which are easily disrupted by stars and the galaxy's tides over Gyr-timescales. The majority of the dark matter by mass is expected to be no longer bound inside miniclusters today~\cite{Tinyakov:2015cgg, Dokuchaev, Kavanagh:2020gcy, DSouza:2024flu, Shen:2022ltx}, although there is expected to be an abundance of stable remnant cores. 

Due to the small internal velocity dispersions of miniclusters, these disrupted structures will not be immediately virialised back into the halo but will instead leave behind tidally elongated streams stretching over distances of order a few parsecs. We therefore expect the post-inflationary axion to modify the statements made above---where now the majority of the dark matter should be made up of hundreds to thousands of streams~\cite{OHare:2023rtm}, as a direct result of this enhanced early-time clustering which confined the initial dark matter into early-forming minihalos as opposed to a smooth density field. The velocity dispersions of these streams in this case, however, are not negligible but are fixed by the internal velocity dispersion of the miniclusters plus an additional heating due to the tidal disruption process. Given the expected minicluster mass function, the velocity dispersion of their streams will be on the order of $\sim$0.01~km/s.

Putting together these two ideas---namely that we expect on the order of $10^{14}$ fine-grained streams locally in a generic CDM cosmology, or $\mathcal{O}(1000)$ in a post-inflationary axion cosmology---we are motivated to take seriously the implications of a dark matter velocity distribution that is not smooth. These distributions are very different from the standard assumption made in the field, and will have important ramifications for axion direct detection experiments---``haloscopes''---in which signals are greatly enhanced if the dark matter velocity distribution contains kinematically cold substructures.

Indeed, the idea that fine-grained substructure could be important for haloscopes is not a new one. Several haloscope collaborations already conduct specialised ``high-resolution'' re-analyses of data from their standard axion searches~\cite{Duffy:2005ab, ADMX:2006kgb, ADMX:2023ctd, ADMX:2024pxg,Hoskins:2011iv,Quiskamp:2024oet,Yi:2025ktq}, partly inspired by models of Sikivie~\cite{Sikivie:1992bk,Sikivie:2001fg,Duffy:2008dk}. The axion signal in a haloscope is oscillatory at a frequency $\omega = m_a(1+v^2/2)$ where $m_a$ is the axion mass and $v$ is drawn from the local velocity distribution. The Fourier power spectrum of this oscillating signal when observed over timescales longer than the coherence time $T_{\rm coh} \sim (m v^2)^{-1}$ is then related to the distribution of axion kinetic energies. This makes the local velocity distribution of dark matter an observable in haloscope experiments, and hints at the futuristic concept of ``axion astronomy'' in which dark matter experiments could be used to measure this astrophysical quantity~\cite{OHare:2017yze, Foster:2017hbq, Knirck:2018knd, Foster:2020fln}, which is impossible to access through any other means. The topic that interests us here is the fact that ultra-cold streams of dark matter would manifest as ultra-narrow features in a haloscope power spectrum. These streams could be so cold that the possible enhancement over the background noise level could be significantly higher than what would be expected from a smooth distribution of dark matter. However, the key idea, around which our study will focus, is that the frequency resolution must be at least comparable to the frequency spread of these features, or no such enhancement will be observed. This is what motivates the development and implementation of the dedicated high-resolution analyses mentioned above.

This study aims to put the scientific motivation for high-resolution axion analyses on a firmer footing by proposing and exploring several plausible models for the degree of fine-grained velocity substructure in axion dark matter. We will estimate the sensitivity of high-resolution analyses of existing and in-preparation haloscopes to these models by exploring two different statistical approaches---both a parametric likelihood-based analysis, and a non-parametric approach which removes the need to assume the precise form of a particular signal model (which is important here as some of our models can only be specified in statistical terms). As our motivation is the pre- and post-inflationary misalignment scenarios for QCD axion dark matter, we will quote numerical results for axion haloscopes operating in the GHz-THz frequency range, as these are targeting axion mass windows which lead to the correct abundance of dark matter in these scenarios~\cite{Gorghetto:2020qws, Buschmann:2021sdq, Saikawa:2024bta, Benabou:2024msj}. That said, our results will also be quoted in terms of their parametric dependence on the main timescale in the problem---the axion's oscillation period, $T_{\rm ax} = 2\pi/m_a$---and so they can be extended to other mass ranges.

The paper is structured as follows. In Sec.~\ref{sec:models} we review the theory required to predict the axion ``lineshape''---the signal observable in an axion haloscope~\cite{Turner:1990qx}---and write down several lineshape models that incorporate different types of fine-grained velocity substructure. Then in Sec.~\ref{sec:stats} we describe two different statistical approaches which we use to estimate the sensitivity of a given haloscope experiment to an axion signal. We then end with Secs.~\ref{sec:miniclusters} and~\ref{sec:finegrained}, which go through the implications of our two most well-motivated models, namely that of the minicluster streams and fine-grained CDM streams, respectively.

\section{Axion haloscope power spectrum}\label{sec:models}
Our first goal is to derive an expression for the expected signal in a generic haloscope experiment searching for wave-like dark matter. Specifically, we wish to derive the ``lineshape'', which encapsulates the experiment's dependence on the local dark matter velocity distribution, $f(\mathbf{v})$. The lineshape, $S(\omega)$, is the power spectrum extracted when one takes a discrete Fourier transform of some time-series measurements of the value of the axion field at a single spatial point. A schematic of how this idea is affected by the choice of velocity distribution is shown in Fig.~\ref{fig:diagram}, which also highlights several important timescales and other quantities introduced in this section.

\subsection{Axion lineshape}\label{sec:lineshape}
To describe the axion field value at a single spatial point, we can adopt a convenient formalism introduced in Ref.~\cite{Foster:2017hbq}, which imagines constructing the full classical field out of some large (but arbitrary) number $N_a$ of real-valued waves:
\begin{equation}
    a_i(v, t)=\frac{\sqrt{2 \rho_{\mathrm{DM}} / N_a}}{m_a} \cos \left[m_a\left(1+\frac{v_i^2}{2}\right) t+\phi_i\right] \, ,
\end{equation}
where $i = 0,..., N_a$ is a label running over individual waves, $v_i \equiv |\mathbf{v}_i|$ are the speeds of each wave, $\phi_i$ are random phases, and $m_a$ is the axion mass. The known value of the dark matter density at our position in the galaxy, $\rho_{\rm DM} \approx 0.3$--$0.5$~GeV/cm$^3$, is what allows us to fix the time-averaged value of the field's energy density. Although it is possible that the local density of dark matter in the solar system is enhanced beyond this value, which is inferred using stellar kinematic data spread across much larger scales---see discussions in the literature such as~\cite{Budker:2023sex}---there is currently no dynamical evidence for any such enhancement. If there are ultra-local enhancements (or suppressions) in the density of dark matter for certain models, then experimental exclusion limits on those models must be rescaled appropriately. However, this scaling is trivial to impose, and so is not our focus here. Rather, we focus on the effects of the velocity distribution which enters through $v_i$. Therefore, throughout this discussion, we will fix $\rho_{\rm DM} = 0.45$~GeV/cm$^3$ for all examples unless otherwise stated. We will further assume that the phases of each wave are uncorrelated with each other, which means that each $\phi_i$ is drawn from a uniform distribution.\footnote{This assumption is equivalent to the statement that the axion density is uniform across the spatial scales being measured. Any $\mathbf{v}$-dependent correlations in phases lead to localised enhancements in the axion energy density. The axion substructures we consider here are overdensities in phase space, not in physical space, which means we will continue to use the assumption of random phases.} 

There are two relevant timescales when describing the axion field that it is useful to define. One is the axion \emph{oscillation period},
\begin{equation}
    T_{\rm ax} \equiv \frac{2\pi}{m_a} = 41 \, {\rm ps} \left(\frac{100\,\upmu {\rm eV}}{m_a} \right),
\end{equation}
which strictly speaking is the \textit{shortest} period with which the axion field can oscillate since it corresponds to $v = 0$. The second timescale is the axion \emph{coherence time}, $T_{\rm coh}$, which gives a measure of the duration over which the axion field oscillations are expected to fall out of phase due to the fact that $v_i$ is drawn from a distribution with a non-zero width. The departure from a monochromatic oscillation is set by the local velocity distribution from which $v_i$ are drawn, and is therefore model-dependent. It is possible to rigorously define\footnote{For instance, if we assume the boosted Gaussian form for the velocity distribution (see Sec.~\ref{sec:fv}), the formal expression for the coherence time becomes $T^{\rm SHM}_{\rm coh} \approx \sqrt{\pi} \operatorname{erf}\left[v_c / \sigma_v\right]/(m_a \sigma_v v_c)$ where $v_c$ is the boost velocity and $\sigma_v$ the one-dimensional velocity dispersion~\cite{Cheong:2024ose}. Computing this with the SHM's parameter values (see Model 0 below) gives us a factor of $T^{\rm SHM}_{\rm coh}/T_{\rm ax} \approx (0.57$--$0.62)\times 10^6$ instead of $10^6$ in Eq.(\ref{eq:coherencetime}) depending on the time of year.} the notion of the coherence time, as in Ref.~\cite{Cheong:2024ose}, however, for the discussion here, we only need to define a typical timescale for an observation above which the velocity distribution should become important. Since our point of comparison will be the conventionally assumed Standard Halo Model (SHM), we will define the coherence time to be the heuristic value usually quoted in the literature, which is,
\begin{equation}\label{eq:coherencetime}
    T^{\rm SHM}_{\rm coh} \equiv 10^6 T_{\rm ax} = 41 \, \upmu{\rm s} \left(\frac{100\,\upmu {\rm eV}}{m_a} \right) \, .
\end{equation}

\begin{figure}
    \centering
    \includegraphics[width=0.99\linewidth]{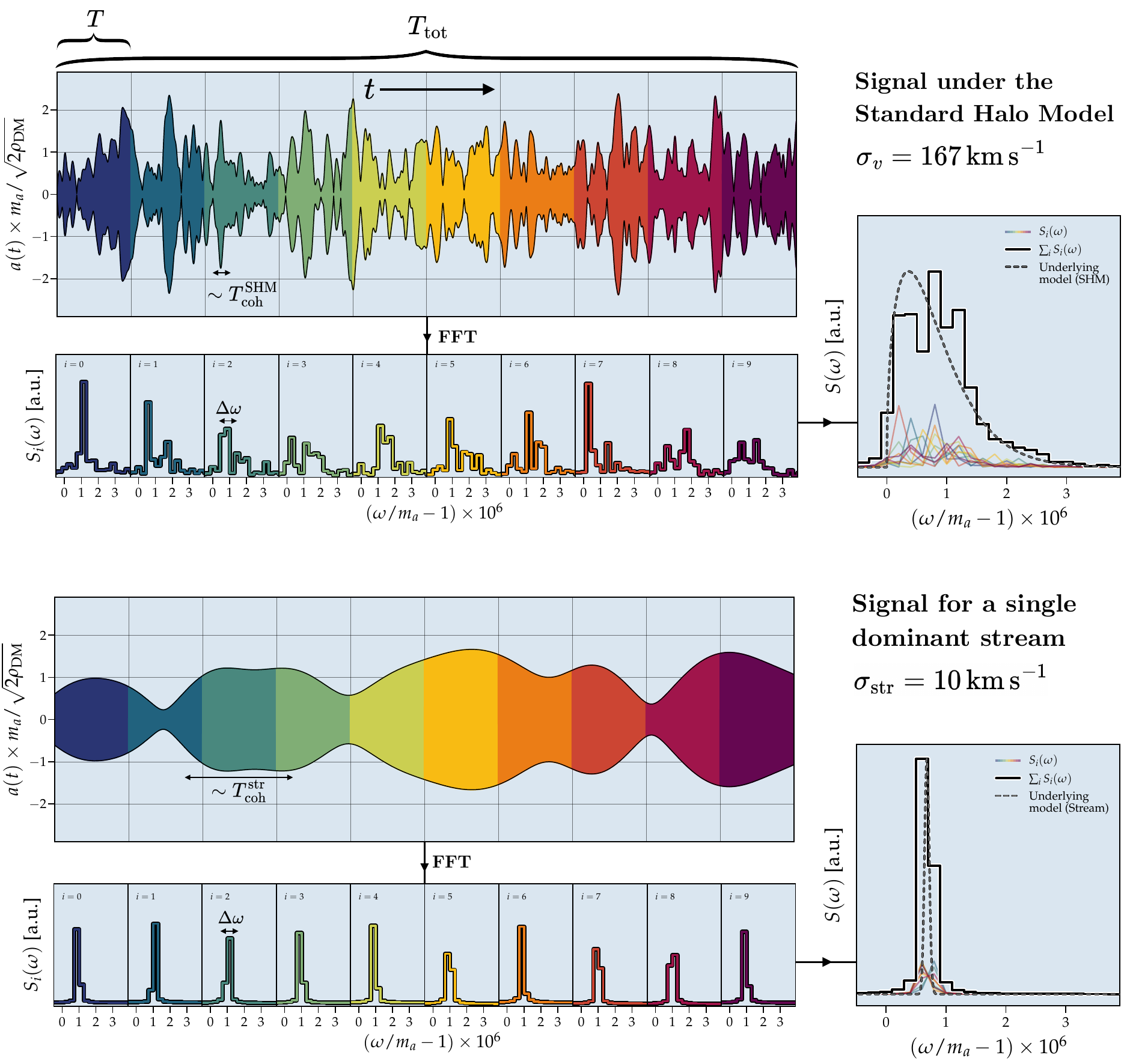}
    \caption{Illustration of the axion field evolution, and the definition of various timescales relevant to this study. The axion field oscillates very rapidly with a period of $\sim 2\pi/m_a$, which is captured by the coloured shading. The amplitude of those oscillations varies over the longer timescale of $T^{\rm SHM}_{\rm coh} \sim 10^6\times 2\pi/m_a$, as can be seen in the panels showing $a(t)$, which we normalise by the time-averaged oscillation amplitude, $\sqrt{2\rho_{\rm DM}}/m_a$. The axion \textit{lineshape}, $S(\omega)$, is measured by taking a Fourier transform of this oscillating signal over some integration time $T$. The frequency resolution of $S(\omega)$ is $\Delta\omega = 2\pi/T$. A full power spectrum for the signal observed over some longer time $T_{\rm tot}$ can then be obtained by summing the individual power spectra, $S(\omega) = \sum_i S_i(\omega)$. The figure is shown twice for two extreme assumptions for the underlying velocity distribution: the Standard Halo Model (SHM) and a distribution composed of a single stream with dispersion $\sigma_{\rm str} = 10$~km~s$^{-1}$. One can see that the presence of a kinematically cold feature like a stream effectively extends the coherence time $T_{\rm coh}^{\rm str}/T_{\rm coh}^{\rm SHM} \sim \sigma_v/\sigma_{\rm str}$.}
    \label{fig:diagram}
\end{figure}

Next, we must introduce the local speed distribution, $f(v)$. To simplify the sum, axions with similar speeds---i.e.~$v_i$ falling within some narrow speed bin with width $\Delta v$---can be grouped together so that the axion field is expressed as a sum over velocity bins,
\begin{equation}
    a(t)=\frac{\sqrt{\rho_{\mathrm{DM}}}}{m_a} \sum_j \alpha_j \sqrt{f\left(v_j\right) \Delta v} \cos \left[m_a\left(1+\frac{1}{2}v_j^2\right) t+\phi_j\right] \, ,
\end{equation}
where $\phi_j$ is another randomly drawn phase, and the sum over $j$ extends across the bins in speed between $v=0$ to $v=v_{\rm max}$. We take the latter to be $v_{\rm max} = 800$~km/s to be in the ballpark of the particles travelling at the galactic escape speed receiving the largest possible boost into the laboratory rest-frame. All reasonable choices for $f(v)$ are exponentially suppressed at large $v$, so the exact choice of $v_{\rm max}$ has a negligible impact on our results as long as it is larger than around three times the Earth's galactic velocity of $\sim 250$~km/s. The random amplitude $\alpha_j$ is drawn from a Rayleigh distribution,
\begin{equation}
    P\left[\alpha_j\right]=\alpha_j e^{-\alpha_j^2 / 2} \, ,
\end{equation}
which follows directly from the assumption that the $\phi_i$ phases are uncorrelated. We then consider making measurements of the axion field value at regular intervals in time $t_n = n\Delta t$, where $n = 0,..., T/\Delta t$ for some integration time $T$. For the cases we are interested in here (axion haloscopes probing the axion-photon coupling), we express these measurements as,
\begin{equation}
    y(t_n) = \sqrt{\mathcal{A}} \sum_j \alpha_j \sqrt{f\left(v_j\right) \Delta v} \cos \left[m_a\left(1+\frac{1}{2}v_j^2\right) t_n +\phi_j\right] \, .
\end{equation}
The only assumption we have made is that the experimental signal depends linearly on the axion field value.\footnote{Counter-examples to this include experiments searching for interactions that depend quadratically on the axion field~\cite{Kim:2022ype, Beadle:2023flm, Kim:2023pvt}, or axion-fermion couplings which depend on $\nabla a$~\cite{Gramolin:2021mqv, Berlin:2023ubt, Lisanti:2021vij, Foster:2023bxl, JacksonKimball:2017elr, Garcon:2019inh, Lee:2022vvb, Gao:2022nuq}. We defer an exploration of these cases to future studies.} The generic quantity $\mathcal{A}$, discussed in more detail below, encapsulates all constants of proportionality and their relevant dimensions needed to calculate the amplitude of whatever experimental signal is measured. This will include the axion-photon coupling $g_{a\gamma}^2$, the parameters associated with the experiment, and it will also absorb the dependence on the axion field amplitude $\sqrt{\rho_{\rm DM}}/m_a$.

We are then interested in the discrete Fourier transform of the timestream data $y(t_n)$, or more specifically, the power spectrum of that Fourier transform as a function of frequency $\omega$, defined as,
\begin{align}
    S(\omega)&\equiv\frac{\Delta t^2}{T} \bigg| \sum_{n=0}^{N-1} y(t_n) e^{-i \omega t_n} \bigg|^2 \, \\
   &=  {\cal A} \bigg|\, \sum_j \sum_n \alpha_j \sqrt{\frac{f\left(v_j\right) \Delta v}{T}}
\times \Delta t \cos \left[\omega_j n \Delta t+\phi_j\right] e^{-i \omega n \Delta t}\bigg|^2.
\end{align}
In earlier studies~\cite{Foster:2017hbq}, the procedure taken from this point was to work in the limit $T\to \infty$---which is to say that $T$ can be taken to be much longer than the other relevant timescale in the problem, $T_{\rm coh}$. In this limit, the term after the $\times$ sign in the equation above becomes a Dirac delta function, $\delta(\omega_j - \omega)$, with $\omega_j = m_a(1+v_j^2/2)$. This delta function singles out a single frequency contribution from the sum over $j$ and leads to the well-known relationship between the signal lineshape and the speed distribution: $S(\omega) \propto f(v)/v$. 

However, this limit is inappropriate for the cases we are interested in here. Our signal lineshape will consist of narrow features in frequency-space which, as shown in Fig.~\ref{fig:diagram}, can be thought of as features with very long coherence times.\footnote{For clarity, we will not use the language of coherence times to define these features, instead keeping the discussion focused around the velocity distribution, which remains well-defined even in the case where there are multiple distinct components of the dark matter distribution in phase space.} Another way of saying this is that we want to account for the case where our frequency bin width $\Delta \omega = 2\pi/T$ may be of comparable size or wider than the width of a particular feature in $S(\omega)$. With that in mind, the terms inside the sum can be massaged to obtain,
\begin{equation}
\begin{aligned}
 \Delta t \sum_{n=0}^{N-1} \cos \left[\omega_v n  \Delta t+\phi_j\right] e^{-i \omega n  \Delta t} 
=  \frac{ \Delta t}{2}\left(e^{i \phi_j} \frac{1-\exp \left[i\left(\omega_v-\omega\right) T\right]}{1-\exp \left[i\left(\omega_v-\omega\right)  \Delta t\right]}\right. 
 \left.+e^{-i \phi_j} \frac{1-\exp \left[-i\left(\omega_v+\omega\right) T\right]}{1-\exp \left[-i\left(\omega_v+\omega\right)  \Delta t\right]}\right) \\
\approx  \frac{e^{i\left(\phi_j+\left(\omega_v-\omega\right) T / 2\right)}}{2}\left(\frac{\sin \left[\frac{1}{2}\left(\omega_v-\omega\right) T\right]}{\frac{1}{2}\left(\omega_v-\omega\right)}\right. 
 \left.+e^{-i\left(2 \phi_j+\omega_v T\right)} \frac{\sin \left[\frac{1}{2}\left(\omega_v+\omega\right) T\right]}{\frac{1}{2}\left(\omega_v+\omega\right)}\right) \, .
 \end{aligned}
\end{equation}
The terms involving $\omega_v + \omega$ relate to negative values of $\omega_v$, which are not present here so we discard them. The expression can be further simplified if we assume that $(\omega_v-\omega)\Delta t \ll 1$, which will be the case here. To see why, consider the largest relevant values of $\omega_v -\omega$, which will be on the order of the width of the axion lineshape: $\omega_v -\omega \sim m_a v^2 \sim 10^{-6} m_a$. So this regime is appropriate as long as we consider experiments taking measurements at a rate faster than once per million axion oscillation cycles. Making this approximation, we get,
\begin{equation}
     \Delta t \sum_{n=0}^{N-1} \cos \left[\omega_v n  \Delta t+\phi_j\right] e^{-i \omega n  \Delta t}  \approx \frac{1}{2}e^{i\left(\phi_j+\left(\omega_v-\omega\right) T / 2\right)} \frac{\sin \left[\frac{1}{2}\left(\omega_v-\omega\right) T\right]}{\frac{1}{2}\left(\omega_v-\omega\right)} \, ,
\end{equation}
giving us,
\begin{equation}
\begin{aligned}
    S(\omega) &=  \mathcal{A} \left| \sum_j \alpha_j \sqrt{\frac{f\left(v_j\right) \Delta v}{T}}
\times  \frac{1}{2} e^{i\left(\phi_j+\left(\omega_v-\omega\right) T / 2\right)} \frac{\sin \left[\frac{1}{2}\left(\omega_v-\omega\right) T\right]}{\frac{1}{2}\left(\omega_v-\omega\right)}\right|^2 \\
 &=  \frac{\mathcal{A}T}{4} \left| \sum_j \alpha_j e^{i\phi_j} \sqrt{f(v_j)\Delta v}\, {\rm sinc}\left( \frac{1}{2}(\omega_v - \omega)T\right)\right|^2 \, ,
\end{aligned}
\end{equation}
where we redefine $(\phi_j + (\omega_v - \omega)T/2) \to \phi_j$ since both quantities are randomly drawn phases from a uniform distribution. Evaluating this for a set of randomly drawn $\alpha_j$ and $\phi_j$ will give one possible realisation of the axion lineshape $S(\omega)$, where $\omega$ must be taken at intervals defined by $\omega_k = 2\pi k /T$ for some integer $k$. This means that the value of $S(\omega)$ contains the power in the field in the frequency range $[\omega - \Delta \omega/2,\omega+\Delta \omega/2]$ where $\Delta \omega = 2\pi/T$ is the frequency resolution.

Due to the dependence on $|\sum_j\alpha_j|^2$, the values of $S(\omega)$ at a given $\omega$ will be exponentially distributed with mean,
\begin{equation}
    \langle S(\omega) \rangle = \frac{\mathcal{A}T}{2} \sum_j \left[ \sqrt{f(v_j) \Delta v} \,{\rm sinc}\left( \frac{1}{2}(\omega_v - \omega)T\right) \right]^2 \, .
\end{equation}
Since we introduced $\Delta v$ to be some arbitrarily small binwidth for grouping an arbitrarily large number of axion waves together, we are free to express this sum as an integral,
\begin{equation}\label{eq:signal_final}
    \langle S(\omega) \rangle = \frac{\mathcal{A}T}{2} \int_0^\infty {\rm d}v f(v)\, {\rm sinc}^2\left( \frac{1}{2}(\omega_v - \omega)T\right),
\end{equation}
where we define $\omega_v = m_a(1+v^2/2)$ as a shorthand for the frequency corresponding to a particular speed $v$. The intuition for this expression follows from the definition of the discrete Fourier transform: since we are splitting the power in the signal into equal-sized bins in frequency, the signal in each bin is then the convolution of whatever the signal's distribution in frequency is (dictated by $f(v)$) with a sinc$^2$ window function (which is the squared Fourier transform of a top-hat function). 

This more general form for the signal can be compared to the result used in Ref.~\cite{Foster:2017hbq}, which worked in the limit where $T \to \infty$, 
\begin{equation}\label{eq:signal0}
    \left\langle S (\omega)\right\rangle_{T\to\infty}=\left.\mathcal{A} \frac{\pi f(v)}{m_a v}\right|_{v=\sqrt{2 \omega / m_a-2}}.
\end{equation}
This result straightforwardly follows from Eq.(\ref{eq:signal_final}) in that same limit because the sinc$^2$ term becomes a delta function that singles out the speed corresponding to $\omega_v$. However, as we mentioned above, this simplified expression is not appropriate for our study since we wish to explore cases where the width of some component of the signal may be comparable to, or narrower than, the frequency bin width. In appendix~\ref{app:lineshape}, we show comparisons between the different calculations of the signal lineshape, as well as an analytic approximation to Eq.(\ref{eq:signal_final}) that greatly accelerates its computation at the cost of some accuracy.
 
On top of the measured signal $S(\omega)$, there will be experimental noise. We make the simplifying assumption here that noise is exponentially distributed and frequency independent, with some mean value $\mathcal{B}$,
\begin{equation}
    \langle \mathcal{B}(\omega)\rangle \equiv \mathcal{B}.
\end{equation}
As shown in Ref.~\cite{Foster:2017hbq}, the value of the combined signal+background, $S(\omega) + \mathcal{B}(\omega)$ power, is also exponentially distributed because the two are summed at the level of the timestream data, not at the level of the power spectrum. The assumption of an exponential background is justified given the expected behaviour of thermal noise in existing haloscopes which follows a Gaussian distribution in the time domain~\cite{ADMX:2024pxg, Salemi:2021gck, HAYSTAC:2018rwy} (see also the discussion in Appendix A of~\cite{Foster:2017hbq}); while the assumption of frequency-independent noise is appropriate as long as we can expect there to be no large variations in the noise over the extremely narrow frequency window containing the signal which has a bandwidth of $\sim (m_a,3\times 10^{-6} m_a)$. In any case, departures from these assumptions are straightforward to model by defining some function $\mathcal{B}(\omega)$, but for simplicity and to keep our discussion more general, we choose to fix only one background parameter, $\mathcal{B}$.

\subsection{Velocity distributions}\label{sec:fv}
Equation~\eqref{eq:signal_final} shows how the mean signal lineshape $\langle S(\omega)\rangle$ is related to the assumed dark matter speed distribution $f(v)$, which in turn comes from the velocity distribution via an integral over all incident angles: $f(v) = \int v^2 f(\mathbf{v}) {\rm d}\Omega$. Having established this relationship, we will now write down several models for $f(\mathbf{v})$ that capture various possible levels of substructure that could plausibly be present in the local velocity distribution.

To avoid overcomplicating our velocity distribution (both in terms of the discussion here as well as its computation), we choose to build up our more detailed models out of Gaussian components. To that end, we first define a generic Gaussian velocity distribution with two free parameters, a central velocity $\mathbf{v}_c$ and a one-dimensional isotropic dispersion $\sigma_v$,
\begin{equation}
    f(\mathbf{v}; \mathbf{v}_c,\sigma_v) = \frac{1}{(2\pi \sigma_v)^{3/2}} \exp \left(-\frac{|\mathbf{v} - \mathbf{v}_c|^2}{2\sigma_v^2} \right).
\end{equation}
This is a simple building block that can be used to construct much more complicated velocity distributions while remaining analytically tractable, and avoids needing to describe features such as the high-speed cut-off due to the escape velocity, anisotropies, or deviations from Gaussianity, none of which are easily measurable from the axion lineshape alone. The speed distribution associated with this velocity distribution is,
\begin{equation}\label{eq:gaussian_fv}
    f(v \,;  \mathbf{v}_c,\sigma_v) = \frac{v}{\sqrt{2\pi} v_c \sigma_v} \left(e^{-(v^2 + v_c^2 - 2 v v_e)/2\sigma_v^2} - e^{-(v^2 + v_c^2 + 2 v v_e)/2\sigma_v^2} \right) \, .
\end{equation}
which is what we insert into Eq.(\ref{eq:signal_final}). We will build up more complicated models by summing together multiple speed distributions of this form with different parameter values. Note that $\langle S(\omega)\rangle$ is linear in $f(v)$, so we can sum these components also at the level of the power spectrum. See appendix~\ref{app:lineshape} for approximate analytic results for $\langle S(\omega)\rangle$ for this general form for the velocity distribution. 

\begin{figure}
    \centering
    \includegraphics[width=0.99\linewidth]{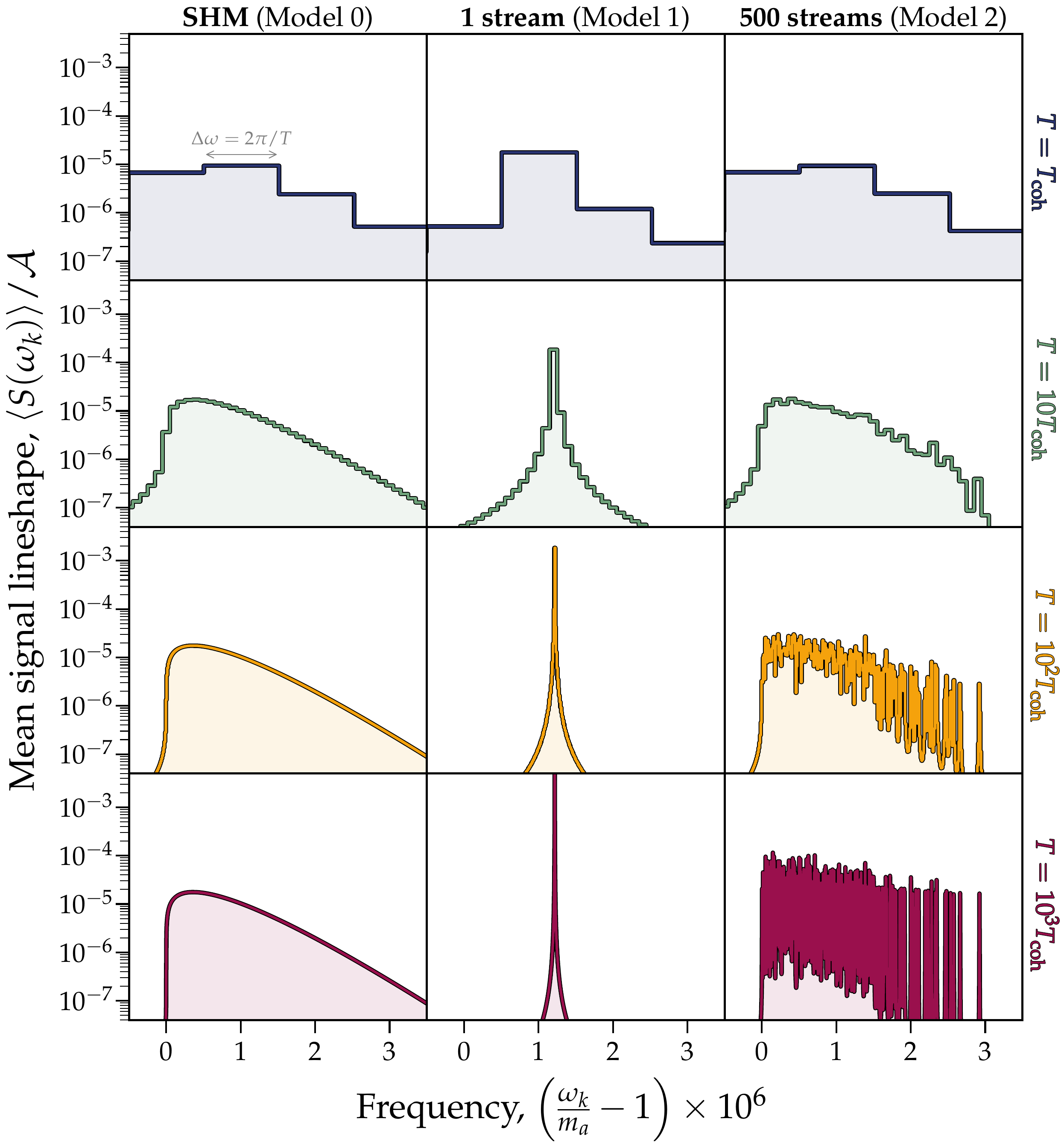}
    \caption{Mean power spectral densities normalised by the signal amplitude $\mathcal{A}$, for some example models introduced in this section. Reading the columns horizontally, we compare three model lineshapes: the Standard Halo Model (SHM), a single stream with $\sigma_{\rm str} = 1$~km/s, and a model composed of 500 streams with $\sigma_{\rm str} = 1$~km/s. Reading from top to bottom, each row compares the lineshapes for increasing total integration time $T$ in units of the axion coherence time under the SHM, $T^{\rm SHM}_{\rm coh} = 10^6 T_{\rm ax}$. Note that the fact that the power spectrum is given by the convolution of the dark matter speed distribution with a sinc$^2$ function [Eq.(\ref{eq:signal_final})] causes power to appear at frequencies smaller than the axion mass in the case where the frequency resolution is comparable to the width of the signal in frequency space.}
    \label{fig:Distributions_vs_Model}
\end{figure}

We now list several concrete models, each of which is constructed out of the velocity distribution in (\ref{eq:gaussian_fv}):
\begin{itemize}
    \item {\bf Model 0: The Standard Halo Model}. Our zeroth-order baseline model is the velocity distribution associated with the Standard Halo Model (SHM). The SHM is an isothermal sphere with a $\rho(r)\sim r^{-2}$ density profile (or a logarithmic potential). The distribution function that is consistent with this potential can be found by solving Poisson's equation. It is defined by a single temperature (i.e.~velocity dispersion, $\sigma_v$), that is also tied to the radius-independent speed of a circular orbit $v_{\rm circ} = \sqrt{2}\sigma_v$, which is a feature of the logarithmic potential. This is the conventionally assumed lineshape for direct dark-matter detection analyses. The velocity distribution is written in terms of our generic Gaussian as,
    \begin{equation}
        f_0(v) = f(v ;-\mathbf{v}_{\rm lab},167 \, {\rm km/s}) \, ,
    \end{equation}
    where we centre the distribution on $\mathbf{v}_c = -\mathbf{v}_{\rm lab}$ due to the fact that we observe dark matter after a boost into the laboratory frame which is travelling at a velocity $\mathbf{v}_{\rm lab}$ with respect to the galactic centre.
    \item {\bf Model 1: Single stream}. This is a simple example which we will use to build intuition for our later results, although it is not expected to be a very realistic model for the local dark matter velocity distribution. It consists of the same distribution function as the SHM but with $\mathbf{v}_c \to \mathbf{v}_{\rm str} - \mathbf{v}_{\rm lab}$ and $\sigma_v \to \sigma_{\rm str}$.
    \begin{equation}
        f_1(v ;  \mathbf{v}_{\rm str},\sigma_{\rm str}) = f(v ; \mathbf{v}_{\rm str}-\mathbf{v}_{\rm lab}, \sigma_{\rm str}).
    \end{equation}
    The model has four free parameters: the components of the galactic-frame velocity $\mathbf{v}_{\rm str}$ and the stream dispersion $\sigma_{\rm str}$. We will be interested in cases where $\sigma_{\rm str} \ll 167 \, $ km/s.
    \item {\bf Model 2: Many equal-density streams}. This is the simplest extension of Model 2 that considers the entire distribution as being composed of some $N_{\rm str}$ streams of equal density and velocity dispersion,
    \begin{equation}
        f_2(v ;  N_{\rm str}, \sigma_{\rm str}) = \sum_{i=1}^{N_{\rm str}} \frac{1}{N_{\rm str}} f_1(v ;  \mathbf{v}^i_{\rm str},\sigma_{\rm str}) \, .
    \end{equation}
    Note that we do not set $\mathbf{v}^i_{\rm str}$ as free parameters for this model because we will assume that each stream's $\mathbf{v}^i_{\rm str}$ is drawn from the Standard Halo Model's velocity distribution, $f_0(v)$.
    \item {\bf Model 3: Stream ensemble}. This is a more complicated but more realistic extension of Model 2, which now allows each stream to have different constituent densities and velocity dispersions:
    \begin{equation}
        f_3\big(v ; P[\sigma_{\rm str},\rho_{\rm str}] \big)= \sum_i^{N_{\rm str}} \frac{\rho^i_{\rm str}}{\rho_{\rm DM}} f_1(v ;  \mathbf{v}^i_{\rm str},\sigma^i_{\rm str}).
    \end{equation}
    Here instead we no longer have a single set of free parameters but rather a probability distribution from which the values of $\sigma_{\rm str}$ and $\rho_{\rm str}$ are drawn. To define $P$, we will take inspiration from simulations. We will use this model to describe the lineshape from disrupted miniclusters, where $P[\sigma_{\rm str},\rho_{\rm str}]$ can be estimated semi-analytically from the minicluster mass function; and we will use it to create a model for the ultra-fine-grained substructure in the halo due to phase mixing. We will define these cases more concretely in Secs.~\ref{sec:miniclusters} and~\ref{sec:finegrained}, after we have gained some intuition by studying Models 0--2.
    
\end{itemize}
Some examples of the resulting mean signal power spectra, $\langle S(\omega)\rangle $ for Models 0, 1, and 2 are shown in Fig.~\ref{fig:Distributions_vs_Model}. We compare the signal shapes as a function of the integration time, $T$. In particular, this plot already reveals a key idea that will be important for the later discussion: notice that the multi-stream model for a large value of $N_{\rm str}$ is indistinguishable from the SHM at low-resolution (short $T$), but becomes drastically different once the frequency resolution, $\Delta \omega$, is comparable to the frequency spread of a given stream.

Before continuing, we point out that it is also possible that there will be broader features present in the velocity distribution due to the late infall and accretion of dark matter subhalos. These have been the subject of numerous studies in the literature e.g.~Refs.~\cite{Freese:2003tt, OHare:2014nxd, Foster:2017hbq, OHare:2018trr, Knirck:2018knd, OHare:2019qxc, Ko:2023gem}, but in the case of the axion, these structures are less relevant. Although the dark matter fractions due to accreted subhalos are not known precisely at present, they are unlikely to reach more than a few percent, and the typical velocity dispersions of accreted dark matter subhalos will be at the level of tens of km/s typical of dwarf galaxies, meaning they lead to very minor enhancements to the lineshape (we make this statement quantitative in some of our later results). So we neglect any discussion of these possible substructures here and focus on the \textit{fine-grained} structure of the halo, which is likely to be much more important and about which we can make more confident statements.

\subsection{Daily modulation}\label{sec:modulation}
\begin{figure}
    \centering
    \includegraphics[height=0.47\linewidth]{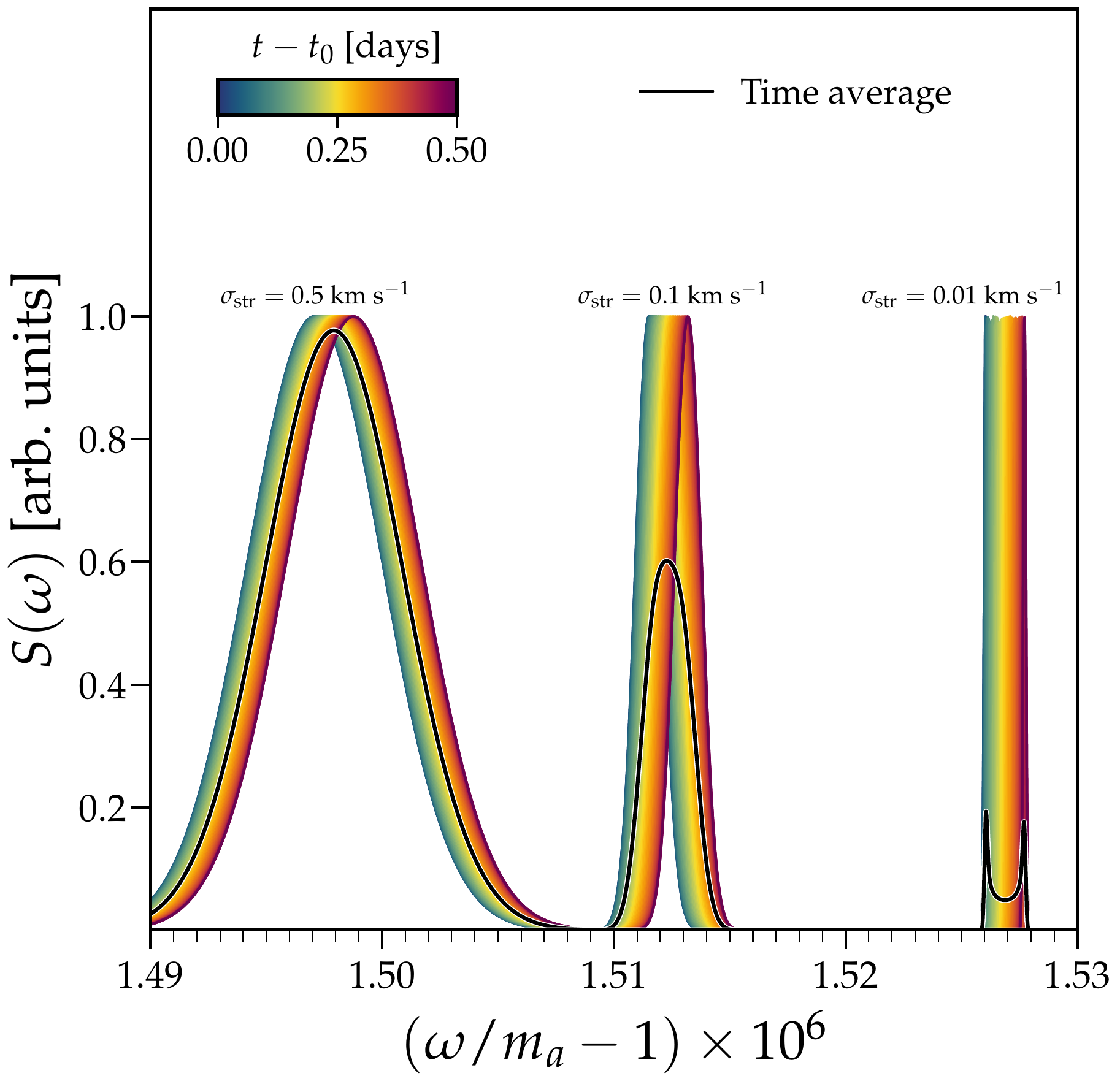}
    \includegraphics[height=0.47\linewidth]{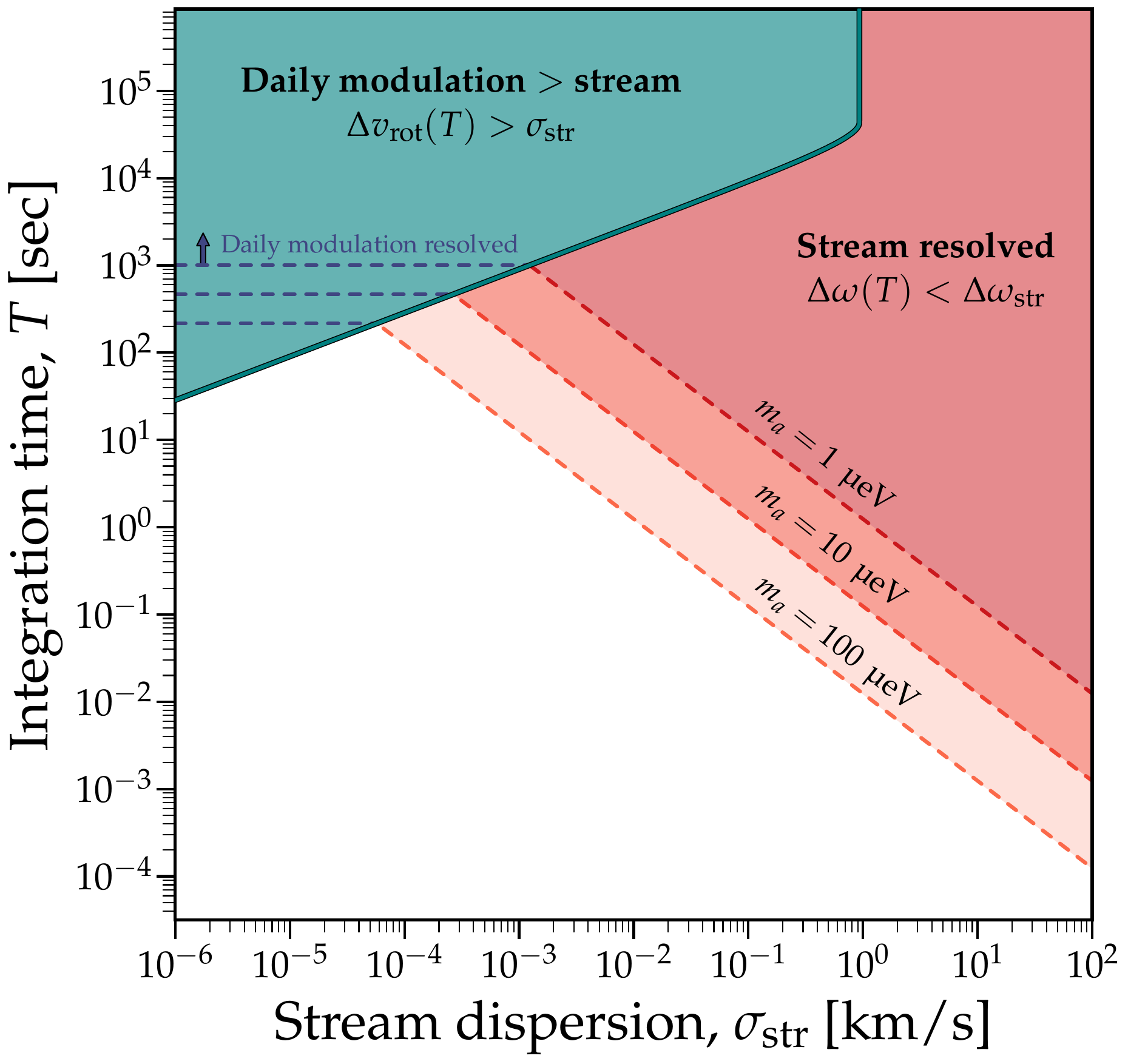}
    \caption{{\bf Left:} Impact of the daily modulation on the signal lineshape for an arbitrary stream with three possible dispersions $\sigma_{\rm str}$, demonstrating the severe impact of the modulation on very narrow streams. The black line shows the lineshape averaged over 0.5 days, which we see is suppressed when the modulation in the lab velocity (which is equal to $2v_{\rm rot}\approx 1$~km~s$^{-1}$ here) is much larger than the stream's dispersion. {\bf Right:} Parameter space over which daily modulations are expected to modify the axion lineshape. When the integration time $T$ and the velocity dispersion of a stream $\sigma_{\rm str}$ lie within the teal region, the daily modulation of the stream becomes larger than the width. On the other hand, only in the red regions (which depend on the axion mass) is the integration time long enough to resolve the stream to begin with. If we lie outside of these regions, we can safely neglect the daily modulation as an important effect; however, when inside both of them, we must account for it, as discussed in Sec.~\ref{sec:modulation}.}
    \label{fig:dailymodulation}
\end{figure}
The final detail of the axion signal we must account for in certain cases is its time dependence. This time dependence is a feature of all of our models because the velocity by which axions are boosted into the laboratory frame, $\mathbf{v}_{\rm lab}$, is time dependent. Specifically, there is an expected \textit{annual} modulation in the size of $\mathbf{v}_{\rm lab}$ of order $30$ km/s, due to the varying orbital velocity of the Earth around the Sun. We will not consider total measurement timescales longer than a few hours in this study and so we will neglect any further discussion of the annual modulation, suffice to say that it is easily observed in a post-discovery setting once an axion signal is repeatedly measured over the course of a year---doing so would provide a further observational handle on the velocity distribution, as studied in Ref.~\cite{OHare:2017yze}. Rather, our focus will be on the second modulation of around $\sim 0.5$~km/s, which is a \textit{sidereal} variation in the speed distribution due to the rotational motion of the laboratory about the Earth's axis. 

Although the sidereal modulation is usually irrelevant for dark matter searches, it is important when considering extremely cold substructures where $\sigma_{\rm str}$ is very small. In particular, the sidereal modulation can become an $\mathcal{O}(1)$ effect at the level of $S(\omega)$ if the variation in the central velocity of a given substructure due to $\mathbf{v}_{\rm lab}(t)$ is larger than both the spectral resolution \textit{and} the width of the substructure in velocity, i.e.~when $\sigma_{\rm str}$ is small but $T$ is large. 

We now write all velocities in a detector-centred coordinate system with axes aligned along the North, West, and Zenith cardinal directions (we refer the interested reader to the appendix of Ref.~\cite{Mayet:2016zxu} for the relevant rotation matrices we use to transform between galactocentric and laboratory-fixed coordinate systems). The lab velocity can be written in the following form, which pulls out the daily modulating part:
\begin{equation}
    \mathbf{v}_{\rm lab}(t) = \mathbf{v}_{\rm lab}^0 - v_{\rm rot}\cos{\lambda}\cos{\bigg(\frac{2\pi t}{{\rm 1 \,day}}\bigg)}\hat{\mathcal{W}} \, ,
\end{equation}
where $\lambda$ is the latitude of the experiment on the Earth, $\hat{\mathcal{W}}$ is a unit vector pointing due West, and $v_{\rm rot} = 0.46$~km/s.

For concreteness, we will fix $\lambda = 45^\circ$ and set $\mathbf{v}_{\rm lab}^0 = (218, -119, 21)$~km/s to be the Earth's velocity on March 8, as this gives a speed very close to the annual average~\cite{Evans:2018bqy}. We have assumed for convenience that measurement times, $t$, begin from a time where the rotational velocity is a maximum, allowing us to neglect the additional temporal phase that would otherwise appear in the expression above. None of our results are affected by these choices.

We expect daily modulation to be important for substructures when,
\begin{equation}\label{eq:dailymod_cond1}
    {\rm max}\left[v_{\rm rot}(1 - \cos(2\pi t/{\rm 1 day})\right] \gtrsim \sigma_{\rm str}
\end{equation}
where $t \in [0,T]$. The largest the left-hand side of this inequality can be is $2 v_{\rm rot}$ when $T = 0.5$~days. This can be compared to the value of $T$ required to resolve the stream in the lineshape, i.e.
\begin{equation}\label{eq:dailymod_cond2}
    2\pi/T \lesssim m_a v_{\rm str} \sigma_{\rm str} \, .
\end{equation}
The effect on a stream's lineshape when the first of these two conditions is not met is to broaden it, such that it has a width $\Delta v_{\rm rot}(T)\sim v_{\rm rot}(1-\cos(2\pi T/{\rm 1\,day}))$. This is demonstrated in the left-hand panel of Fig.~\ref{fig:dailymodulation}, which shows how the daily modulation affects the lineshape for three representative values of $\sigma_{\rm str}$. As can be seen in the narrowest stream case, the daily modulation effect dominates over the underlying lineshape. So when we integrate the distribution over the measurement time $T$, we need to account for the varying $v_{\rm lab}(t)$, which in turn requires that we transform all of our stream velocities $v_{\rm str}$ from the galactic coordinate system in which they are defined, into the laboratory coordinate system~\cite{Mayet:2016zxu, Knirck:2018knd}.

In the right-hand panel of Fig.~\ref{fig:dailymodulation}, we show various regimes in the space of values of $T$ and $\sigma_{\rm str}$ where the daily modulation is important. Since we have introduced a new timescale into the problem, there is now a dependence on the axion mass we need to keep track of when showing these regimes as a function of $T$. Smaller axion masses have a longer $T^{\rm SHM}_{\rm coh}$ and so require larger values of $T$ for either the stream or the daily modulation to be resolved, as can be seen in Fig.~\ref{fig:dailymodulation}.

Note that in our later results, we will rarely consider durations longer than a few hours. That said, accounting for the daily modulation correction will become essential when we consider the fine-grained streams discussed in Sec.~\ref{sec:finegrained} whose velocity dispersions are effectively negligible compared to $\Delta v_{\rm rot}(T)$.

\section{Statistical tests}\label{sec:stats}
Having now introduced several models for fine-grained dark matter substructure, we must now develop some statistical methodology to test the extent to which these models influence a haloscope's sensitivity to the axion. We will explore two possible methodologies in this work, one of which requires that we define a model to test the data against, and a second, non-parametric approach, which would not require any model assumptions and mimics the strategy of identifying and re-scanning high-significance power excesses, which is adopted by many resonant haloscope experiments. The parametric test we adopt is the powerful profile likelihood ratio test, which we expect to outperform the non-parametric test in terms of sensitivity. However, as we will discuss further below, some of the models we are entertaining here can come with an unmanageable number of free parameters, so given the large degree of uncertainty in the exact realisation of a multi-stream distribution that might be present in the vicinity of the Earth, a non-parametric test may be more robust.

\subsection{Profile Likelihood Ratio test}
We begin by constructing a likelihood function for a set of data given some model that is expected to describe that data. We assume that the collected data consists of Fourier-transformed time domain signal, resulting in a power spectrum $P_{\rm obs}(\omega_k)$ where $\omega_k = 2 \pi k/T$ are the central values of the frequency bins, which are separated by $\Delta \omega = 2\pi/T$ where $T$ is the duration of the timestream used to construct the power spectrum, which we will refer to as the \emph{integration} time. We assume that the signal being examined is contained within frequency bins, $k_i$ and $k_f$. Note that our signal model Eq.(\ref{eq:signal_final}) leads to the presence of non-zero signal amplitude in frequencies $\omega<m_a$ and $\omega>(m_a + v_{\rm max}^2/2)$, due to the convolution with the sinc$^2$ function. This can be observed in Fig.~\ref{fig:Distributions_vs_Model}. Although the amplitude in those bins is highly suppressed, to ensure that we do not miss too much of the signal power when $T$ is small, we choose $k_i$ and $k_f$ to lie at values of $\omega$ that are twenty frequency bins smaller and greater than the expected minimum and maximum frequencies, respectively.

Given that we anticipate some experiments will require long measurement times relative to the timescale $T_{\rm ax}$, data storage limitations will require breaking down these measurements to create and then stack together multiple lower-resolution power spectra. So we will set $T_{\rm tot}$ to be the total measurement time, to be then broken down into $N_{\rm PS} = T_{\rm tot}/T$ individual power spectra from samples of duration $T$. We label the central time for each of these power spectra as $t_j$ and compute the signal for that time bin as the average,
\begin{equation}
    \langle S(t_j,\omega_k)\rangle = \frac{1}{T} \int_{t_j- T/2}^{t_j+T/2} \textrm{d}t~\langle S(t,\omega_k)\rangle \, ,
\end{equation}
where we make explicit the time-dependence in $S(\omega)$ due to the time-dependent boost velocity present in the speed distribution, ~$\mathbf{v}_c(t)$ as discussed in Sec.~\ref{sec:modulation}. The full observed data is then the value of the power spectral density across all time and frequency bins, $\mathbf{d} = \{ P_{\rm obs}(t_j,\omega_k)\}$. Given that the signal+background power is exponentially distributed in each time and frequency bin, the likelihood for this data given some model $\mathcal{M}$ with parameters $\boldsymbol{\theta} = \{\mathcal{A},\mathcal{B}\}$ is then:
\begin{equation}
    \mathcal{L}(\mathbf{d}|\mathcal{M}, \boldsymbol{\theta})=\prod_{j = 1}^{N_{\rm PS}}\prod_{k=k_i}^{k_f} \frac{1}{P_{\rm exp}(t_j,\omega_k;\boldsymbol{\theta})} e^{-P_{\rm obs}(\omega_k)/ P_{\rm exp}(t_j,\omega_k;\boldsymbol{\theta})}.
\end{equation}
The second product runs over frequency bins $\omega_k = 2\pi k/T$ between some initial $k_i$ and some final $k_f$, which are chosen to lie within the vicinity of the axion frequency being tested. The expected signal+background power is
\begin{equation}
    P_{\rm exp}(t_j,\omega_k;\mathcal{A},\mathcal{B}) = \langle S(t_j,\omega_k;\mathcal{A}) \rangle + \mathcal{B}.
\end{equation}

\noindent We can now construct the profile likelihood ratio (PLR) which compares the maximum likelihood under the signal+background model $\mathcal{M}_{S+B}$ with the background-only model $\mathcal{M}_B$. The former has parameters $\{\mathcal{A},\mathcal{B}\}$ and the latter only $\mathcal{B}$,
\begin{equation}
\Lambda\left(\mathcal{A}\right)=2\left[\ln \mathcal{L}\left(\mathbf{d} \mid \mathcal{M}_{S+B},\left\{\mathcal{A}, \hat{\mathcal{B}}\right\}\right)-\ln \mathcal{L}\left(\mathbf{d} \mid \mathcal{M}_B, \hat{\mathcal{B}}\right)\right],
\end{equation}
where $\hat{\mathcal{B}}$ is the value of $\mathcal{B}$ that maximises each likelihood under their respective model assumptions---the former is the signal+background model at a fixed value of $\mathcal{A}$ and the second is the background-only model where $\mathcal{A}=0$.
We then define the test statistic for discovery (quantifying the evidence in favour of the signal+background hypothesis when tested against the background-only hypothesis),
\begin{equation}
    {\rm TS} = \Lambda(\hat{\mathcal{A}}),
\end{equation}
where $\hat{\mathcal{A}}$ is the value of $\mathcal{A}$ that maximises $\Lambda$.
Given that the null hypothesis is a special case of the alternative hypothesis under the condition $\mathcal{A}=0$, Wilk's theorem holds (or Chernoff's theorem~\cite{Chernoff:1954eli, Algeri:2019arh} if $\mathcal{A}$ is chosen to be strictly positive). This implies that $\Lambda$ follows a $\chi_1^2$ distribution and so the local significance that signal discovery can be claimed is then simply $\sqrt{{\rm TS}}$. We note that this is the \textit{local} significance, and for experiments that are searching for candidate axion signals over some large number of frequency bins, the required significance value will necessarily need to be amplified according to the number of independent local tests to account for the look-elsewhere effect, as discussed in Ref.~\cite{Foster:2017hbq}. Since this number will depend strongly on the scan strategy adopted by a given experiment, in the discussion in this section we will simply fix a particular threshold of $\sqrt{{\rm TS}} = 5$ for a 5$\sigma$ local discovery significance as a general benchmark to fix the amplitude of our example signals to a level that is generally expected to be detectable. Adjusting this value will enable us to invert these expressions and determine the required values of $\mathcal{A}/\mathcal{B}$ and $T$ needed to generate a detectable signal at that level of local significance. Since our goal is to compare models of the lineshape, we are more interested in ratios of values of $\mathcal{A}/\mathcal{B}$ and $T$ for different models rather than their absolute values, so this arbitrary fiducial choice for the required significance is not critical to the present discussion.

Inputting now our functional forms for the signal and background lineshape models, we have for the likelihood ratio:
\begin{equation}
    \Lambda\left(\mathcal{A}\right)=2 \sum_{j=1}^{N_{\rm PS}}\sum_{k=k_i}^{k_f}\left[P_{\rm obs}(t_j,\omega_k)\left(\frac{1}{\mathcal{B}}-\frac{1}{P_{\rm exp}(t_j,\omega_k;\mathcal{A},\mathcal{B})}\right)-\ln \frac{P_{\rm exp}(t_j,\omega_k;\mathcal{A},\mathcal{B})}{\mathcal{B}}\right].
\end{equation}
We have simplified this expression by noticing that the MLEs for $\mathcal{B}$ can be approximated as e.g.~$\hat{\mathcal{B}} = \sum_k \sum_j P_{\rm obs}(t_j,\omega_k)/N_{\rm PS}N_k \approx \mathcal{B}$ under $\mathcal{M}_B$, for example. This holds in all the cases we consider here because the noise level will dominate in (almost) every frequency bin: $S(t_j,\omega_k)\ll \mathcal{B}$. This remains true even in our multi-stream examples because, although streams may appear above the noise in a handful of bins, the \textit{fraction} of the total power contained in these bins is very small---this is because the large power enhancements from the streams can only emerge when the number of bins is very large.

A particularly informative mock dataset to consider now is that of the ``Asimov dataset'' which is engineered to exactly match the expectation in every bin for a chosen set of `true' parameters, i.e.~we set $P_{\rm obs}(t,\omega) = P_{\rm exp}(t,\omega;\mathcal{A}_{\rm true},\mathcal{B}_{\rm true})$. In the case of a discovery of a signal with signal amplitude $\mathcal{A}_{\rm true}$, we have that $\hat{\mathcal{A}} = \mathcal{A}_{\rm true}$, and so the discovery test statistic under the Asimov dataset is,
\begin{equation}\label{eq:TS_asimov}
    \widetilde{{\rm TS}} = 2 \sum_{j=1}^{N_{\rm PS}}\sum_{k=k_i}^{k_f}\left[\left(\frac{P_{\rm exp}(\omega_k;\mathcal{A},\mathcal{B})}{\mathcal{B}}-1\right)-\ln \frac{P_{\rm exp}(t_j,\omega_k;\mathcal{A},\mathcal{B})}{\mathcal{B}}\right] \, .
\end{equation}
Values of statistics under the Asimov dataset will asymptote towards the median of their distribution in the limit of large numbers of data points, making it a useful test case that circumvents the need for extensive Monte Carlo simulations that would usually be needed to find the distribution of TS. To find the amplitude $\mathcal{A}_{5\sigma}$ for which 50\% of experiments could obtain a (local) $5\sigma$-significance discovery or better, we must satisfy the equation,
\begin{equation}
    5 = \sqrt{\tilde{\Lambda}(\mathcal{A}_{5\sigma})} \, ,
\end{equation}
where the Asimov PLR statistic $\tilde{\Lambda}$ in this expression is the value evaluated for the dataset $P_{\rm obs} = P_{\rm exp}(\mathcal{A}_{5\sigma},\mathcal{B})$.

For reference in later sections, we will note in passing here that the same formalism can be used to derive expected upper limits by defining a set of Asimov data where $\mathcal{A}_{\rm true} = 0$. To estimate the median one-sided limit on $\mathcal{A}$ to some confidence level (CL) e.g.~95\%, one would then solve the equation for the exclusion test statistic, which has a similar but subtly different form,
\begin{equation}\label{eq:exclusion}
     \tilde{\Lambda}(\mathcal{A}_{95\%}) - \tilde{\Lambda}(0) = -[\Phi^{-1}(0.95)]^2 \simeq -2.71 \, ,
\end{equation}
where $\Phi^{-1}$ is the inverse cumulative distribution function of the Gaussian. What we are interested in here is the enhanced discovery potential of haloscopes due to fine-grained substructure, so for that reason, we will focus the discussion in this section on the discovery test statistic, TS. Due to the similarity in their expressions, most of the insights gained from our results will extend qualitatively to the strength of upper limits as well. That said, we caution that it is a dangerous practice to simply conjecture the existence of some arbitrarily prominent peak in the lineshape just to set stronger limits on the axion-photon coupling, $g_{a\gamma}$. The degree of substructure in the lineshape is completely speculative at present, whereas we \textit{do} have a priori expectations for the value of $g_{a\gamma}$ for the QCD axion as well as the \textit{total} dark matter density, $\rho_{\rm DM}$. So we would advocate that the appropriate strategy is for experimental collaborations to choose the most conservative option for the lineshape (the SHM) when reporting upper limits, while still leaving open the possibility of signal-enhancing substructure by running high-resolution analyses with the purpose of potentially encountering a candidate signal that can be confirmed after subsequent interrogation with more data. 

\subsubsection{Halo integrals}
\begin{figure}
    \centering
    \includegraphics[width=0.8\linewidth]{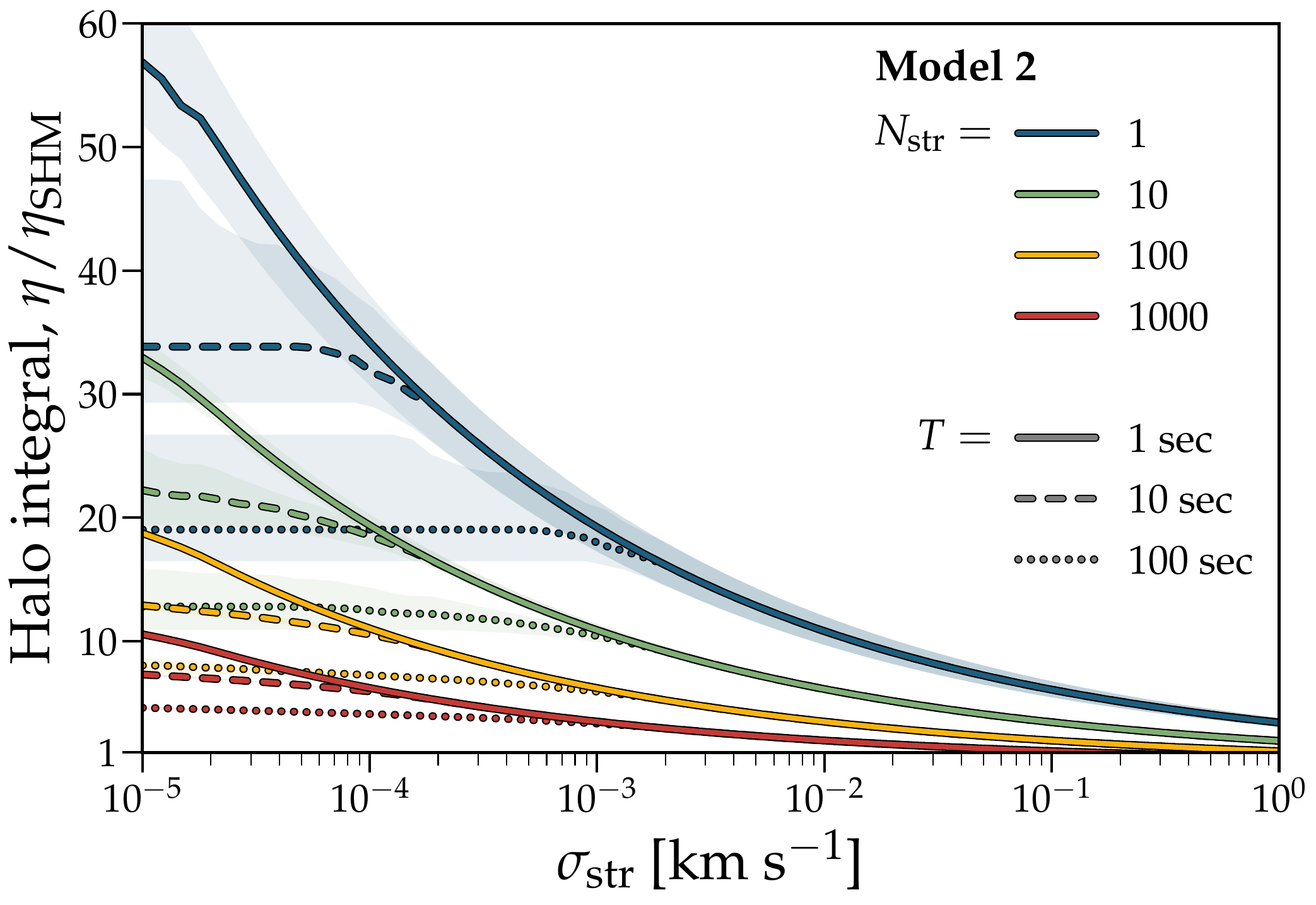}
    \caption{The halo integral, $\eta$---as defined in Eq.(\ref{eq:halointegral})---for various stream model lineshapes, relative to the value for the SHM, $\eta_{\rm SHM}\approx 27.7$. The halo integral quantifies how the sensitivity of a haloscope depends on the axion lineshape. For example, a value of $\eta/\eta_{\rm SHM} = 10$ would correspond to a lineshape that would give a factor of 10 enhancement in sensitivity to $g_{a\gamma}$ compared to the SHM. We show how this quantity depends on the number of streams present in the lineshape, each as a function of the velocity dispersion of the streams, $\sigma_{\rm str}$. The streams in each case are sampled randomly given the model parameters---the lines correspond to the median value of $\eta$ with the shaded region enclosing 68\% of possible lineshapes for that model. To highlight the effect of the daily modulation, we also plot the halo integral for three integration times. The daily modulation prevents the sensitivity enhancement from continuing to rise towards very small values of $\sigma_{\rm str}$, which is due to the effect highlighted in Fig.~\ref{fig:dailymodulation}.}
    \label{fig:HaloIntegrals}
\end{figure}
We have introduced the PLR test to demonstrate the impact that the velocity distribution has on the sensitivity of an experiment. But before evaluating the test numerically, we can gain some intuition for the impact of $f(v)$ by taking the limiting case where the discrete sum above can be turned into an integral. Let us assume $T$ is sufficiently long, or equivalently $\Delta 
\omega$ sufficiently small, so that there are negligible bin-to-bin variations due to the shape of $f(v)$. When this is the case, we can safely turn the $k$ sum into an integral over $v$. If we further assume\footnote{Note that this assumption is not valid in the cases where the daily modulation effect discussed in Sec.~\ref{sec:modulation} is important.} that $f(v)$ does not change substantially over the course of $T_{\rm tot}$, then we can simply replace the sum with a multiplicative factor of $N_{\rm PS} = T_{\rm tot}/T$. Applying these approximations results in the following form for the Asimov test statistic,
\begin{equation}\label{eq:TS_asimov_limit}
    \widetilde{{\rm TS}} \approx \left(\frac{\mathcal{A}}{\mathcal{B}}\right)^2\frac{T_{\rm tot} \pi}{2m_a} \int {\rm d}v \frac{f(v)^2}{v} \, ,
\end{equation}
which was the limiting case derived in Ref.~\cite{Foster:2017hbq}. The squared velocity distribution appearing here arises after expanding the logarithm in Eq.(\ref{eq:TS_asimov}), assuming that $\langle S(\omega)\rangle\ll \mathcal{B}$. Note that in subsequent results, we will always compute the full test statistic from Eq.\eqref{eq:TS_asimov}. This analytic form is helpful for gaining an intuition into how the sensitivity to axions depends on the assumed form for the local velocity distribution---which is the focus of our study.

We now see more clearly how the test statistic, which encapsulates the sensitivity of the experiment to a given model, depends upon the choice of lineshape. To give some insight into the role played by the velocity distribution, independent of other experimental factors like $\mathcal{A}$ and $T$, we first show in Fig.~\ref{fig:HaloIntegrals} what we refer to as the ``halo integral'' for axion haloscopes~\cite{Foster:2017hbq, Cheong:2024ose}. This quantity is defined to be, 
\begin{equation}\label{eq:halointegral}
    \eta = \left[ \int {\rm d}v \frac{f(v)^2}{v} \right]^{1/4} \, .
\end{equation}
The reason for defining this quantity can be seen by comparing with Eq.(\ref{eq:TS_asimov_limit})---this is how the experimental discovery power depends upon the speed distribution in the large $T$ limit. The reason for taking the $1/4$ power is because ${\rm TS} \propto \mathcal{A}^2 \propto g_{a\gamma}^4$, so we can interpret ratios of $\eta$ for different lineshape assumptions to correspond to sensitivity enhancements in terms of axion-photon coupling. 

We show the ratio between the halo integral for various lineshape models and the halo integral for the SHM. The latter permits an analytic expression,
\begin{equation}
        \eta_{\rm SHM} = \left[ \int {\rm d}v \frac{f_0(v)^2}{v} \right]^{1/4} = \left[ \frac{{\rm erf}(v_{\rm lab}/\sigma_v)}{\sqrt{4\pi}\sigma_v v_c} \right]^{1/4} \approx 27.7\, .
\end{equation}
Figure~\ref{fig:HaloIntegrals} plots the value of $\eta/\eta_{\rm SHM}$ as a function of $\sigma_{\rm str}$, assuming Model 2 (multiple streams of equal density). Generally, as the velocity dispersion of the streams decreases, we expect the signal in the stream's corresponding frequency bins to rise further above the noise, and so the sensitivity to improve. The improvement is most drastic for the extreme (albeit unlikely) case that the distribution is dominated by a single or a few streams. For larger $N_{\rm str}$ though, the fixed total dark matter density of $\rho_{\rm DM} = 0.45\,{\rm GeV\,cm^{-3}}$ distributed across more peaks and so the lineshape becomes smoother. In the limit of $N_{\rm str}\to \infty$ the distribution would become exactly smooth and converge on the SHM, $\eta \to \eta_{\rm SHM}$---this limit emerges by construction because we are drawing each value of $\mathbf{v}_{\rm str}$ randomly from the SHM.

In Fig.~\ref{fig:HaloIntegrals} we also demonstrate the effect of the daily modulation that was discussed in Sec.~\ref{sec:modulation}. To be clear, we are not accounting for the finite frequency resolution in this plot, we are merely calculating the value of $\eta$. However, it is still possible to take the time average of $\eta$ over some duration $T$, and doing so will isolate the effect of the daily modulation, which is to limit the sensitivity enhancement that can be gained from the streams below a certain value of $\sigma_{\rm str}$. This can be seen most clearly for the $N_{\rm str} = 1$ case where the sensitivity plateaus for values of $\sigma_{\rm str}$ below $10^{-3}$~km/s for $T = 100$~seconds because the daily modulation, $\Delta v = |v_{\rm lab}(0) - v_{\rm lab}(T)|$ is larger than $\sigma_{\rm str}$ for those values. Note that since we are talking about the halo integral here, not the lineshape, this figure applies to any value of the axion mass; however, the required values of $T$ needed to discover a given axion signal \textit{do} depend upon the axion mass being probed.

\subsubsection{Scaling of the test statistic}
The ability to discover axion dark matter is encoded in the value of the test statistic~Eq.\eqref{eq:TS_asimov} which depends on the signal-to-noise figure of merit $\mathcal{A}/\mathcal{B}$, as well as the integration time $T$. To assess realistic discovery potentials of current and future experiments, we need to write down reasonable values of $\mathcal{A}/\mathcal{B}$ so that we can find the values of $T$ required to reach values of $g_{a\gamma}$ expected for the QCD axion. For haloscopes that exploit a resonance to detect the axion, the general schematic forms for $\mathcal{A}$ and $\mathcal{B}$ will be the following:
\begin{equation}
    \begin{aligned}
{\cal A}  & =g_{a \gamma \gamma}^2 \frac{\rho_{\mathrm{DM}}}{m_a} \kappa Q B^2 V C,  \\
{\cal B} & =k_B T_s,
\end{aligned}
\end{equation}
where $g_{a\gamma}$ is the axion-photon coupling, $B$ is the magnetic field, $V$ is the volume, and $Q$ is the quality factor of the resonance; while $\kappa$ and $C$ are $\mathcal{O}(1)$ factors that depend on the experiment's readout and geometry respectively. For the background, we denote $T_s$ as the total noise in the system, including the physical temperature of the experiment and the noise added in the amplification stage. For the post-inflationary scenario (see Section \ref{sec:miniclusters}), the mass range of interest is $m_a\in (50, 500) \upmu$eV~\cite{Gorghetto:2020qws, Buschmann:2021sdq, Saikawa:2024bta, Benabou:2024msj}, targeted by ORGAN~\cite{McAllister:2017lkb, Quiskamp:2022pks}, QUAX~\cite{Alesini:2019ajt, Alesini:2020vny}, ALPHA~\cite{Lawson:2019brd, Millar:2022peq} and CADEx~\cite{Aja:2022csb}, however this expression equally applies for all resonant cavities searching for axion dark matter, such as ADMX~\cite{ADMX:2021nhd}, CAPP~\cite{Lee:2022mnc}, HAYSTAC~\cite{HAYSTAC:2020kwv}, TASEH~\cite{TASEH:2022vvu, TASEH:2022noe}, GrAHal~\cite{Grenet:2021vbb} and RADES~\cite{CAST:2020rlf} (see e.g.~Ref.~\cite{AxionLimits} for a summary). From inspecting Eq.\eqref{eq:TS_asimov_limit} we notice that TS depends specifically on the combination $\mathcal{A}/\mathcal{B}$, which has dimensions of time$^{-1}$. This ratio effectively scales the signal-to-noise ratio without reference to any integration times, so it will form our figure of merit when describing a specific experiment. For broadband or semi-broadband experiments like DMRadio~\cite{DMRadio:2022pkf}, MADMAX~\cite{Beurthey:2020yuq, MADMAX:2024sxs} and BREAD~\cite{GigaBREAD:2025lzq, BREAD:2021tpx} the figure-of-merit requires some modification to account for the way the signal is detected in those cases (e.g.~by replacing the quality factor with the area of the dish), but the general parametric scaling of the equivalent figure of merit remains the same so we do attempt an exhaustive accounting of every $\mathcal{A}/\mathcal{B}$ for each case in this work.

To see how the $\mathcal{A}/\mathcal{B}$ figure of merit compares to the more conventional notion of signal-to-noise, we can once again inspect the limiting form of the test statistic, Eq.\eqref{eq:TS_asimov_limit}, for long integration times, which becomes,
\begin{equation}
    {\rm TS}\approx \left(\frac{\cal A}{\cal B}\right)^2\frac{\pi T_{\rm tot}\eta[f(v)]^4}{2m_a}.\label{eq:TS_a2}
\end{equation}
This can then be compared with the Dicke radiometer equation, which is used in the resonant cavity haloscope literature to define the signal-to-noise ratio (SNR),
\begin{equation}
    {\rm SNR} = \frac{\cal A}{\cal B} \sqrt{T T_{\rm coh}}\,,
\end{equation} 
where $\mathcal{A}/\mathcal{B}$ are defined in an identical way as above. To see how this is related to TS, consider the halo integral for the SHM velocity distribution $f_0(v)$, 
\begin{equation}
    \eta[f_0(v)]^4 = \int {\rm d} v \frac{f_0(v)^2}{v}=\frac{\operatorname{erf}\left[v_{\mathrm{lab}} / \sigma_v\right]}{\sqrt{4\pi} \sigma_v v_{\mathrm{lab}}} \approx 5.9\times 10^5 = 0.59 \,\frac{m_a T^{\rm SHM}_{\rm coh}}{2\pi},
\end{equation}
where we have inserted our parameter values for the SHM to derive the numerical factor, and in the final step we recall we are adopting the heuristic definition of the coherence time used in the literature,\footnote{Note that in the haloscope literature this is often expressed as a ``linewidth'', i.e.~$\Delta \nu = 10^{-6}\nu_{\rm ax}$} $T^{\rm SHM}_{\rm coh} = 10^6 T_{\rm ax}$. Then, Eq.\eqref{eq:TS_a2} becomes
\begin{equation}
    {\rm TS} = \frac{0.59}{4}\left( \frac{\cal A}{\cal B}\right)^2 T T^{\rm SHM}_{\rm coh},
\end{equation}
leading to 
\begin{equation}
    {\rm SNR} \approx 2.6\sqrt{\rm TS} \, .
\end{equation} 
This shows that the test statistic can be understood as quantifying the signal-to-noise of the experiment, however it also shows why the full expression of the test statistic is more useful, as it leaves the precise dependence on the lineshape explicit, as opposed to introducing the (often ambiguously defined) coherence time.

We can now use this to examine how an experiment's sensitivity to $g_{a\gamma}$ depends on $T$ and the lineshape model. If we fix a required value of TS needed to claim a discovery, the expression above can be inverted to reveal that the sensitivity to the signal parameter should scale as ${\cal A} \propto 1/\sqrt{T}$. Given that ${\cal A}\propto g_{a\gamma}^2$ we then recover the well-known sensitivity-time scaling for haloscopes in the $T\gg T_{\rm coh}$ limit: $g_{a\gamma}\propto T^{-1/4}$. We expect this scaling to be relatively generic and independent of the lineshape as long as the lineshape is resolved. However, because of this latter requirement, the exact value of $T$ where this scaling takes over will depend on how narrow or broad the lineshape is, as we will now show. For shorter integration times, i.e.~$T<T_{\rm coh}$, the expected scaling is different, $g_{a\gamma} \propto T^{-1/2}$, as has been pointed out in several other related contexts previously~\cite{Budker:2013hfa,Foster:2017hbq,Dror:2022xpi}.

To make our discussion concrete, we will specify a few examples based on two different experiments targeting different motivated axion mass windows. Our first example is the well-known resonant cavity haloscope ADMX, which is targeting the pre-inflationary window around $\mathcal{O}(\upmu{\rm eV})$ axion masses. We fix our figure-of-merit to be similar to the specifications of their Run-1C cavity~\cite{ADMX:2021nhd}, which was also used for a high-resolution analysis in Ref.~\cite{ADMX:2024pxg},
\begin{equation}
\begin{aligned}
    \left(\frac{\cal A}{\cal B}\right)_{\rm ADMX}= 9.1\, {\rm Hz} \,&\bigg(\frac{C_{a\gamma}}{C^{\rm KSVZ}_{a\gamma}} \bigg)^2 \bigg(\frac{\rho_{\rm DM}}{0.45\,{\rm GeV\,{\rm cm}^{-3}}} \bigg) \bigg( \frac{1\,\upmu{\rm eV}}{m_a}\bigg) \\ \, &\times \bigg(\frac{\kappa}{0.5}\bigg)\bigg(\frac{Q}{80,000}\bigg) \, \bigg( \frac{B}{7.5\,{\rm T}}\bigg)^2 \, \bigg( \frac{V}{136\,\ell}\bigg) \bigg( \frac{C}{0.4}\bigg)\bigg(\frac{600 \,{\rm mK}}{T_s} \bigg).\label{eq:ADMX_AB}
\end{aligned}
\end{equation}
We have written the figure of merit here in terms of what would be required to detect the QCD axion. The dimensionless QCD axion photon coupling, $C_{a\gamma}$ is related to the dimensionful coupling $g_{a\gamma}$ as,
\begin{equation}
g_{a \gamma} \equiv \frac{\alpha}{2 \pi} \frac{C_{a\gamma}}{f_{a}}=2.0 \times 10^{-16} \,\mathrm{GeV}^{-1}\,C_{a \gamma} \bigg(\frac{m_{a}}{\upmu \mathrm{eV}}\bigg) \, ,
\end{equation}
where $f_a$ is the Peccei-Quinn scale, $\alpha \approx 1/137$ is the fine-structure constant. For the common benchmark Kim-Shifman-Vainstein-Zakharov (KSVZ) QCD axion model we have $|C^{\rm KSVZ}_{a\gamma}| = 1.92$.

Although we have fixed these values with ADMX in mind for the sake of concreteness, broadly speaking, all cavity haloscopes can be described by the same expression. We would find very similar results if we swapped out values for their experimental parameters as long as we stayed within the $\mathcal{O}(1-10~\upmu{\rm eV})$ mass window. For much larger masses, however, the small volumes required to maintain a resonance at the axion's frequency cause the sensitivity to suffer greatly, making the standard resonant cavity approach impractical.

To have another point of comparison that applies to higher axion masses, we consider the plasma haloscope ALPHA, which is targeting the post-inflationary axion mass range. Using specifications for ALPHA Phase I from Ref.~\cite{Millar:2022peq} (cf.~equation (41) of that paper), we find,
\begin{equation}
\begin{aligned}
    \left(\frac{\cal A}{\cal B}\right)_{\rm ALPHA} = 19.0\, {\rm Hz} \,&\bigg(\frac{C_{a\gamma}}{C^{\rm KSVZ}_{a\gamma}} \bigg)^2 \bigg(\frac{\rho_{\rm DM}}{0.45\,{\rm GeV\,{\rm cm}^{-3}}} \bigg) \bigg( \frac{100\,\upmu{\rm eV}}{m_a}\bigg) \\ \, &\times \bigg(\frac{\kappa}{0.5}\bigg)  \bigg(\frac{Q}{10^4}\bigg) \, \bigg( \frac{B}{13\,{\rm T}}\bigg)^2 \, \bigg( \frac{V}{0.29\,{\rm m}^3}\bigg) \bigg( \frac{C}{0.7}\bigg) \bigg(\frac{4.0 \,{\rm K}}{T_s} \bigg) \, .\label{eq:ALPHA_AB}
\end{aligned}
\end{equation}
In the discussion below, we will mostly quantify our results in terms of ratios of these figures of merit between choices for the lineshape, however one may use these values to straightforwardly convert those ratios into absolute terms, e.g.~values of $C_{a\gamma}$ and $T$.

\begin{figure}
    \centering
    \includegraphics[width=0.49\linewidth]{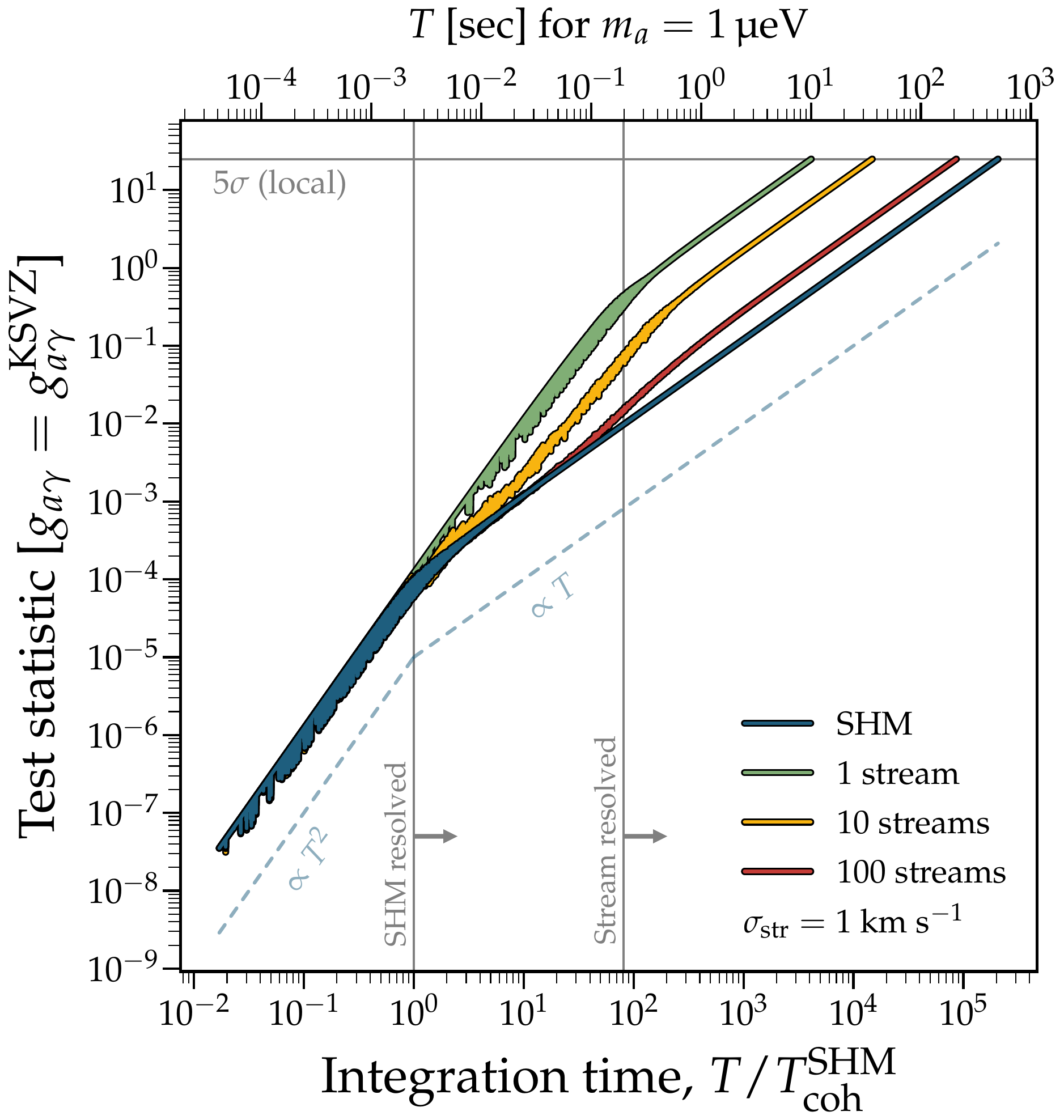}
    \includegraphics[width=0.49\linewidth]{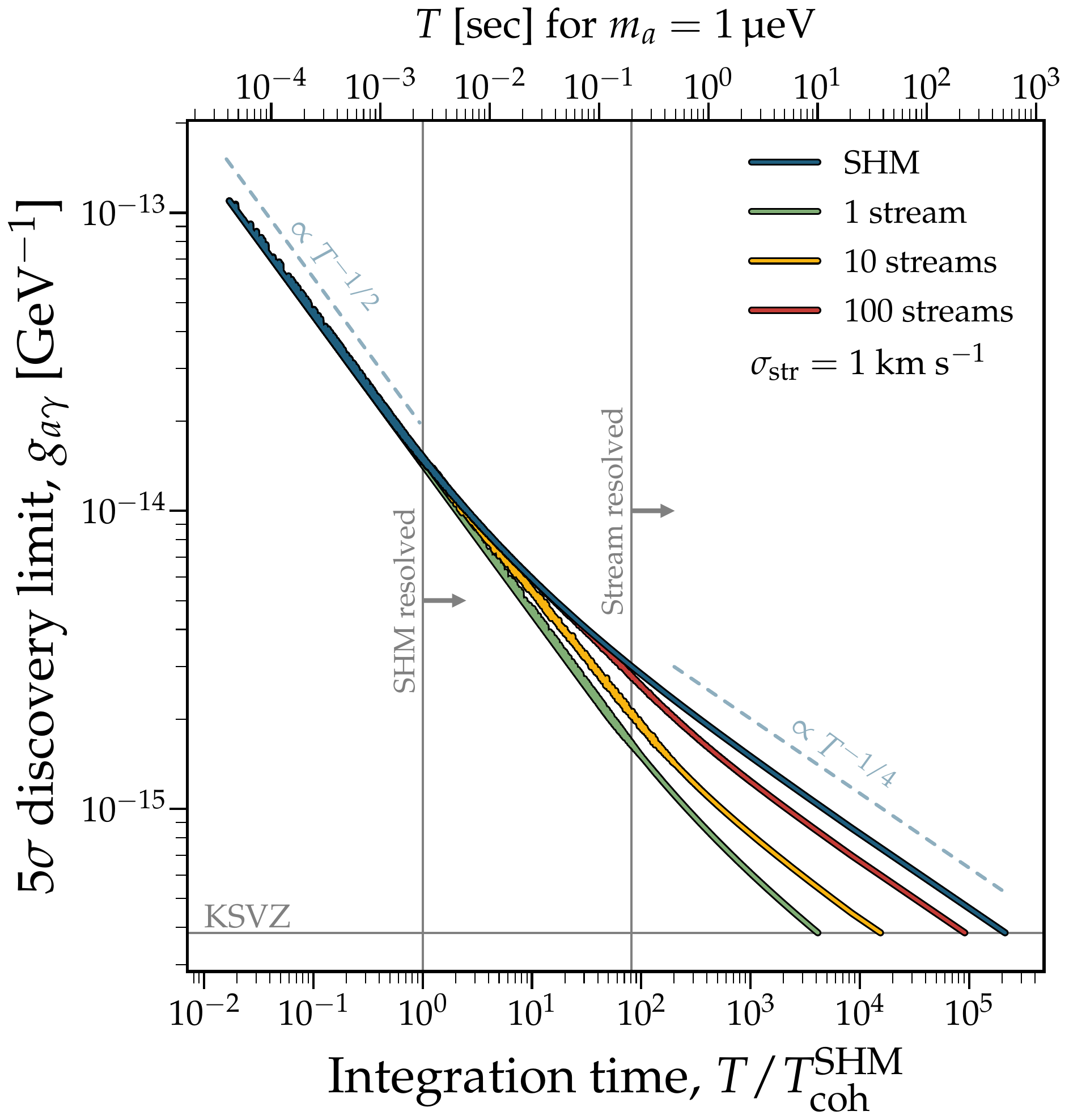}
    \caption{Scaling of the discovery test statistic ({\bf left}) or equivalently the sensitivity to $g_{a\gamma}$ ({\bf right}) as a function of the integration time, $T$, assuming an ADMX-like experimental figure of merit $\mathcal{A}/\mathcal{B}$ as in Eq.(\ref{eq:ADMX_AB}). In the left-hand panel, we fix the photon coupling to the KSVZ model and plot the value of $\widetilde{{\rm TS}}$ versus $T$ up to the value required for the median experiment to claim a $5\sigma$ discovery, $\widetilde{{\rm TS}} = 25$. On the right-hand panel, we fix the value of $\widetilde{{\rm TS}} = 25$ and plot the value of the axion-photon coupling that can be discovered at that significance level. Each plot contains several lines for different lineshape assumptions, where we highlight the comparison between the standard assumption of the SHM (dark blue) and various stream models (green, yellow, red). In this case, we adopt Model 2 (see Sec.~\ref{sec:fv}) with a varying number of streams $N_{\rm str} = 1,\,10,\,100$ all with $\sigma_{\rm str} = 1$~km~s$^{-1}$. We choose this simplified model solely to highlight the way the lineshape influences how the sensitivity scales with $T$. When features in the lineshape are resolved, we expect the asymptotic $g_{a\gamma} \propto T^{-1/4}$ to take over. This occurs at higher values of $T$ in the stream models because the features are narrower in frequency space than the SHM's lineshape and so require a higher frequency resolution before they appear. We note that the rapid fluctuations in the lines, which are most pronounced at smaller $T$, are not due to numerical noise but are actual variations in the median sensitivity which occur due to the lineshape being split across coarse frequency bins.}
    \label{fig:TS_versus_T}
\end{figure}

\begin{figure}
    \centering
    \includegraphics[width=0.49\linewidth]{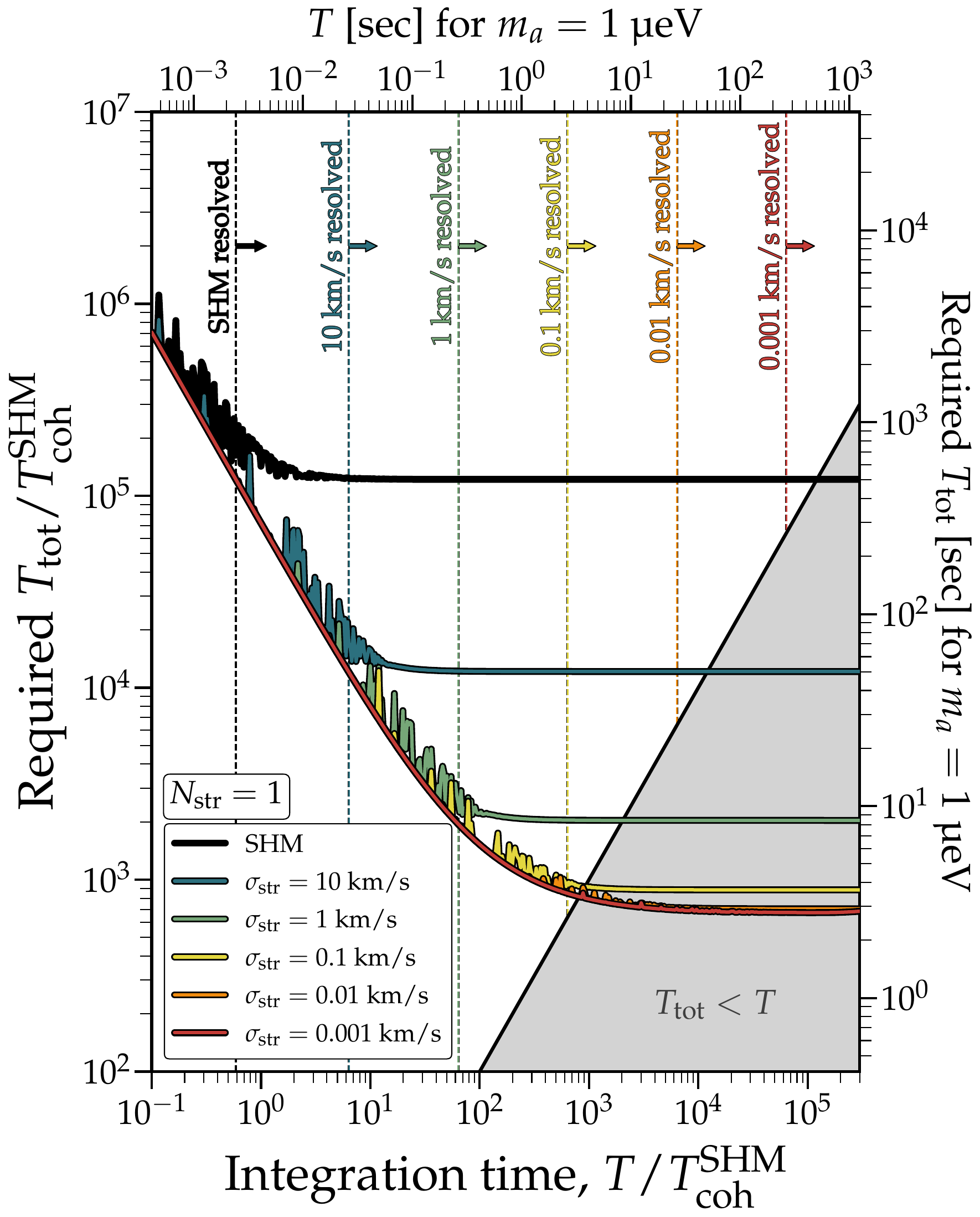}
    \includegraphics[width=0.49\linewidth]{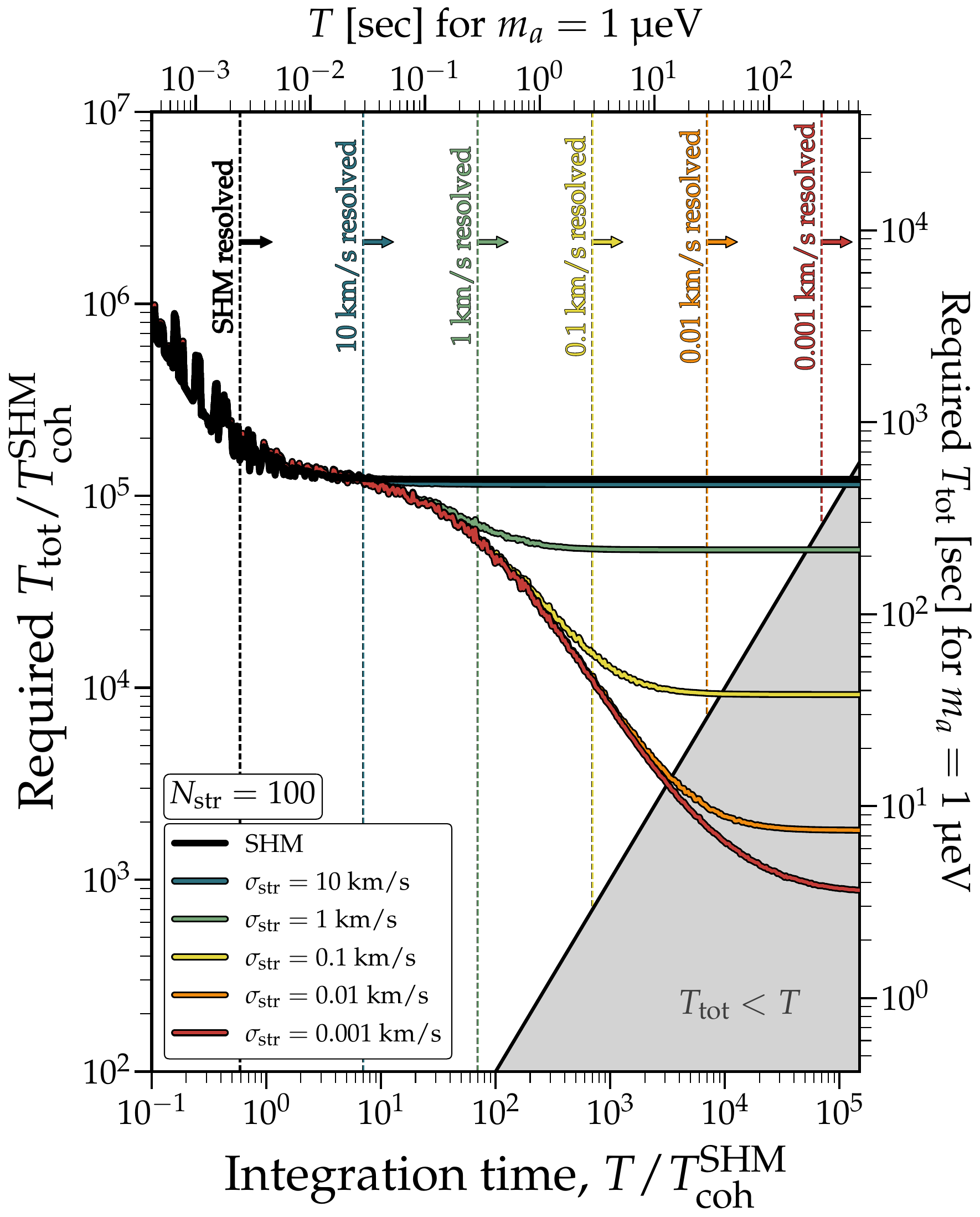}
    \caption{Required \textit{total} measurement time, $T_{\rm tot}$ needed to obtain a median $5\sigma$ local significance discovery of an axion signal as a function of the \textit{individual} integration times $T$ used to construct each of the $N_{\rm PS} = T_{\rm tot}/T$ power spectra. As in the previous figure, we assume an ADMX-like sensitivity figure of merit for $\mathcal{A}/\mathcal{B}$. The left-hand panel is for Model 1 (single stream) while the right-hand panel is for Model 2 (multiple streams, where $N_{\rm str} = 100$), and we plot several lines for different choices for the stream velocity dispersion, $\sigma_{\rm str}$. The grey region in the bottom right-hand corner denotes where the required $T_{\rm tot}$ is smaller than $T$. When the lines cross into this region, it simply means that $>5\sigma$ discovery is already achievable from a single power spectrum of duration $T$. The plot demonstrates that the stream models require less total data to detect the axion, but that the integration time needs to reach a critical value dictated by the stream's velocity dispersion to obtain the corresponding sensitivity enhancement.}
    \label{fig:T_versus_Ttot}
\end{figure}

With these ingredients in hand, we can now demonstrate how the presence of narrow features in the lineshape can enhance the discoverability of the axion. Firstly, in Fig.~\ref{fig:TS_versus_T} (left), we show how the value of TS improves as a function of the integration time. We first highlight that the expected scaling of ${\rm TS}\propto T$ emerges only at large enough values of $T$. When $T$ is small, the scaling is ${\rm TS}\propto T^2$. The cross-over occurs when $T$ is larger than the coherence time, or in other words, when the lineshape is resolved. This then explains the scaling for the other lineshape choices, which in this case is when the stream is resolved. This can be seen in Fig.~\ref{fig:TS_versus_T} where the transition to the latter scaling takes over when the stream width in frequency space is larger than the frequency resolution $\Delta \omega = 2\pi/T$. The largest enhancement is seen when a single stream is present in the distribution, as this generates the lineshape with the largest overall amplitude and hence signal-to-noise. When there are much larger numbers of streams, e.g.~100, the lineshape appears smooth for low values of $T$, and so the value of TS is closer to the expectation under the SHM.

For comparison, we show in the right-hand panel of Fig.~\ref{fig:TS_versus_T} what is essentially the same information but presented in an alternative way. Rather than fixing the axion-photon coupling and looking at how the value of TS varies with the $T$, here we fix the required value of TS for a discovery, i.e.~${\rm TS} = 5\sigma$, and solve for the value of $g_{a\gamma}$, which could be discovered in a given integration time. Recall that we have constructed this example so that the sensitivity curves all cross the ${\rm TS} = 25$ for a KSVZ coupling for this axion mass. The same behaviour as a function of $T/T_{\rm coh}^{\rm SHM}$ would be seen for any other choice of $m_a$, although of course the corresponding values of $T$ would change because $T_{\rm coh}^{\rm SHM} \propto m_a^{-1}$.

Following on from this result, we now move to Fig.~\ref{fig:T_versus_Ttot}. The previous figure dealt with the case where $N_{\rm PS} = 1$ in the sum in Eq.(\ref{eq:TS_asimov}), i.e. $T_{\rm tot} = T$. Now we highlight the importance of the frequency resolution for unlocking the sensitivity enhancement that comes from the streams, by allowing both $T$ and $T_{\rm tot}$ to vary. Specifically, what we show is the required value of $T_{\rm tot}$ to discover the KSVZ axion at $5\sigma$ local significance given ADMX-like experimental parameters, as a function of $T$. To find this, we first fix a value of $T$ and then ask how many power spectra of duration $T$ does one need to collect to build up to the required ${\rm TS} = 25$. The first point to emphasise is that the expected behaviour under the SHM should be that the required $T_{\rm tot}$ is independent of $T$. In other words, dividing up some total duration $T_{\rm tot}$ into different power spectra does not lead to any gains in sensitivity. Inspecting the SHM line in Fig.~\ref{fig:T_versus_Ttot} we see that this is indeed the case \textit{as long as $T>T^{\rm SHM}_{\rm coh}$}, which we understand to be simply a manifestation of the two scaling regimes for the sensitivity shown in the previous figure.

We can now inspect how this idea changes when we consider the other models shown in Fig.~\ref{fig:T_versus_Ttot}. Here we explore stream models with varying $\sigma_{\rm str}$ and either $N_{\rm str} = 1$ or $N_{\rm str} = 100$ for the left and right-hand panels of Fig.~\ref{fig:T_versus_Ttot} respectively. This example shows that the degree to which the total measurement time $T_{\rm tot}$ is broken up into multiple lower-resolution power spectra \textit{does} matter. The fact that lines continue to descend to lower values of $T_{\rm tot}$ for increasing $T$ in those cases tells us that measuring higher-resolution power spectra allows for a discovery to be made with \textit{less} total data. In the most extreme cases when $\sigma_{\rm str}$ is small, the amount of data required can be cut by almost three orders of magnitude, if the frequency resolution were extremely high. However, the trend does not persist to arbitrarily high values of $T$. Once all of the streams are resolved, i.e. $\Delta \omega$ is smaller than both the width and spacing of the streams in frequency space, then the lines plateau, for the same reason as the line plateaus in the SHM case---one does not gain any statistical power by analysing the lineshape at a higher resolution than is necessary to resolve all its features. 

These examples serve to illustrate the impact that fine-grained substructure has on the sensitivity of haloscopes. We emphasise that the velocity distributions we have used to illustrate these ideas at this point have been purposely oversimplified. For instance, the cases in which a single stream dominates the distribution are probably very unlikely to apply to the dark matter in the solar neighbourhood. However, in subsequent sections, when we discuss more physically-motivated examples for distributions composed of streams, we will see that the distributions with a large number of $>$100 components are not terribly approximated by our Model 2, and so the intuition gained here is useful.

\subsection{Extreme value statistic}\label{sec:EVS}
Although the likelihood framework is powerful, it has a conceptual flaw when applied to some of the scenarios we will explore in later sections. The discovery of an axion signal using the PLR test asks the question of whether the data is better described by a given signal+background model compared to the background-only null hypothesis. This is a useful methodology if there is some trust that, broadly speaking, the signal model being used would describe the data well if a signal were present. It is possible to introduce more flexibility into a model by adding more parameters---for example, one could float the values of $\sigma_v$ and $v_c$ etc.~in the lineshape. However, the amount of data needed to just make a detection claim is not enough to also give sensitivity to these finer details of the lineshape, so in practice, it is more convenient to simply fix an assumption about the lineshape.\footnote{In fact, depending on the experiment, some groups may opt for even simpler assumptions, like approximating the axion signal as a delta function in frequency or as having a Breit-Wigner-like resonance profile.} 

The axion velocity distributions we are proposing here could lead to drastically different signal shapes, but at the same time, modelling those shapes would involve introducing potentially hundreds of nuisance parameters. This not only makes the PLR test impractical, but it is also not a fair representation of how testing for the presence of these substructures should work in practice. Given that a large signal excess due to a stream could appear anywhere in the lineshape, a better approach in practice would be to analyse the data at a high-enough resolution to potentially reveal a stream of a given width, and then test for their presence individually. We want to assess the extent to which this procedure could increase the chances of a real candidate axion signal emerging, thereby accelerating us towards a discovery. Our results for the PLR approach are useful for illustrating the enhancement of the axion signal in the presence of fine-grained substructure; however, it comes with a downside, which is that these enhancements can only be taken quantitatively if we imagine a scenario in which all of the many features of these models are measured through some parameter inference process, which is perhaps better thought of as a \textit{post}-discovery scenario.

Given this issue, we will now explore a \textit{non-parametric} statistical test that does not require any assumptions to be made about the axion lineshape, at the cost of being less sensitive than constructing a full likelihood. The upside, however, is that this approach will mimic aspects of a very common practice employed by resonant axion haloscopes called re-scans. This is a procedure in which the frequencies of any high-significance excesses above the expected noise level observed during a scan over a range of axion masses are recorded and re-measured later to check if they persist (as would a real signal) or not (as would e.g.~transient radio interference). In the case of a lineshape made up of many streams, there is an enhanced probability for high-significance upward-fluctuations above the noise that would be localised to a single frequency bin in a high-resolution scan. So the test we introduce here pertains to the statistics of those large single-bin fluctuations if an axion signal is present, given some a priori knowledge of the expected statistical distribution of the noise.

We will explore an approach based on a simple extreme value statistic (EVS), namely the maximum value of the power observed in a pre-defined window of frequency bins. Specifically, if we have some data $P_{\rm obs}(\omega_k)$, we define the EVS to be,
\begin{equation}
    P_{\rm max} \equiv {\rm max}(P_{\rm obs}(\omega_k)) \,.
\end{equation}
The distribution of this statistic will be dictated by the number of frequency bins contained in the window, which is a choice that needs to be made when conducting the test. For a given value of the axion mass, there is a natural choice for this window, which is for $k$ to run between only the frequency bins that correspond to physical axion velocities, i.e. between $\omega = m_a$ and $\omega = m_a(1+v_{\rm max}^2/2)$. For simplicity, in the following expressions, we will label $k$ as starting on the bin whose central frequency is closest to $m_a$. As in the PLR test, we are imagining that the test would be repeated over a range of masses during the haloscope's full scan, but that the mass is taken to have a fixed value when each one of these tests is performed.

If each value of $P_{\rm obs}(\omega_k)$ is drawn from an exponential distribution with a mean value of $P_{\rm exp}(\omega_k) = \langle S(\omega_k)\rangle + {\cal B}$, then the cumulative distribution function for the statistic $P_{\rm max}$ can be computed straightforwardly to be
\begin{equation}
    F(P_{\rm max};{\cal A},{\cal B})=\prod_{k=1}^{N_{\rm bins}}\left(1-e^{-P_{\rm max} / P_{\rm exp}(\omega_k;{\cal A},{\cal B})}\right) \, ,
\end{equation}
while the probability distribution function is,
\begin{align}
    f(P_{\rm max})=\frac{{\rm d}}{{\rm d} P_{\rm max}} F(P_{\rm max})&=\frac{{\rm d}}{{\rm d} P_{\rm max}}\left[\prod_{k=1}^{N_{\rm bins}}\left(1-e^{-P_{\rm max} / P_{\rm exp}(\omega_k)}\right)\right] \\
    &=\sum_{j=1}^{N_{\rm bins}}\left(\frac{e^{-P_{\rm max} / P_{\rm exp}(\omega_j)}}{P_{\rm exp}(\omega_j)} \prod_{k \neq j}\left(1-e^{-P_{\rm max} / P_{\rm exp}(\omega_k)}\right)\right) \, .
\end{align}
We can also straightforwardly incorporate a procedure in which, to reduce data storage requirements, power spectra are computed for integration times of length $T$, but the total measurement time is larger, i.e.~$T_{\rm tot} = N_{\rm PS}T$. If we assume the maximum power excess is searched for across each of these $N_{\rm PS}$ power spectra, the cumulative distribution function for $P_{\rm max}$ is written instead as,
\begin{equation}
    F(P_{\rm max};{\cal A},{\cal B})=\left(\prod_{k=1}^{N_{\rm bins}}\left(1-e^{-P_{\rm max} / P_{\rm exp}(\omega_k;{\cal A},{\cal B})}\right) \right)^{N_{\rm PS}} \, ,\label{eq:cdf_n}
\end{equation}
although we will keep $N_{\rm PS}=1$ for this discussion.

For the case where the expected power does not depend on frequency, e.g.~under the background-only model $P_{\rm exp}(\omega) = {\cal B}$, the cumulative distribution function simplifies to:
\begin{equation}
    F(P_{\rm max};{\cal A}=0, {\cal B}) = \left(1-e^{-P_{\rm max}/{\cal B}}\right)^{N_{\rm bins}} \, ,
\end{equation}
while the probability distribution under the background-only model is
\begin{equation}
    f(P_{\rm max};{\cal A}=0, {\cal B})=\frac{N_{\rm bins}}{{\cal B}}\left(1-e^{-P_{\rm max} / {\cal B}}\right)^{N_{\rm bins}-1} e^{-P_{\rm max} / {\cal B}} \, .
\end{equation}

The fact that the null hypothesis distribution of the statistic is analytic allows us to straightforwardly compute percentiles for $P_{\rm max}$. Let us define some threshold as the value of $P_{\rm max}$ below which a fraction $1-p$ of the distribution lies under the background-only model. This value can be found by inverting the expression for $F$ above to find:
\begin{equation}
    P_{\rm max}^{1-p} \equiv -{\cal B} \ln \left(1-(1-p)^{1 / N_{\rm bins}}\right) \, .
\end{equation}

In order to establish a comparison with the Asimov PLR test statistic, we calculate the expected significance of some signal with $\mathcal{A} \neq 0$ by determining the median value of $P_{\rm max}$, which is found by solving $F(P_{\rm 50};{\cal A}) = 0.5$, i.e. the cumulative distribution of $P_{\rm max}$ when a signal is present. We can then use this value to compute the $p$-value from the cumulative distribution of $P_{\rm max}$ when the signal is absent,
\begin{equation}
    1-p = F(P_{50} ; {\cal A}=0).
\end{equation}
Unfortunately, $P_{\rm 50}$ does not yield an analytic solution in this case, but can be obtained numerically by solving:
\begin{equation}
    \sum_{k=1}^{N_{\rm bins}} \ln\left(1-e^{-P_{\rm max}/P_{\rm exp}(\omega_k;{\cal A},{\cal B})}\right) = - \ln 2 \, .
\end{equation} 
We then use $p$-value to define a $Z$-score defined in the usual way, i.e.~a Gaussian distributed variable found $Z$ standard deviations above the mean would have a one-sided probability equal to the same $p$-value. This is calculated as $Z = \Phi^{-1}(1-p)$, where $\Phi^{-1}$ is the inverse cumulative distribution function of the Gaussian. 

\begin{figure}
    \centering
    \includegraphics[width=0.49\linewidth]{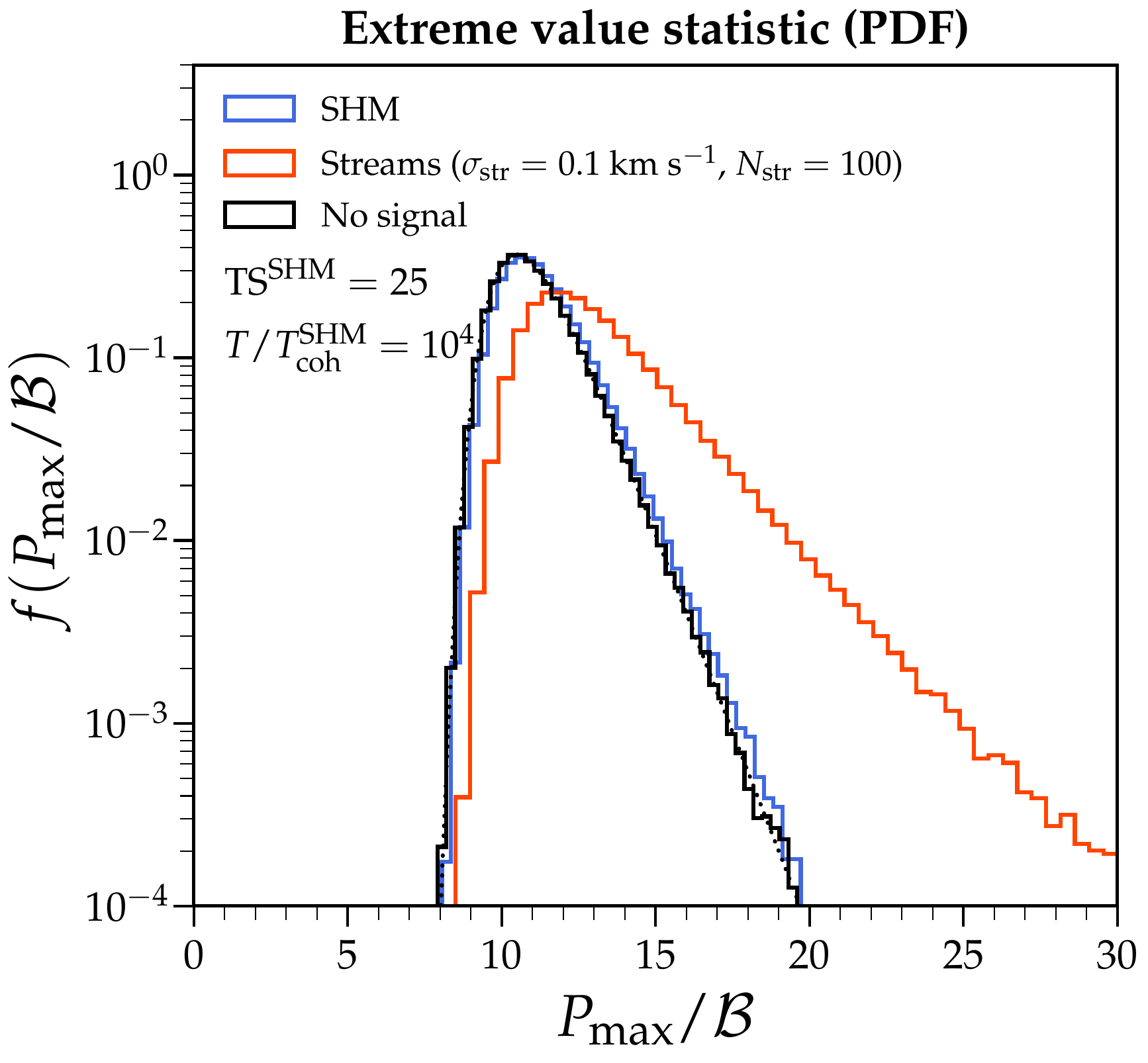}
    \includegraphics[width=0.49\linewidth]{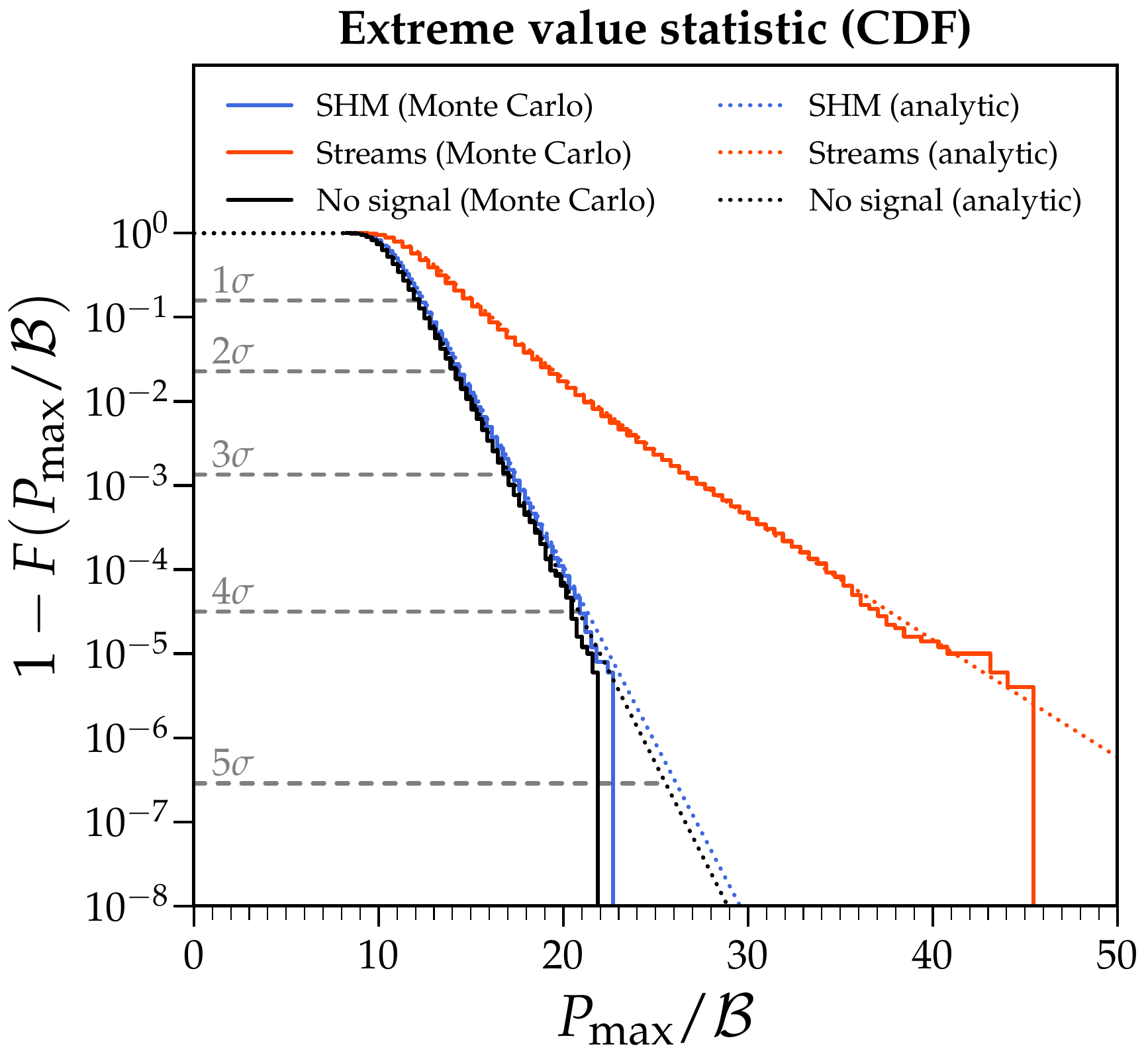}
    \caption{Probability distribution function (\textbf{left}) and cumulative distribution function (\textbf{right}) for the extreme value statistic $P_{\rm max}$ under various models. We fix the value of ${\cal A}/{\cal B}$ such that the PLR test statistic under the SHM returns a $5\sigma$ median significance for the same integration time, $T = 10^4 T_{\rm coh}^{\rm SHM}$. The stream model (orange) is constructed from 100 streams with velocity dispersion $\sigma_{\rm str} = 0.1$~km/s. The distributions calculated from 500,000 Monte-Carlo tests performed on simulated data are shown as histograms, while the analytic result is shown as a dotted line. For the cumulative distribution function, we overlay several significance levels defined via equivalent Gaussian one-sided $p$-values, also known as the $Z$-score. We construct a statistical test using $P_{\rm max}$ as the test statistic, where the only assumption made is that the data is exponentially distributed with mean $\mathcal{B}$ if no signal is present.}
    \label{fig:EVS_MonteCarlo}
\end{figure}

\subsubsection{Scaling of the EVS test}

To gain a visual sense for how this statistic behaves, and why it is useful for our situation, Fig.~\ref{fig:EVS_MonteCarlo} shows the probability distribution function and cumulative distribution function for $P_{\rm max}$. We compare three cases: when there is no signal present, $\mathcal{A} = 0$; when there is a signal present described by Model 0 (the SHM); and when a signal is present described by Model 2 (multiple streams) with $N_{\rm str}=100$ and $\sigma_{\rm str}=0.1$ km/s). We choose $T$ to be $10^4$ SHM coherence times, and fix the value of $\mathcal{A}/\mathcal{B}$ such that the PLR test applied on the SHM case would return a $5\sigma$ local discovery significance. If $m_a = 1~\upmu$eV, this would coincide with a coupling of $C_{a\gamma} = 1.85\times C_{a\gamma}^{\rm KSVZ}$ for our ADMX-like experiment benchmark. 

Even though the \textit{total} power contained in the signal is exactly the same for both the SHM and stream cases in this figure, the EVS distribution for the SHM is almost identical to the distribution when no signal is present, while the stream case is drastically different. This figure demonstrates that the EVS test is particularly sensitive to signals that can have large fluctuations over the noise level, even if those fluctuations are rare.

\begin{figure}
    \centering
    \includegraphics[width=0.8\linewidth]{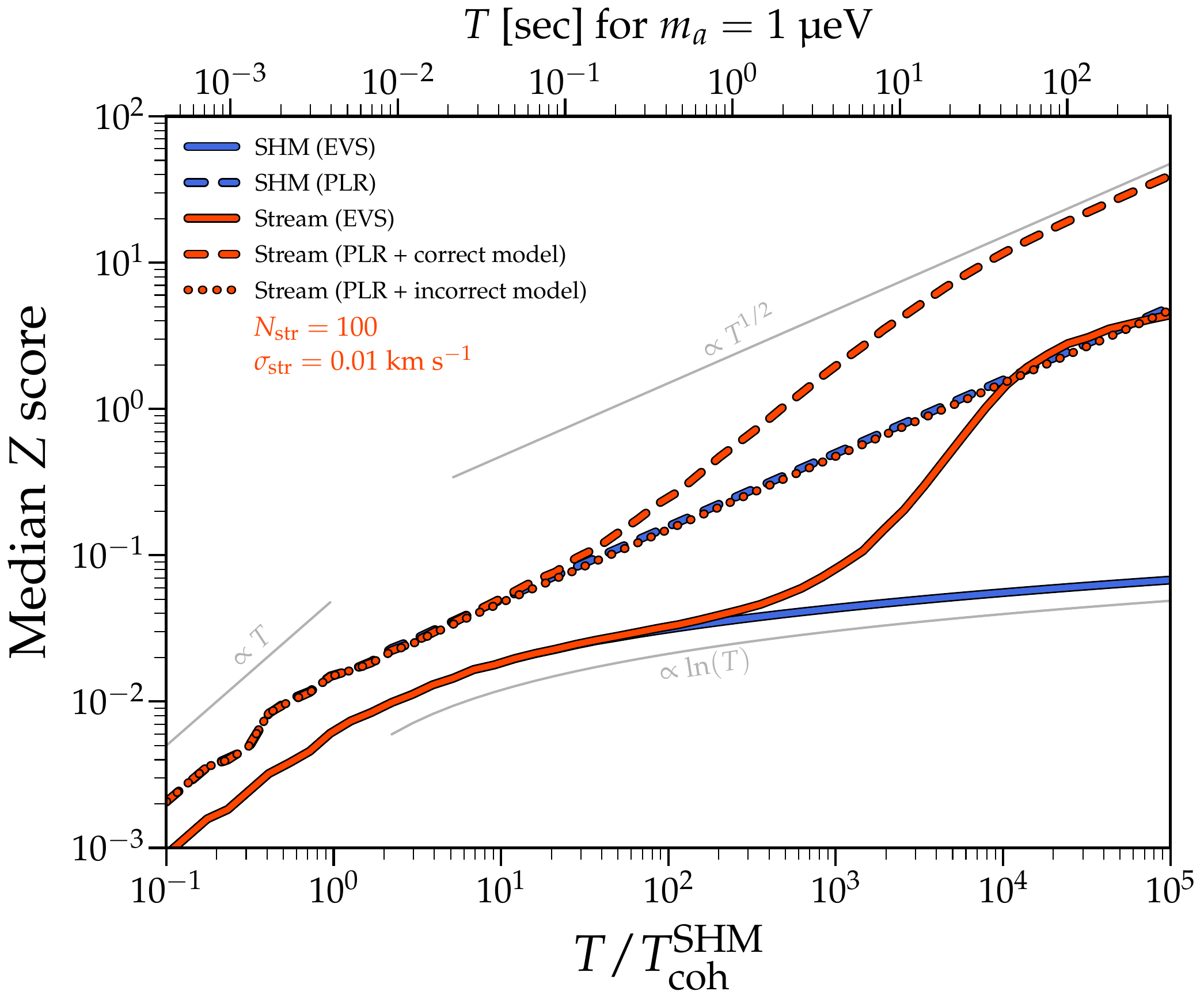}
    \caption{Median Z-score (significance) as a function of integration time, comparing various statistical tests. All lines in blue are when the tests are applied on data described by the SHM, whereas the lines in orange are where a multi-stream model (Model 2) with $N_{\rm str} = 100$ and $\sigma_{\rm str} = 0.01$~km~s$^{-1}$ describes the data. The dashed lines in both cases represent the median discovery significance obtained by applying the PLR test to the data, assuming the correct model. For comparison, the dotted orange line shows the result when the PLR test is used on the data containing streams, but the model assumption being made is \textit{incorrect}---namely, we assume the SHM describes the data. This line shows that the gain in sensitivity due to streams in the PLR test is only achieved if the data is tested against a model containing streams. In contrast, we show as solid lines the result from applying the EVS test, where, in this case, an enhancement in sensitivity due to the presence of streams is obtained without needing to specify a model to describe the signal. In this example, we have normalised the signal-to-noise figure of merit here so that a $5\sigma$ significance discovery of the axion when the SHM is true would be obtained for $T = 10^5 T_{\rm coh}^{\rm SHM}$. The slow (logarithmic) scaling of the EVS test significance as a function of $T$ make this test insensitive for smooth lineshape models, but may allow for detection of very narrow features in the lineshape that do not have a priori known frequencies or amplitudes.}
    \label{fig:EVS_Zscore}
\end{figure}

To make this idea more quantitative, in Fig.~\ref{fig:EVS_Zscore}, we compare how the median significance of the SHM and stream signal models scales, while also comparing the scaling of the statistical power of the PLR and EVS tests. In particular, we illustrate a point we made earlier by performing the PLR test on data that is described by the stream lineshape, but while assuming (i.e.~incorrectly) that the SHM describes the data. This is shown as the orange dotted line in the figure, which should be compared with the orange dashed line that we obtain by applying the test with the \textit{correct} model assumption. This shows that the sensitivity enhancement due to the presence of streams is only obtained when the data is tested against a model that is able to fit those streams.

When we apply our EVS test however (solid lines), we observe the large enhancement in sensitivity in the stream case (orange solid line) compared to the SHM case (blue solid line). Note that although we plot the $Z$ score values for the PLR and EVS tests on the same figure, the emphasis here is on the scaling rather than the numerical values because the meaning of significance in these two cases is slightly different. Our PLR test pertains to the discovery of a signal given a model, so the significance refers to the extent to which the data prefers the signal hypothesis over the background-only hypothesis. In the EVS test, however, the significance refers to how unlikely the maximum power value measured in the data would be if it were just distributed exponentially with mean $\mathcal{B}$. That said, because the EVS test incorporates much less information, we do expect it to be less powerful than the PLR test. Indeed, the plot reveals the significance scales logarithmically at large $T$, although the scaling for values of $T$ where the lineshape (or features in the lineshape) are not resolved is similar to the PLR test. This suggests that the optimal value of $T$ where this approach is most effective is when the streams are only just resolved, e.g.~when,
\begin{equation}
  m_a |\mathbf{v}_{\rm lab} - \mathbf{v}_{\rm str}|\sigma_{\rm str} \approx 2\pi/T  \, ,
\end{equation}
For larger values of $T$, the stream is then split across multiple frequency bins, which suppresses the possible value of $P_{\rm max}$ that could be obtained while, at the same time, increasing the total number of bins, weakening the test.

\begin{figure}
    \centering
    \includegraphics[width=0.8\linewidth]{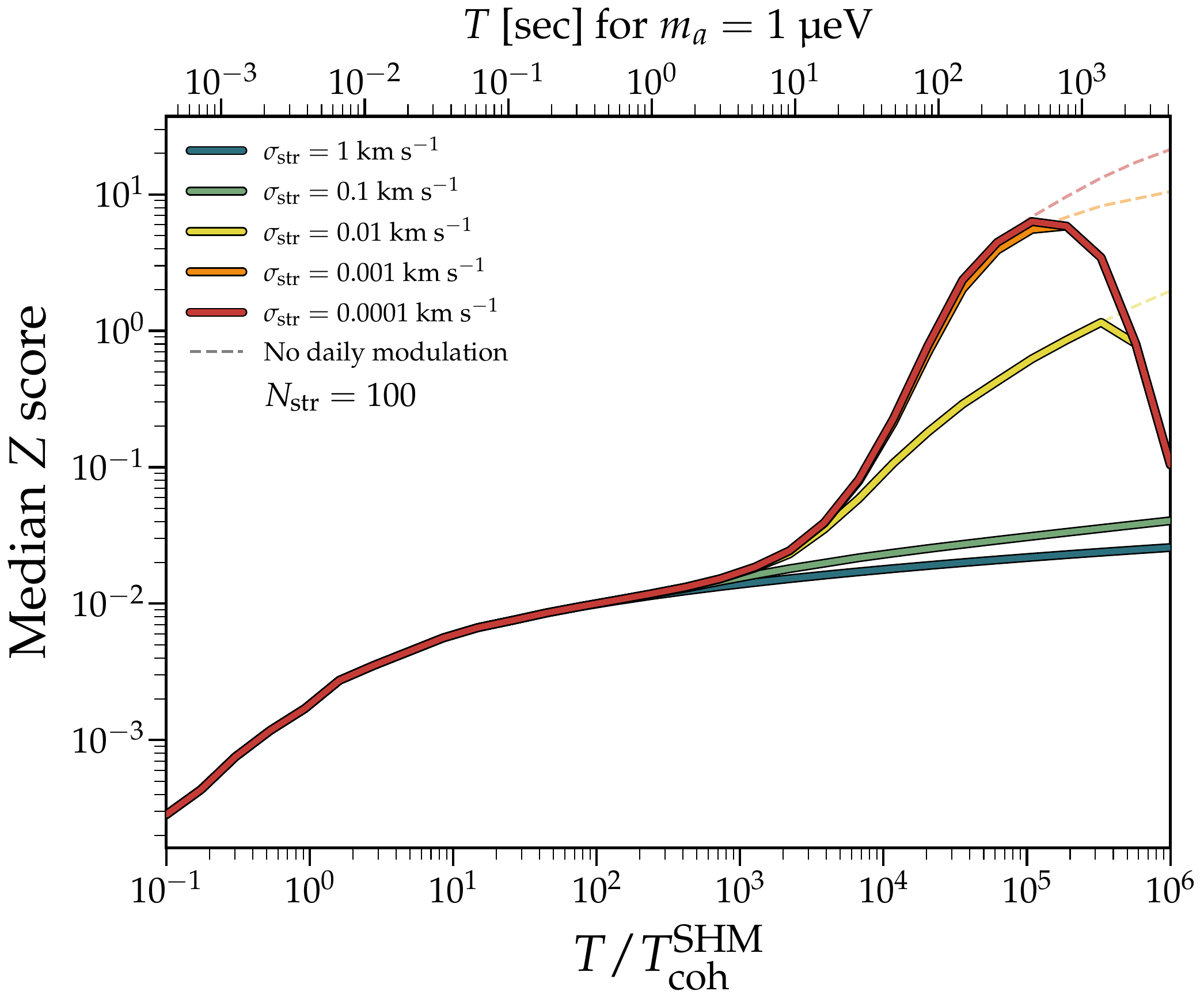}
    \caption{As in Fig.~\ref{fig:EVS_Zscore}, but now showing the median Z-score for the EVS test applied on a signal lineshape described by Model 2 for varying $\sigma_{\rm str}$, keeping the number of streams fixed at 100. For the narrowest velocity dispersion cases shown here, this plot reveals that the sensitivity enhancement is further limited by the daily modulation. The dashed lines for each colour show the result of the same calculation if we ignore the time-dependence in $\mathbf{v}_{\rm lab}$. The optimal value of $T$ to choose in these cases is then the value where the frequency shift due to daily modulation matches the frequency resolution.}
    \label{fig:EVS_Significance_stream_comparison}
\end{figure}

To clarify this latter point further, we present Fig.~\ref{fig:EVS_Significance_stream_comparison}, which shows similar information to the previous figure, but now comparing the performance of the test on Model 2 for streams of varying width. For the large values of $\sigma_{\rm str}$, we indeed see the significance approach a logarithmic scaling once the streams are resolved. Additionally, for the smallest $\sigma_{\rm str}$ cases, we observe another interesting effect at large $T$: the significance value begins to decline above a value of $T$ where the streams should not yet be fully resolved. The reason for this becomes clear when we show the result we obtain if we ignore the daily modulation (i.e.~keep $v_{\rm lab}(t)$ fixed). This reveals the impact of the discussion in Sec.~\ref{sec:modulation} and uncovers another limitation in the performance of any statistical test on very narrow streams. If the frequency shift due to daily modulation over some duration $T$ is larger than the frequency spread intrinsic to the stream, then the daily modulation will be the limiting factor in the performance of our test as opposed to $\sigma_{\rm str}$. In this case, the optimal value of $T$ is instead found by satisfying,
\begin{equation}\label{eq:Topt_dailymod}
    m_a |\mathbf{v}_{\rm lab} - \mathbf{v}_{\rm str}|\Delta v_{\rm rot}(T) \approx \frac{2\pi}{T} \, ,
\end{equation} 
where $\Delta v_{\rm rot}(T) = \big| |\mathbf{v}_{\rm lab}(0) - \mathbf{v}_{\rm str}| - |\mathbf{v}_{\rm lab}(T) - \mathbf{v}_{\rm str}|\big|$ is the speed shift due to daily modulation (as long as $T<0.5$~days). We will see both of these limiting factors become important in different ways over the next two sections.

\section{Minicluster streams}\label{sec:miniclusters}
So far, we have established several methodologies for evaluating the sensitivity enhancement in axion haloscopes due to dark matter substructure and have illustrated them using toy models. Our goal for the remainder of this paper is to quantify these prospects within more realistic scenarios. Over the next two sections, we will explore two physically-motivated examples of lineshapes composed of fine-grained streams. The primary reason for introducing the alternative EVS test in the previous section is that these models will predict there to be very large numbers of substructures with varying expected densities, velocities and dispersions that we may only be able to specify at a statistical level. 

The first of these models describes the streams from disrupted miniclusters anticipated in a post-inflationary axion cosmology. Here, the relevant axion mass range is in the approximate range $m_a \sim (50,500)~\upmu$eV~\cite{Buschmann:2019icd, Buschmann:2021sdq, Gorghetto:2018myk, Gorghetto:2020qws, Kim:2024wku, Saikawa:2024bta, Benabou:2024msj}. The second example, presented in Sec.~\ref{sec:finegrained}, will describe the much larger number of \textit{ultra}-fine-grained streams expected in any halo forming from the collapse of a smooth initial density sheet---this case we expect to apply to CDM generically, but here we will use it as an example for axion masses outside the post-inflationary prediction, for example those which may only be produced with the correct cosmological dark matter abundance in a \textit{pre}-inflationary scenario. The correct abundance can be obtained for any $m_a \lesssim $~meV (up to fine-tuning of the initial misalignment $|\theta_i| \to \pi$), but we will use fiducial values around $m_a \sim 1$-$50\,\upmu$eV so as to match the range of existing resonant cavity haloscope experiments.

\subsection{Lineshape construction}
We start with the post-inflationary case, in which the expected present-day substructure in the dark matter distribution comes from the streams left behind after the tidal destruction of axion miniclusters. These streams are ultimately a relic of the large field inhomogeneities that arise after the spontaneous symmetry breaking phase transition that produces the axion. These inhomogeneities are left behind in our observable Universe in this scenario specifically because inflation has already occurred by the time the phase transition takes place. The Peccei-Quinn phase transition in this case occurs at temperatures around $T\sim f_a\gtrsim 10^{10}$ GeV, which is the epoch when a network of cosmic strings is produced. The cosmic string network then enters a scaling regime while radiating massless axion waves until it collapses around the QCD phase transition when those axions become massive. What is left behind is a distribution of dark matter with small-scale inhomogeneities which make a head start on gravitational collapse around matter-radiation equality. This sequence of events is what results in the early formation of small-scale dark matter minihalos, or miniclusters.

Several numerical simulations have been performed to address the post-inflationary scenario, modelling the evolution of global string networks~\cite{Fleury:2015aca, Gorghetto:2018myk, Buschmann:2021sdq, Saikawa:2024bta}, the string-wall network collapse~\cite{ Klaer:2017ond, Vaquero:2018tib, Buschmann:2019icd, Gorghetto:2020qws, OHare:2021zrq}, the formation and evolution of miniclusters~\cite{Zurek:2006sy, Eggemeier:2019jsu, Xiao:2020pra, Eggemeier:2022hqa, Pierobon:2023ozb} and the minicluster tidal disruption into streams~\cite{Kavanagh:2020gcy, Shen:2022ltx, Dandoy:2022prp, OHare:2023rtm, DSouza:2024flu, DSouza:2024uud}. Informed by the latest simulations, we expect the local density of axions to be made up of three distinct populations: \begin{itemize}
    \item \emph{Minivoids}. The diffuse background of cold axions, which are diffuse and never cluster into minihalos. This distribution nevertheless is still part of the virialised galactic halo today and so can be modelled with a Maxwellian speed distribution (i.e. Model 0), but with a suppressed density fraction, $\rho_{\rm void}/\rho_{\rm DM} <1$. N-body simulations of axion minicluster formation found that the typical ambient background dark matter density is around $\rho_{\rm void}/\rho_{\rm DM} \sim 0.08$~\cite{Eggemeier:2022hqa}, which we adopt as a fiducial value.
    \item \emph{Minicluster streams}. The majority of the mass of dark matter is initially clustered in structures of up to $\sim 10^{-6}~M_\odot$ with NFW profiles and $\lesssim$ mpc virial radii. Their concentrations make those at the high-mass end of the mass function highly susceptible to tidal disruption, with numerical simulations and semi-analytic modelling generally agreeing that most of the axions by mass will be stripped~\cite{DSouza:2024flu, DSouza:2024uud}. These axions will then be left behind in the form of tidal streams which will retain a similar velocity dispersion to their original host, but will be spread out over a length given by $L \sim \sigma_{\rm str}t_{\rm disr}$ where $t_{\rm disr}$ is the time since disruption. The streams are expected to have radii around a mpc and lengths around 0.1~pc, which would make their experimental signals reasonably persistent on the timescales we are interested in here\footnote{The typical duration a given stream will persist in the lineshape will be ${\cal O}$(months-years).}. Their combined density is expected to almost saturate the total average dark matter density of $\rho_{\rm DM}$, which is measured astronomically on much larger scales ($>100$~pc) than the typical inter-minicluster spacing. This tidal debris will make up the bulk of our lineshape model, as we detail further below.
    \item \emph{Miniclusters}. The smallest and most concentrated miniclusters---particularly those which have not undergone many mergers and so retain their highly cusped initial density profiles~\cite{Eggemeier:2024fzs}---are resilient to tidal disruption. These will be present in the galaxy even at the present day, potentially in large numbers, due to the bottom-heavy mass function for miniclusters. However, they make up a very small fraction of the total dark matter mass and also fill a negligible amount of the volume, implying an exceedingly small encounter rate in direct detection experiments. For this reason, we do not include any surviving miniclusters in our signal model, although an encounter with one is of course a possibility~\cite{Dandoy:2023zbi}.
\end{itemize}

We will construct models for the axion lineshape in the post-inflationary scenario in a probabilistic manner. Once we specify several model parameters---like the minicluster mass function, concentration parameters, level of heating due to disk shocks etc.---we are able to estimate the expected number of streams overlapping at a position inside the galaxy as well as their expected densities and dispersions. Unfortunately, the input parameters describing the initial population of miniclusters are highly uncertain, as we will now discuss. 

Generally speaking, the mass function will be of a Schechter-like form \begin{equation}
    \frac{{\rm d}n}{{\rm d}\log_{10} M} \propto M^{-\gamma}\exp\left(-\frac{M}{M_{\rm max}}\right) \, ,
\end{equation} 
where $\gamma$ is a power law index and $M_{\rm max}$ is a cut-off in the minicluster mass distribution representing the largest mass a minicluster can grow to before it begins to be tidally stripped during infall into the halo and through stellar encounters. The index $\gamma$ has been determined through semi-analytical modelling~\cite{Fairbairn:2017sil} to be $\gamma = 0.5$, which broadly agrees with the index found in N-body simulations~\cite{Pierobon:2023ozb}. However, there are still large uncertainties inherent in the lattice simulations of the cosmic-string/domain wall collapse that are fed into gravitational simulations of minicluster formation, which then propagate into large uncertainties in the mass function. See, for example, Ref.~\cite{Pierobon:2023ozb}, which shows that the mass function can shift in both overall mass scale and in the slope depending on the early-Universe simulation results. We deal with this in our analysis by treating $\gamma$ and $M_{\rm max}$ as parameters we will sample probabilistically within some prior range as summarised in Eq.(\ref{eq:minicluster_prior}) below.

Once the mass function has been specified, we can then sample minicluster masses from it. Each sampled mass $M$ can be assigned an NFW virial radius $R$, which we parameterise as follows, 
\begin{equation}
R = R_{10} \left( \frac{M}{10^{-10}~M_{\odot}} \right)^{1/3} \, ,
\end{equation} 
where the free parameter $R_{10}$ is a proxy for the concentration of the minicluster just prior to the start of tidal stripping. This parameter is similarly uncertain, so we select it over a prior range $R_{10}\in (0.1,1)$ mpc (see Ref.~\cite{OHare:2023rtm} for more discussion). These parameters also dictate the initial minicluster velocity dispersion,
\begin{equation}
    \sigma_{\rm mc} = \sqrt{\frac{\beta G M}{R}} \, ,
\end{equation}
where $\beta=3.54$~\cite{Kavanagh:2020gcy}.

With the original miniclusters' properties in place, we can move on to modelling their streams. Since our goal here is primarily to bracket a reasonable range of lineshapes, we will not perform any simulations of the tidal disruption process as the miniclusters orbit the galaxy, as was done in Ref.~\cite{Kavanagh:2020gcy, OHare:2023rtm}. Instead, we take the insight gained from Ref.~\cite{OHare:2023rtm}, which is that almost all of the axions by mass are stripped from the miniclusters, with only the isolated miniclusters at the low-mass tail of the distribution surviving appreciably. So we make a simplification here and assume that all of the axion miniclusters are stripped within the first two Gyr of the galaxy's life (which roughly brackets the time of the first disk crossing for orbits intersecting the solar position today). Once the axions are unbound from the minicluster, they will continue to orbit around the galaxy on roughly the same orbit their progenitor was on, but their internal velocity spread will cause them to diverge spatially, forming an elongated tidal stream of length,
\begin{equation}
    \ell_{\rm str} \sim 2h\sigma_{\rm mc}  t_{\rm disr} \, ,
\end{equation}
where $t_{\rm disr}$ is the time since the minicluster became unbound. We introduce here a factor $h$ which accounts for the additional heating of the stream due to the tidal shocks felt as the stream passes through the galactic disk, see e.g.~\cite{Ostriker72, Gnedin:1998bp}. For miniclusters of the size we are studying here, where $\sigma_{\rm mc}\sim 0.01$~km~s$^{-1}$, each disk crossing is expected to heat the axions in the stream by a few per cent, which, after $\sim$100 disk crossings for a typical orbit, will amount to an ${\cal O}(1)$ factor for $h$. A more precise estimation of this should be possible through further simulations, but this is currently missing in the literature.

With this knowledge, we can compute the \textit{expected} number of overlapping streams using a simplified version of the approach taken in Ref.~\cite{OHare:2023rtm}. We start by noticing that, on $r\gtrsim$ 100~pc scales in the solar neighbourhood, the density of dark matter should be approximately $\rho_{\rm DM} \sim 0.01~M_\odot~{\rm pc}^{-3}$, as measured via the kinematics of nearby disk stars~\cite{Read:2014qva}. This means there are some very large $N_{\rm mc}$ miniclusters in that volume:
\begin{equation}
    N_{\rm mc} \sim \frac{V_{\rm local}\rho_{\rm mc}}{\langle M\rangle} =\frac{4}{3}\pi r^3 (1-f_{\rm void})\frac{\rho_{\rm DM}}{\langle M\rangle} \sim 10^{14} \, .
\end{equation}
Assuming the miniclusters are fully disrupted into a tidal stream by the present day, each one now occupies a volume,
\begin{equation}
    V_{\rm str} \sim \pi R^2 \ell_{\rm str} \, .
\end{equation}
Since $\ell_{\rm str}\sim 0.1~$pc for typical minicluster parameters, there will be many streams overlapping at each point inside a $100$~pc radius sphere around the Sun. So if we are sampling the distribution of axions at some point, the expected number of overlapping streams at that point is going to be,
\begin{equation}
    \left\langle N_{\mathrm{str}}\right\rangle=\sum_{j=1}^{N_{\mathrm{mc}}} \frac{V_{\mathrm{str}}^j}{V_{\text {local }}} \sim \frac{2 \pi R_{10}^{3 / 2}}{\left(10^{-10} M_{\odot}\right)^{1 / 2}} \sqrt{G \beta} (1-f_{\rm void})\rho_{\mathrm{DM}}t_{\rm disr} \sim 200 \, .
\end{equation}
Note that because we are fixing the total mass of dark matter inside $V_{\rm local}$, the dependence on the minicluster mass drops out of this expression and, by extension, it does not depend on $\gamma$ or $M_{\rm max}$.

Note that this is the \textit{mean} number of streams overlapping at a point. To fix the number that actually overlap for a given hypothetical experiment, we must sample from a Poisson distribution with mean $\langle N_{\rm str}\rangle$. Because $\langle N_{\rm str}\rangle$ is large, we will draw $N_{\rm str}$ from a Gaussian distribution with standard deviation $\sqrt{\langle N_{\rm str}\rangle}$. Once this value is chosen, we then generate $N_{\rm str}$ streams, each with a mass drawn from the mass function weighted by the stream volume $V_{\rm str}$. The logic behind performing a volume-weighted sampling of the miniclusters is that larger streams occupy more space, and so we are more likely to find ourselves inside one. The stream dispersions are then fixed by the model parameters to be $\sigma_{\rm str} = h\sigma_{\rm mc}$. The stream densities will be of order $\rho_{\rm str} \sim M/V_{\rm str}$, but as in Ref.~\cite{OHare:2023rtm} we will apply a simple Gaussian profile along each stream to account for the fact that it will not have perfectly spread out into a homogenous cylinder, given that the mass stripping is gradual and the original minicluster was centrally concentrated.

The full galactocentric-frame velocity distribution for the minicluster scenario is then given by,
\begin{equation}
    f_{\rm mc}(\mathbf{v}) = f_{\rm void}f_0(\mathbf{v}) + \sum_i^{N_{\mathrm{str}}} \frac{\rho_{\mathrm{str}}^i}{\rho_{\mathrm{DM}}} f_1\left(v ; \mathbf{v}_{\mathrm{str}}^i, \sigma_{\mathrm{str}}^i\right) \, ,
\end{equation}
where $f_0(\mathbf{v})$ is the SHM's velocity distribution (Model 0). As in previous sections, the stream velocities $\mathbf{v}_{\rm str}$ are all sampled from the SHM since the miniclusters and their streams will be on typical halo orbits within the Milky Way potential. Note that, unlike our other examples in this study, in this model specifically, we are \textit{not} enforcing that the velocity distribution normalise to unity. This is because the variability in the number of streams overlapping in the solar neighbourhood implies that there are real (albeit small) physical variations in the value of the dark matter density on sub-pc scales. Although this naively would suggest we are allowing for artificial enhancements in the sensitivity compared to the SHM alone, we actually find that the total density fraction from a summed sample of streams (i.e.~$\sum \rho^i_{\rm str}/\rho_{\rm DM}$) is generally slightly smaller than 1---between 0.8--0.9 on average depending on model parameters. So the total signal power in this model turns out to be slightly penalised compared to the baseline SHM where $\rho_{\rm DM} = 0.01~M_\odot~{\rm pc}^{-3}$ constitutes the full measurable density.

\begin{figure}
    \centering
    \includegraphics[height=0.46\linewidth]{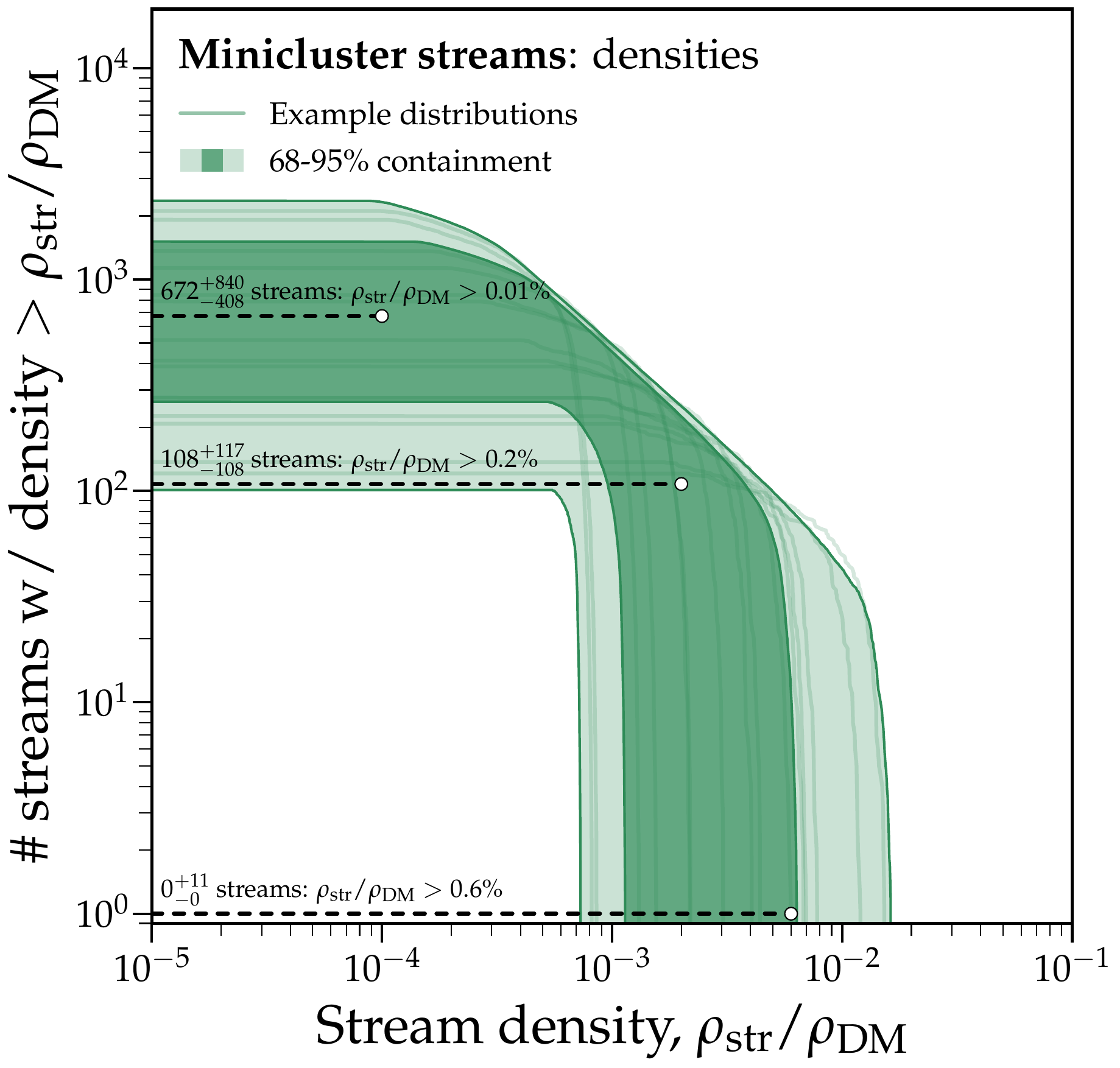}
    \includegraphics[height=0.46\linewidth]{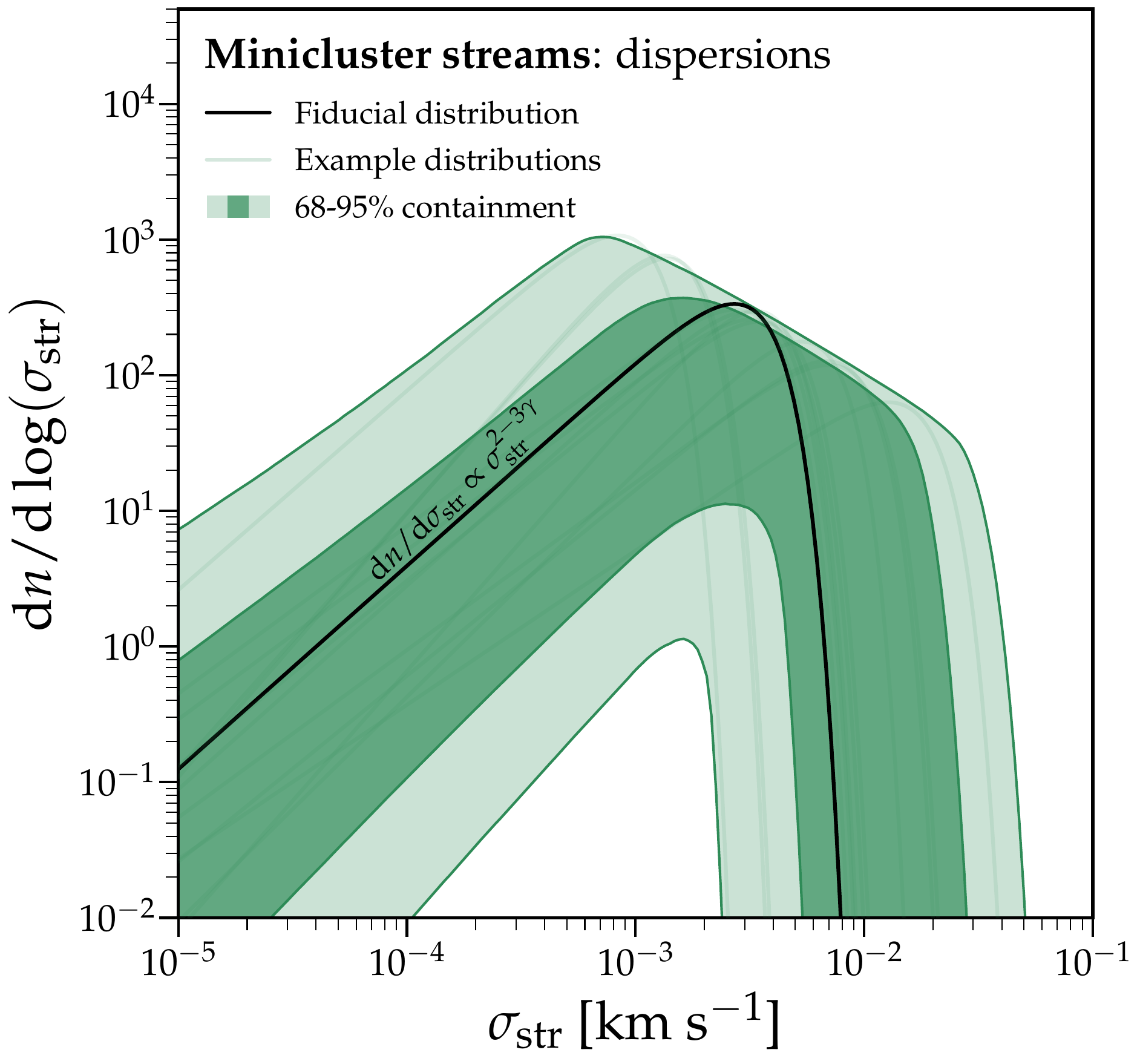}
    \caption{{\bf Left}: Cumulative distribution function for minicluster stream densities after Monte-Carlo generating $10^4$ realisations within the parameter prior ranges in Eq.(\ref{eq:minicluster_prior}). The dark and light shaded regions enclose 68 and 95\% of the possible realisations. A few example cumulative distributions are shown as transparent lines. The quoted stream numbers for the three highlighted density fractions refer to the median and (asymmetric) $1\sigma$ spread around it. {\bf Right}: The distribution of minicluster stream dispersions, $\sigma_{\rm str}$. We highlight our fiducial model, which corresponds to the case where $\gamma = 0.5$ and $M_{\rm max} = 10^{-6}~M_\odot$ as a solid black line, as well as a few alternative distributions. The initial slope of the distribution is related to the slope of the minicluster mass function, $\gamma$.}
    \label{fig:MiniclusterStream_properties}
\end{figure}

With this procedure for generating a local velocity distribution for the minicluster scenario from $f_{\rm mc}{(\mathbf{v}})$, we will now show how this model appears at the level of the lineshape. First, in Fig.~\ref{fig:MiniclusterStream_properties}, we summarise the distributions of the basic stream parameters, namely the expected local stream density fractions and their velocity dispersions. In this example, we have implemented a probabilistic procedure for accounting for the various theoretical uncertainties by sampling model parameters from the following priors,
\begin{align}\label{eq:minicluster_prior}
    \gamma &\in \mathcal{U}(0.3,0.7) \nonumber\\
    \log_{10}\left(\frac{M_{\rm max}}{M_{\odot}}\right)&\in \mathcal{U}(-7,-4.5) \nonumber\\
    \frac{t^i_{\rm disr}}{{\rm Gyr}} &\in \mathcal{U}(10,8) \\
    \frac{R_{\rm 10}}{{\rm mpc}} &\in \mathcal{U}(0.1,1)\nonumber \\
    h^i &\in \mathcal{U}(1,5)  \, ,\nonumber
\end{align}
where we include an index $i$ on the parameters that are sampled at the level of each individual minicluster, as opposed to at the level of the population.

From Fig.~\ref{fig:MiniclusterStream_properties} we see that there are typically several hundred streams observed in each realisation, with a sharply falling density distribution. Most streams make up at least 0.01\% of the total density, with a few rare cases making up to 1\%. The stream dispersions are typically in the range $\sigma_{\rm str}\lesssim 0.01$~km~s$^{-1}$ but have a lopsided distribution that is weighted towards an upper cut-off set by $M_{\rm max}$. This is due to the fact that larger dispersion values are associated with the more massive progenitor miniclusters, which are more prevalent in each sample because they occupy a greater volume.

 \begin{figure}
    \centering
    \includegraphics[width=0.99\linewidth]{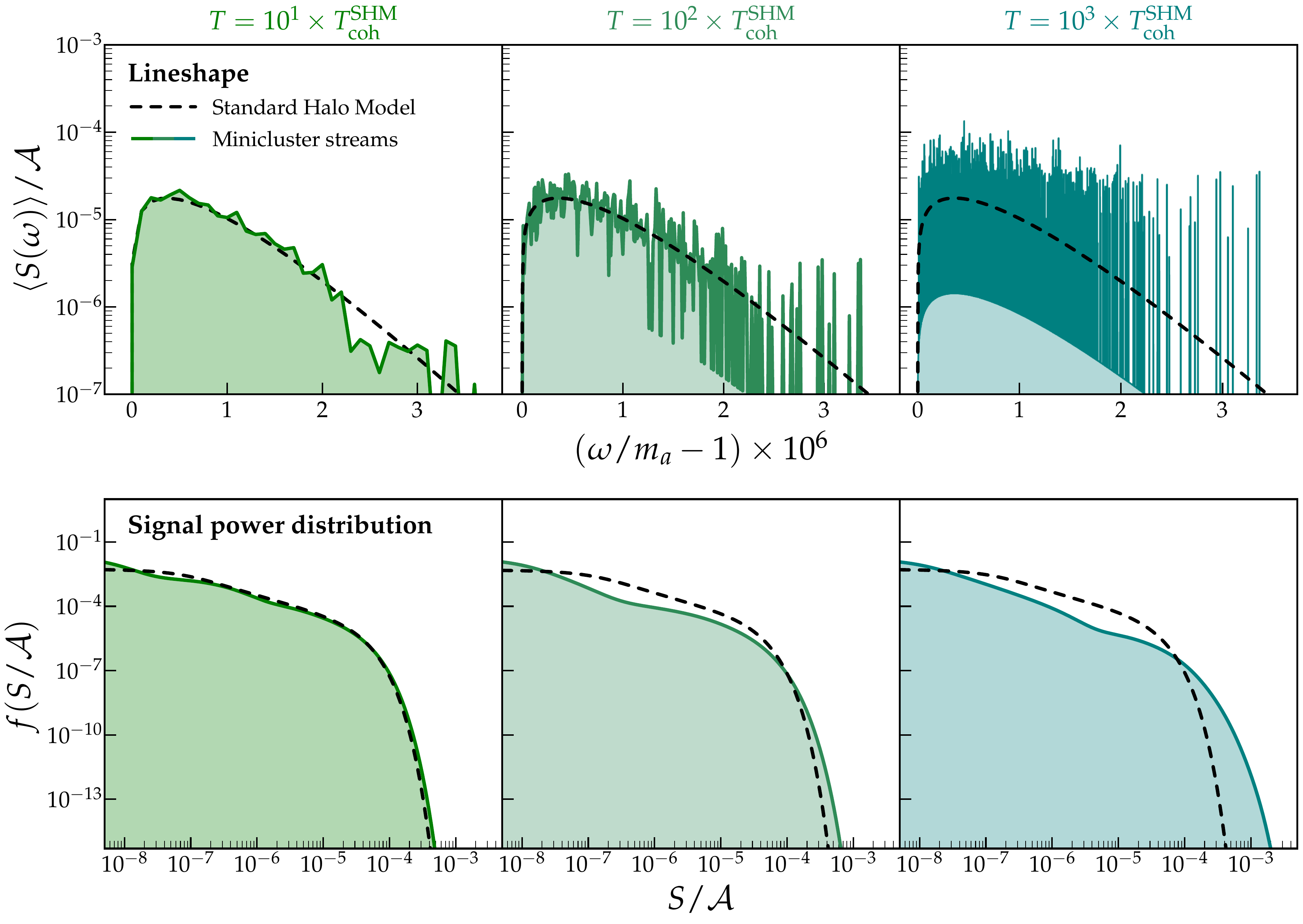}
    \caption{Lineshapes (top panels) and signal power distributions (bottom panels) for a single realisation of a minicluster stream model as described in Sec.~\ref{sec:miniclusters}. The panels from left to right show the lineshape at varying spectral resolution, corresponding to integration durations that are $10$, $10^2$ and $10^3$ times the coherence time of the SHM (which we define to be $T^{\rm SHM}_{\rm coh} = 10^6 T_{\rm ax}$). The black dashed line in both cases shows the SHM lineshape. Note that the top distribution shows the \textit{mean} lineshape $\langle S(\omega)\rangle$ for this realisation of streams, whereas the bottom panels show the full distribution of signal powers accounting for the fact that the values of $S(\omega)$ in each frequency bin are exponentially distributed. The right-most panel suggests that integration times longer than a thousand coherence times are needed to reveal the presence of the streams as large signal-power peaks in the tail of the $f(\mathcal{S})$ distribution---these are the upward fluctuations that will be picked up most readily by the EVS test.}
    \label{fig:MiniclusterStreams_lineshape}
\end{figure}

The distributions in Fig.~\ref{fig:MiniclusterStream_properties} cover the full range of possible stream properties accounting for all model uncertainties. We now take one specific realisation of a local minicluster stream ensemble to create some example lineshapes and signal power distributions shown in Fig.~\ref{fig:MiniclusterStreams_lineshape}. The lineshape plots (upper three panels) show that the streams are expected to be very narrow in frequency space and start to become prominent over the void axion background for integration times longer than $10^3 T^{\rm SHM}_{\rm coh}$. This is also illustrated in the signal amplitude distributions (lower three panels), where a tail of larger signal powers than are expected under the SHM becomes populated for the longer integration times. For low integration times, the two distributions are very similar, which is to be expected by construction because we have a similar total density of axions in both cases, and the distribution of minicluster stream velocities ($\mathbf{v}_{\rm str}$) has been chosen to match the SHM velocity distribution. These lineshapes will form the basis of our statistical analysis detailed in the next section.

\subsection{Haloscope sensitivity}\label{sec:minicluster_sensitivity}

\begin{figure}
    \centering
    \includegraphics[width=0.49\linewidth]{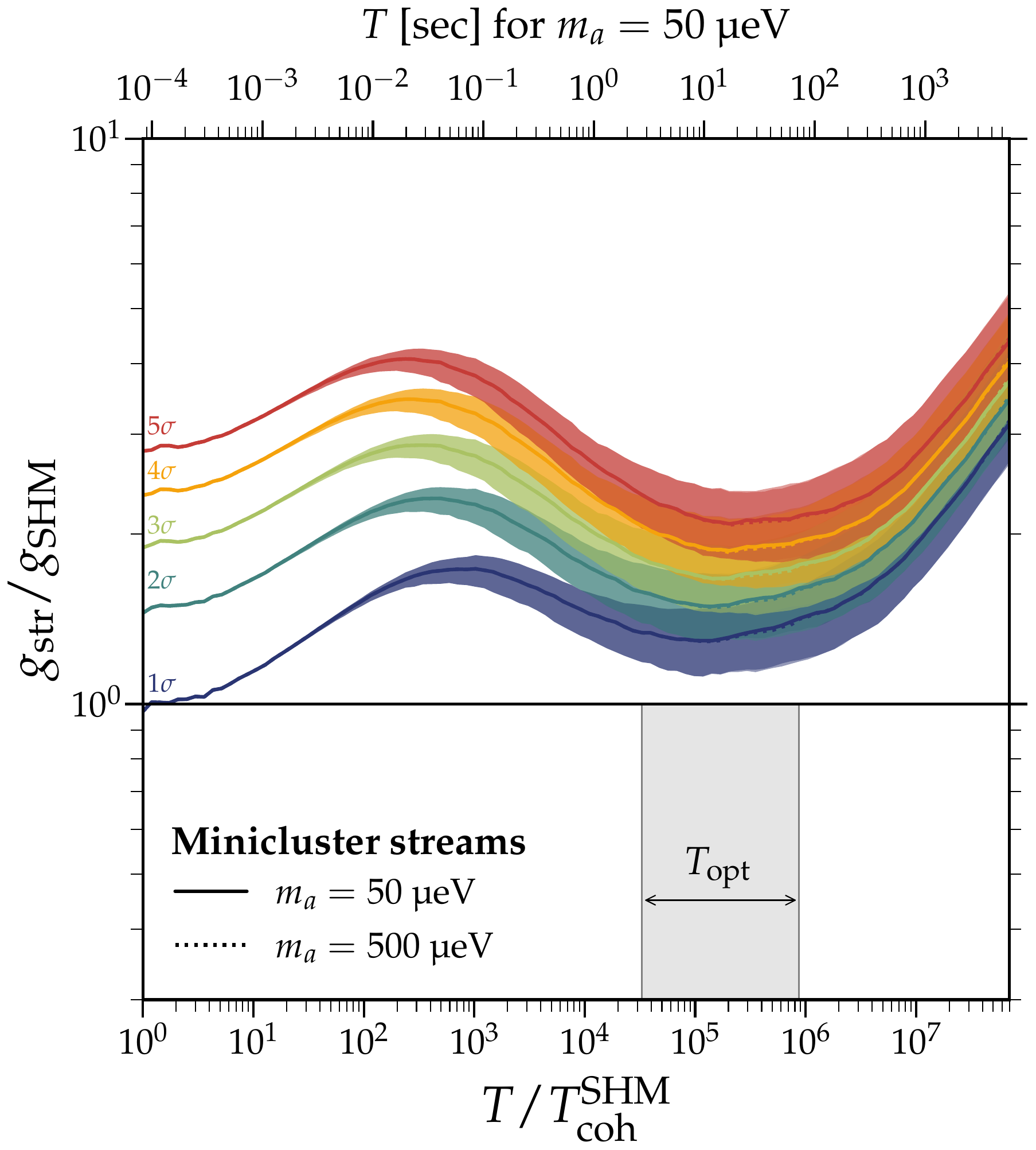}
    \includegraphics[width=0.49\linewidth]{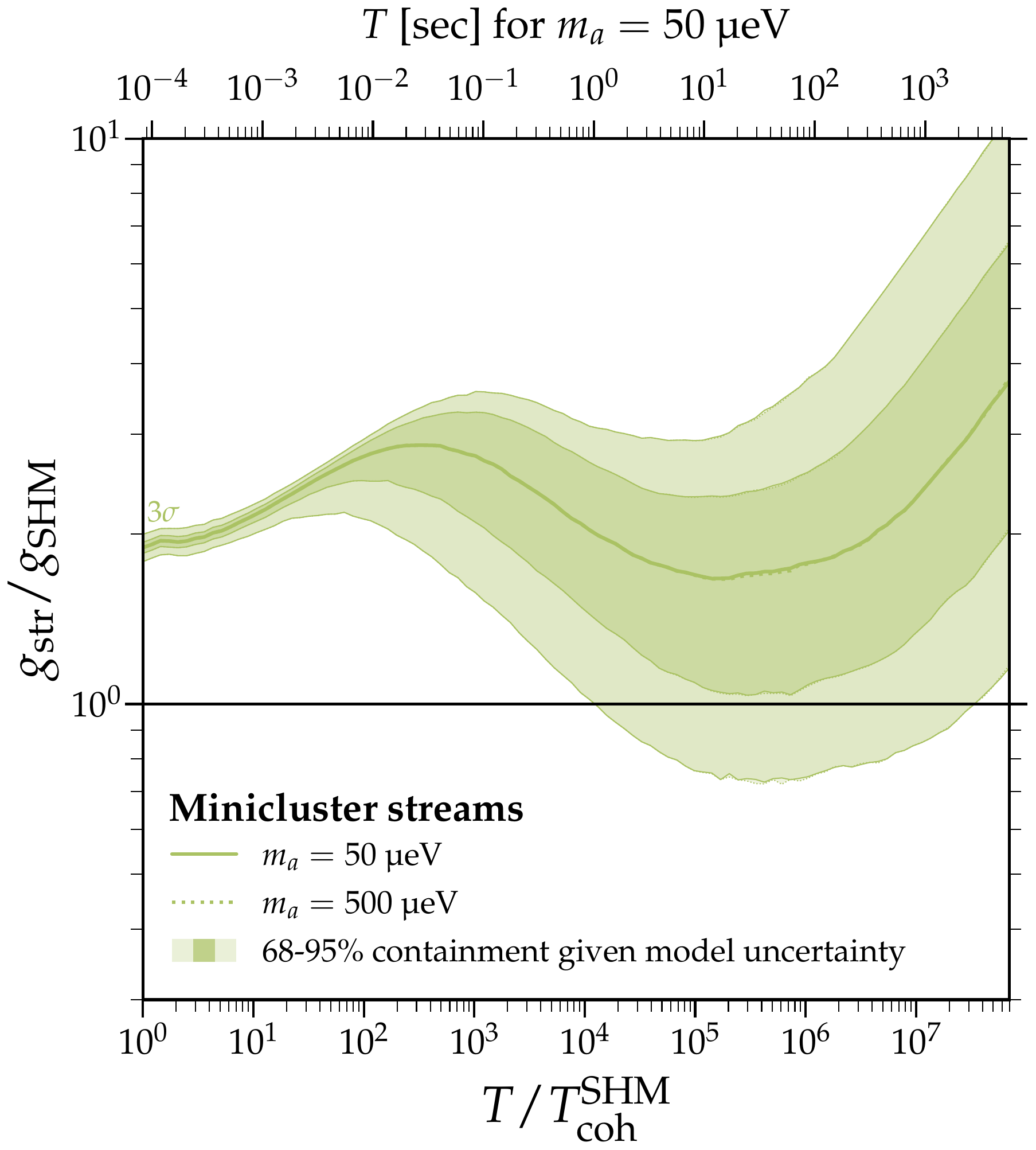}
    \caption{Coupling required for the detection of candidate streams relative to the coupling required to exclude the axion under the SHM. To construct these plots, we fix the value of $\mathcal{A}/\mathcal{B} \propto g^2$ to be what is required to give a 95\% CL exclusion of the axion under the SHM assumption for a given value of $T$. We then ask how large the axion-photon coupling needs to be compared to this value for the median experiment to see an $n\sigma$ significant extreme power fluctuation due to a stream---this is then expressed as the ratio $g_{\rm str}/g_{\rm SHM}$. The left highlights the different significance levels but encloses only 25\% of the model uncertainty for clarity. The right-hand panel, on the other hand, selects only the $3\sigma$ case but shows the full 68 and 95\% containment when sampling over the range of possible minicluster stream models, demonstrating that extended sensitivity ($g_{\rm str}/g_{\rm SHM}<1)$ is possible in some cases, as long as the integration time is chosen optimally, i.e. $T = T_{\rm opt}$---see Eq.(\ref{eq:Topt_miniclusters}). There are solid and dotted lines shown in both plots, corresponding to two different axion masses; however, these lines are almost identical because the dependence on the axion mass drops out for this particular case, as explained further in the text.}
    \label{fig:MiniclusterStreams_PmaxProb_Coupling}
\end{figure}

\begin{figure}
    \centering
    \includegraphics[width=0.69\linewidth]{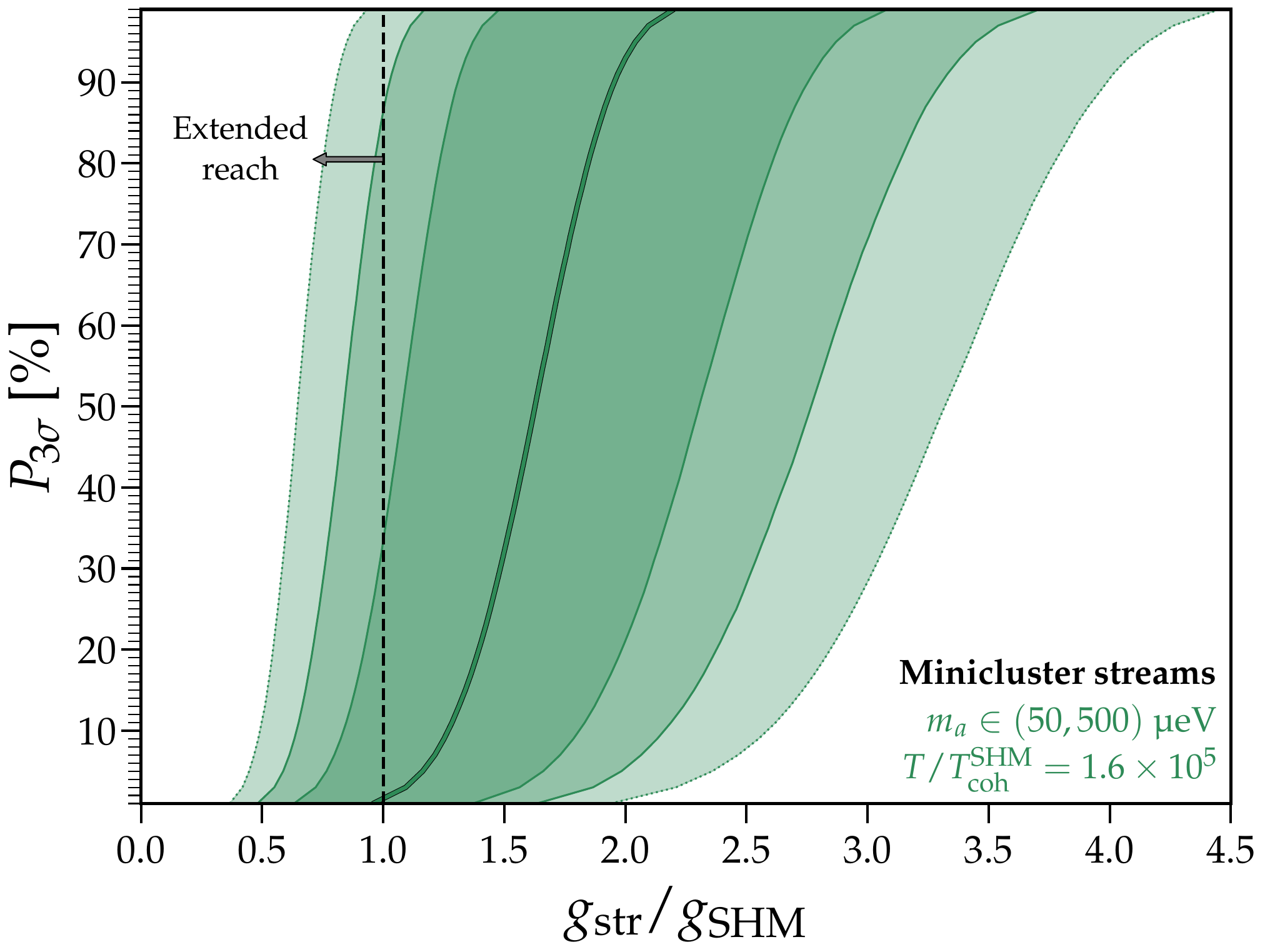}
    \caption{Probability of seeing a $3\sigma$ significant extreme value statistic $P_{\rm max}$ across the axion signal bandwidth as a function of the coupling ratio $g_{\rm str}/g_{\rm SHM}$ as defined in the text. We fix the integration time to the optimal value for this scenario of $T = T_{\rm opt} \approx 1.6\times 10^5 T^{\rm SHM}_{\rm coh}$. The shaded bands contain 68, 95 and 99\% of our models sampled within the parameter priors in Eq.(\ref{eq:minicluster_prior}). As shown in Fig.~\ref{fig:MiniclusterStreams_PmaxProb_Coupling}, there is negligible dependence on the axion mass in this example, so this plot applies to any axion mass for which the minicluster scenario is predicted, i.e.~roughly the range 50 to 500 $\upmu$eV. All cases where this coupling ratio is less than one constitute a scenario in which a real candidate signal would emerge in a high-resolution analysis, even though a low-resolution analysis of the same data would not discover the axion.}
    \label{fig:MiniclusterStreams_ExtendedReach}
\end{figure}

We will now take these simulated lineshape models and quantify the extent to which these could enhance the detectability of an axion signal. For this task, we will apply the EVS test as described in Sec.~\ref{sec:EVS}. To try to make the discussion pointed towards the idea of enhancing the discovery prospects for the axion, we will frame our analysis around the following question: if a haloscope experiment takes data for some duration $T$, for what value of the axion-photon coupling would there be, say, a $50\%$ chance of detecting an $n\sigma$ EVS value (i.e.~a $Z$ score of $n$). We will express the answer to this question as a ratio of couplings, $g_{\rm str}/g_{\rm SHM}$, where $g_{\rm SHM}$ is the value of the axion-photon coupling that would be excluded at 95\% CL under the standard assumption of an SHM lineshape. So if, for example, $g_{\rm SHM} = g_{\rm KSVZ}$, this means the median experiment observing no axion signal after integrating for time $T$ could set a 95\% CL exclusion limit using the PLR test down to the KSVZ line. So that means that if we were to find, for instance, $g_{\rm str} = 0.5g_{\rm KSVZ}$ for one of our stream lineshapes, then this means that an axion that couples \textit{weaker} than the naive sensitivity of the experiment has a chance to be revealed in a high-resolution mode as an $n\sigma$ extreme single-bin fluctuation. This fluctuation alone would then form a strong candidate that would be followed up with further interrogation in a re-scan. So with this in mind, any values of $g_{\rm str}/g_{\rm SHM}<1$ obtained in a high-resolution integration of duration $T$ can be thought of as extending the reach of the haloscope in terms of the axion-photon coupling. We frame the question in this way because it aligns with the motivating philosophy behind high-resolution haloscope analyses---namely that one is gambling on the possibility of there being features in the lineshape that would enhance the signal, but only by analysing existing data in a different way, as opposed to collecting more data. 

So how do we calculate $g_{\rm str}$? The procedure to calculate this extended reach is to first choose a value of $T$ and then solve Eq.(\ref{eq:exclusion}) to find the value of $\mathcal{A}/\mathcal{B}$ that corresponds to a median 95\% CL \textit{exclusion} while assuming that the SHM describes the data. Then, we ask what value of $\mathcal{A}/\mathcal{B}$ is required to obtain a median EVS $Z$-score of $n = 1,...,5$, when the lineshape is described by our minicluster stream model. Because $\mathcal{A}\propto g_{a\gamma}^2$, the square root of the ratio of these two values of $\mathcal{A}/\mathcal{B}$ gives us $g_{\rm str}/g_{\rm SHM}$. We emphasise here that an $n\sigma$ extreme value statistic is \textit{not} the same as a peak in the power spectrum that is $n$ standard deviations above the noise level (i.e.~a \textit{local} power excess). Because the EVS test incorporates the knowledge of how many frequency bins are being tested, it is already naturally corrected for the look-elsewhere effect. As an example, take $T/T_{\rm SHM}^{\rm coh} = 10^4$ (cf.~Fig.~\ref{fig:EVS_MonteCarlo}): a $3\sigma$ EVS value is around $P_{\rm max}/\mathcal{B} \sim 16$, so is effectively a $16\sigma$ local significance excess above the noise.

The left-hand side of Fig.~\ref{fig:MiniclusterStreams_PmaxProb_Coupling} shows the value of $g_{\rm str}/g_{\rm SHM}$ as a function of $T$ for various significance levels, while the right-hand side highlights the $3\sigma$ case so as to capture the full range of possible models in this scenario (we enclose only 25\% of the theoretical uncertainty in the left-hand plot so the different significance lines can be distinguished). The behaviour of $g_{\rm str}/g_{\rm SHM}$ as $T$ increases goes through several regimes which can be understood as follows. For very small values of $T$, the lineshape in both the stream and SHM cases is contained in a small number of bins and is essentially identical. So here the corresponding values of $g_{\rm str}/g_{\rm SHM}$ just reflect the enhancement in the signal strength required to amplify the signal to be $n\sigma\times$ the noise level. As $T$ increases, so too does the number of bins being tested, and so the required signal strength needed to guarantee an $n\sigma$ extreme value statistic must increase to compensate. The reason behind this behaviour is that we are fixing $g_{\rm SHM}$ to be the coupling required to exclude an axion signal at 95\% CL using the PLR test---so the fact that the ratio $g_{\rm str}/g_{\rm SHM}$ is increasing here is a reflection of the fact that the PLR test scales more favourably with $T$ than the EVS test (as shown in Fig.~\ref{fig:EVS_Zscore}). However, when a certain value of $T$ is reached, we see the lines of $g_{\rm str}/g_{\rm SHM}$ turn over and begin to decrease. This reflects the time when the streams are starting to be resolved, and so the EVS test has a better chance of identifying real signal peaks over the noise level. Eventually, the curves reach a minimum, which occurs when the streams are resolved. We consider this to be the \textit{optimal integration time}, $T_{\rm opt}$ for this scenario. We compute it numerically in this example by calculating the average width in frequency space of every stream, weighted by their densities. Heuristically though, this time is essentially when the width of a typical stream matches the frequency resolution, i.e. 
\begin{equation}\label{eq:Topt_miniclusters_step}
    T_{\rm opt} \sim \frac{2\pi}{ m_a |\mathbf{v}_{\rm lab} - \mathbf{v}_{\rm str}|\sigma_{\rm str}} = \frac{T_{\rm coh}^{\rm SHM}}{10^6 |\mathbf{v}_{\rm lab} - \mathbf{v}_{\rm str}|\sigma_{\rm str}} \, ,
\end{equation}
Note that this means that the ratio $T_{\rm opt}/T^{\rm SHM}_{\rm coh}$ turns out to be independent of the axion mass---comparing the solid and dotted lines in Fig.~\ref{fig:MiniclusterStreams_PmaxProb_Coupling}, we see that indeed the results we find for the two mass cases are essentially identical\footnote{The very minor differences between the two axion mass cases around $T_{\rm opt}$ are a result of the daily modulation effect---we deem these differences to be too small to demand much further discussion.}. That said, since $T^{\rm SHM}_{\rm coh}$ depends on $m_a^{-1}$, the actual numerical value of $T_{\rm opt}$ \textit{is} $m_a$ dependent. Using our full numerical evaluation, we find that the optimal time is on the order of a few seconds for a typical mass in the post-inflationary axion mass window,
\begin{equation}\label{eq:Topt_miniclusters}
    T_{\rm opt} \approx 1.6\times 10^5 \,T^{\rm SHM}_{\rm coh} = 6.56 \, {\rm sec} \, \left(\frac{100\,\upmu{\rm eV}}{m_a}\right) \, .
\end{equation}

For values of $T$ larger than $T_{\rm opt}$, we then see that the coupling ratio starts to increase again. This loss in sensitivity in the EVS test can again be understood from the scaling shown in Fig.~\ref{fig:EVS_Zscore}. Here, the streams' lineshapes are now being split across more bins; this reduces the sensitivity of the EVS test, which requires large single-bin fluctuations, as compared to the PLR test, which uses information across all bins. 

Our conclusion here aligns with the general intuition developed in previous high-resolution axion haloscope studies, which is that multiple frequency bin widths should be trialled since the optimal sensitivity will be obtained when the bin width matches the frequency spread of the feature of interest. Here, we place these ideas on more quantitative grounds in the context of a physically motivated model, and find that the optimal integration time for revealing minicluster streams specifically is the one stated above.

To finish this section, we should remark on the numerical values of $g_{\rm str}/g_{\rm SHM}$ found in the plot above. As the right-hand panel shows, these values only dip below 1 for the top $\sim$16\% of models where the maximum stream density happens to be large. We emphasise, though, we are showing the \textit{median} $Z$-scores here, i.e.~the significance reached by at least 50\% of hypothetical experiments. It is, of course, possible for an experiment to get lucky or unlucky. So to illustrate this more clearly, we expand upon the previous figure in Fig.~\ref{fig:MiniclusterStreams_ExtendedReach} where we now fix the integration time to $T_{\rm opt}$, and fix the significance level to $3\sigma$ (which corresponds to a $\sim20\sigma$ local excess when $T =T_{\rm opt}$); only now we are showing the value of $g_{\rm str}/g_{\rm SHM}$ for a range of probabilities. Here we see that even couplings that are as little as half of the sensitivity reached under a standard low-resolution axion search (i.e. $g_{\rm str}/g_{\rm SHM} = 0.5$) have a rare ($\lesssim$ 20\%) but non-negligible chance to come across a large power fluctuation due to a stream. We stress that all that is needed in axion haloscope searches are \textit{candidate} signals, since any interesting excesses can be re-scanned. If one of these rare signals due to a stream happened to be detected, it would persist upon such a re-scan and so lead directly to a confirmed discovery of an axion signal. What we have shown here is that this scenario is possible even if the true axion-photon coupling were a factor of two smaller than the exclusion limit made under the assumption of the SHM lineshape---for reference, the ratio of the KSVZ to DFSZ axion couplings is around 0.4.

\section{CDM fine-grained streams}\label{sec:finegrained}
The next possibility we consider for local fine-grained substructure extends beyond the post-inflationary scenario. In fact, this case should apply to any generic CDM model, including pre-inflationary ALPs and axions, so long as the initial power spectrum for density perturbations is set only by the standard nearly-scale-invariant primordial fluctuations inherited from inflation. This is the fine-grained dark matter substructure alluded to in the introduction (see Fig.~\ref{fig:PhaseMixing}), which emerges purely as a result of the gravitational collapse of a smooth density sheet into modern phase-mixed galactic halos. In this section, we present comparable results to those in the previous section for minicluster streams, after first elaborating on the motivation behind this scenario and explaining how we construct lineshape models to reflect it.

\subsection{Lineshape construction}
In a CDM cosmology, the fine-grained phase space structure of the dark matter distribution at a point in the inner galaxy is expected to be very smooth, but not \textit{arbitrarily} so. This is because of the way in which gravitational collapse into halos proceeds---via the repeated folding of an almost three-dimensional hypersurface embedded in the six-dimensional phase space. In the inner regions of dark matter halos where we reside, the dark matter distribution has undergone a very high degree of phase mixing, which means the phase space will be foliated by an extremely large number of folds of this original sheet. At a single spatial point where we hope to measure the dark matter directly in an experiment, each of these folds manifests as a stream---groupings of dark matter particles travelling along a particular direction. Each stream can be assigned a density set by how much the corresponding part of the phase space has been stretched or squeezed during this folding process. Each one will also have a velocity dispersion set approximately by the initial velocity dispersion of the dark matter particles at production. In the case of GeV-scale thermally-produced relic dark matter, these velocity dispersions are at the level of $\sigma_v \sim 10^{-5}$~km/s, whereas for axions in, say, the pre-inflationary scenario, this initial spread in velocities can be taken to be essentially negligible because the misalignment mechanism populates the zero-momentum mode of the field.

So the ingredients we need to construct a model for this scenario are the expected number and distribution of stream densities at our position in the halo. This question has received some attention in the simulation literature~\cite{Helmi:2002ss,Vogelsberger:2007ny, Dolag:2012uq, Zavala:2013bha, Zavala:2013lia}, but it cannot be answered using N-body treatments alone---additional techniques are required to extrapolate below the particle and spatial resolution of these simulations, which is far too coarse to address the fine-grained structure of the dark matter's phase space. The most useful estimate for our purposes is the study of Vogelsberger \& White~\cite{Vogelsberger:2010gd}. The authors addressed this issue using a technique based on the geodesic deviation equation. The geodesic deviation equation concerns the evolution of small displacements to phase space positions, which, if integrated alongside the equations of motion of particles in an N-body simulation, allows for the reconstruction of the full phase space around each particle. Each simulation particle can then be assigned to a stream with a calculable density. When applied to the Aquarius N-body simulation of a MW-like halo~\cite{Springel:2008cc} the integration of the geodesic deviation equation revealed an expected $10^{14}$ streams locally (i.e.~at the analogous solar radius in their halo), but with a substantial low-density tail---only around a million streams of the densest streams had densities exceeding $10^{-7}$ times the local DM density, and the most massive stream contributes only around 0.1\% of the dark matter.

\begin{figure}
    \centering
    \includegraphics[width=0.89\linewidth]{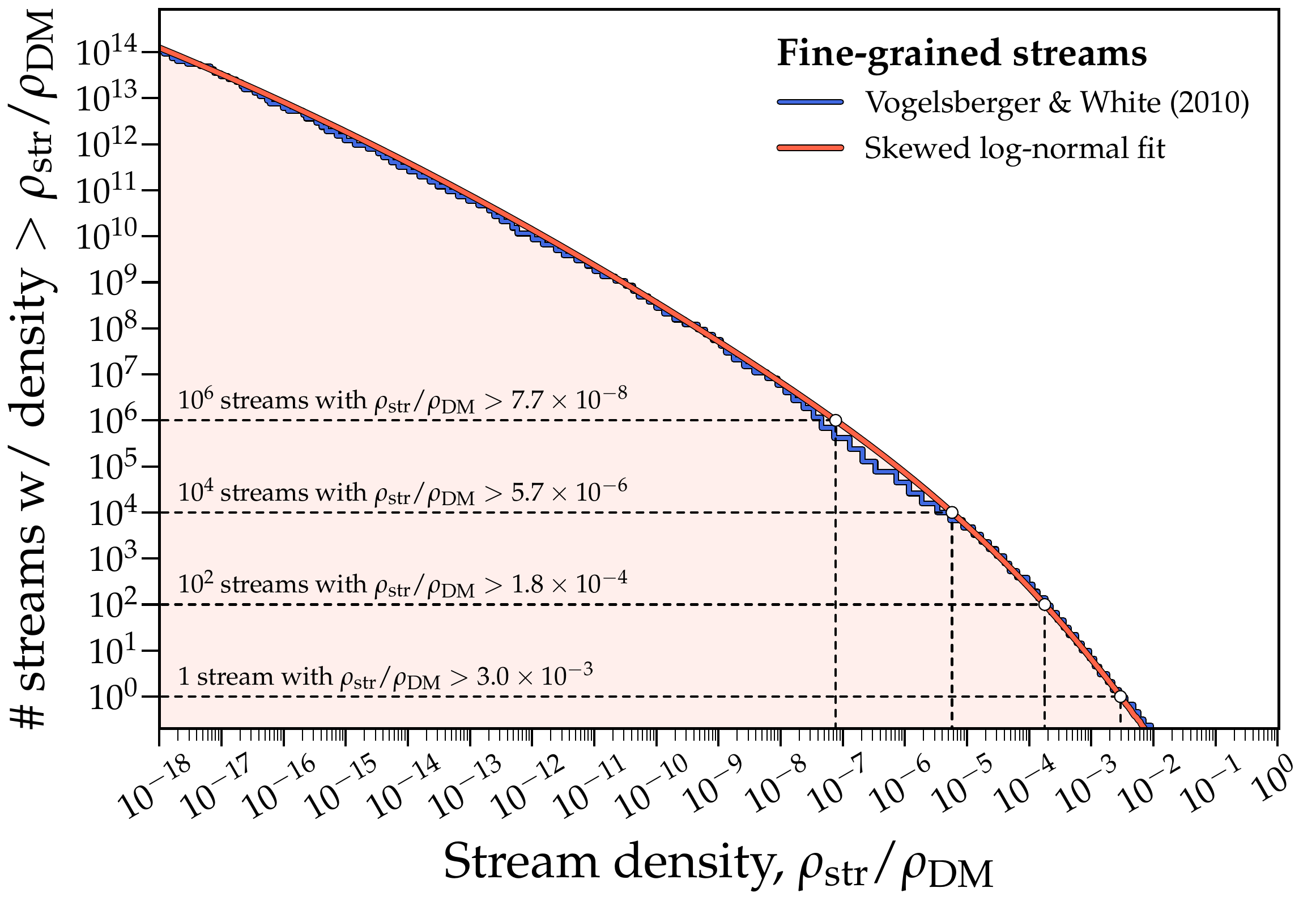}
    \caption{Cumulative distribution of the densities of fine-grained dark matter streams in the solar neighbourhood, where the vertical axis shows the number of streams that individually have local densities exceeding the value indicated by the horizontal axis. The blue line is the result from the simulations of Vogelsberger \& White (digitised from Fig. 9 in Ref.~\cite{Vogelsberger:2010gd}), whereas the orange line shows our analytic fit to a skewed log-normal distribution, which we use as the basis for our model. To aid the discussion in the text, we overlay dashed lines highlighting a few minimum stream densities above which the top one, hundred, ten-thousand, and million streams possess. For example, while there are $\sim 10^{14}$ streams overlapping our position, the single most massive one is expected to make up around 0.1\% of the local density of dark matter.}
    \label{fig:VWstreams_CDF}
\end{figure}

\begin{figure}
    \centering
    \includegraphics[width=0.99\linewidth]{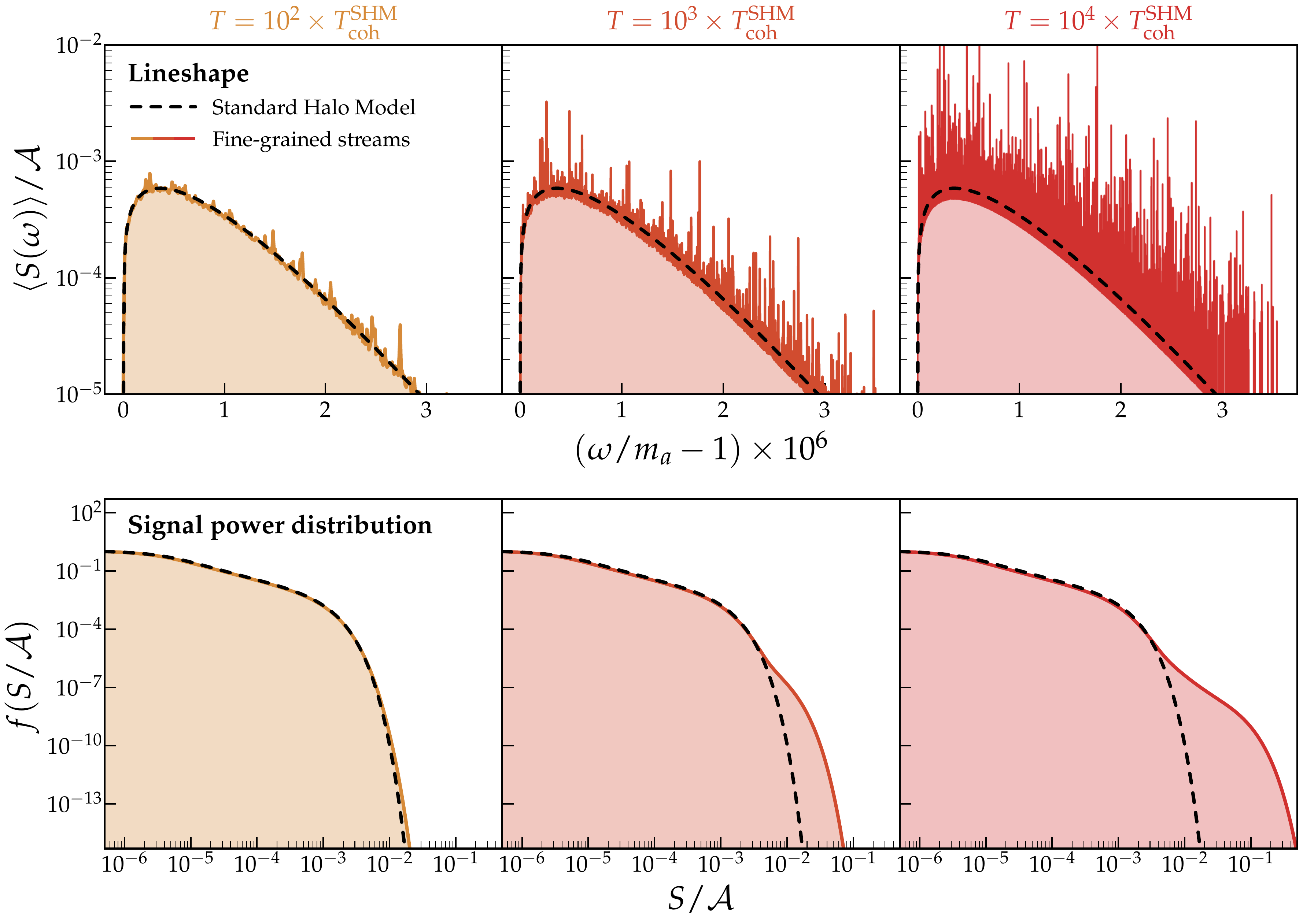}
    \caption{Lineshapes (top panels) and signal power distributions (bottom panels) under the fine-grained streams model described in Sec.~\ref{sec:finegrained}. The panels from left to right show the lineshape at varying spectral resolution, corresponding to integration durations that are $10^2$, $10^3$ and $10^4$ times the coherence time of the SHM. The black dashed line in both cases shows the SHM prediction---both the SHM and fine-grained stream models are normalised to the same local dark matter density. Note that the top distribution shows the mean lineshape $\langle S(\omega)\rangle$ for this realisation of streams, whereas the bottom panels show the full distribution of signal powers accounting for the fact that each $S(\omega)$ is exponentially distributed.}
    \label{fig:VWstreams_lineshape}
\end{figure}

We construct a model inspired by these results by first digitising Figure 9 of Ref.~\cite{Vogelsberger:2010gd}---this provides a mass-weighted distribution of stream densities per simulation particle at the solar radius in that simulated halo. We then convert that into a cumulative distribution function for stream densities, which is shown in Fig.~\ref{fig:VWstreams_CDF} in such a way that we can interpret the plot as the number of streams that \textit{individually} have a density larger than some fraction of the total dark matter density at our position (this is equivalent to the left-hand plot of Fig.~\ref{fig:MiniclusterStream_properties} we showed in the minicluster case). We find a reasonable analytic fit using a skewed log-normal distribution, which is what we will adopt here.\footnote{Specifically we use, $f(\rho) =$ \texttt{scipy.stats.skewnorm.pdf(log(rho),-2.10,loc=-11.37,scale=7.15)/rho}}.

It is highly impractical to simulate a lineshape by summing $10^{14}$ streams, but we also find it to be unnecessary in practice. The low-density tail of the distribution sums to form a distribution that is essentially indistinguishable from a perfectly smooth distribution, with only a fraction of the high-density tail of the distribution standing out as identifiable peaks. So to make the calculation tractable, we build a model out of $N_{\rm str}$ high-density streams on top of a smooth SHM distribution to approximate the sum of $\sim 10^{14} - N_{\rm str}$ low-density streams, i.e.
\begin{equation}
    f_{\rm fine}(\mathbf{v},t) = \left(1-\sum_{i=1}^{N_{\rm str}} \frac{\rho_{\rm str}^i}{\rho_{\rm DM}}\right)f_0(\mathbf{v},t) + \sum_{i=1}^{N_{\rm str}} \frac{\rho_{\rm str}^i}{\rho_{\rm DM}} \delta(|\mathbf{v}_{\rm lab}(t)-\mathbf{v}_{\rm str}^i|) \, ,
\end{equation}
The $N_{\rm str}$ values of $\rho^i_{\rm str}/\rho_{\rm DM}$ are drawn from the distribution shown in Fig.~\ref{fig:VWstreams_CDF} but with a minimum cut-off set by where the chosen value of $N_{\rm str}$ intersects the curve. As in all previous examples, the stream velocities, $\mathbf{v}_{\rm str}^i$, are drawn from the SHM velocity distribution so that the streams still reflect a self-consistent galactic halo when combined together. In the following analysis, we set $N_{\rm str} = 10^4$ as we find that larger values than this do not change any of our results despite increasing the computation time.

This model is very similar to our minicluster case, but with one key difference. We model the fine-grained streams as $\delta$-functions in velocity centred at their streaming velocities, as opposed to the Gaussian assumption made previously. The reason for doing this is that the fine-grained streams will have such narrow velocity dispersions that their frequency spread in the lineshape will be dominated entirely by the frequency shift due to daily modulation for all values of the integration times that are relevant here. A second advantage of this approximation is that it also simplifies the integral over $v$ in the calculation of the mean lineshape, which is now written as follows,
\begin{equation}
    \langle S(\omega)\rangle  =  \left(1-\sum_{i=1}^{N_{\rm str}} \frac{\rho_{\rm str}^i}{\rho_{\rm DM}}\right)\langle S_0(\omega)\rangle + \frac{1}{T}\int_0^T \textrm{d}t\int_0^\infty\textrm{d}v \sum_{i=1}^{N_{\rm str}} \frac{\rho_{\rm str}^i}{\rho_{\rm DM}} \delta(|\mathbf{v}_{\rm lab}(t)-\mathbf{v}_{\rm str}^i|) \,{\rm sinc}^2\left(\frac{1}{2}(\omega_v - \omega)T \right) \, ,\
\end{equation}
where $\langle S_0(\omega)\rangle$ is the mean lineshape for the SHM velocity distribution.

We present visualisations of the lineshape and signal amplitude distributions in Fig.~\ref{fig:VWstreams_lineshape}, in an identical way to the minicluster case we showed in Fig.~\ref{fig:MiniclusterStreams_lineshape}. We show the lineshape and $S$ distributions for three integration times, demonstrating how $T$ needs to be larger than $10^3 T_{\rm coh}^{\rm SHM}$ in this case for the large single-bin power excess to emerge. This is roughly a factor of ten larger than the integration times needed for the large minicluster stream peaks to emerge---this approximate observation will be confirmed more quantitatively in the following results.

\subsection{Haloscope sensitivity}

\begin{figure}
    \centering
    \includegraphics[width=0.49\linewidth]{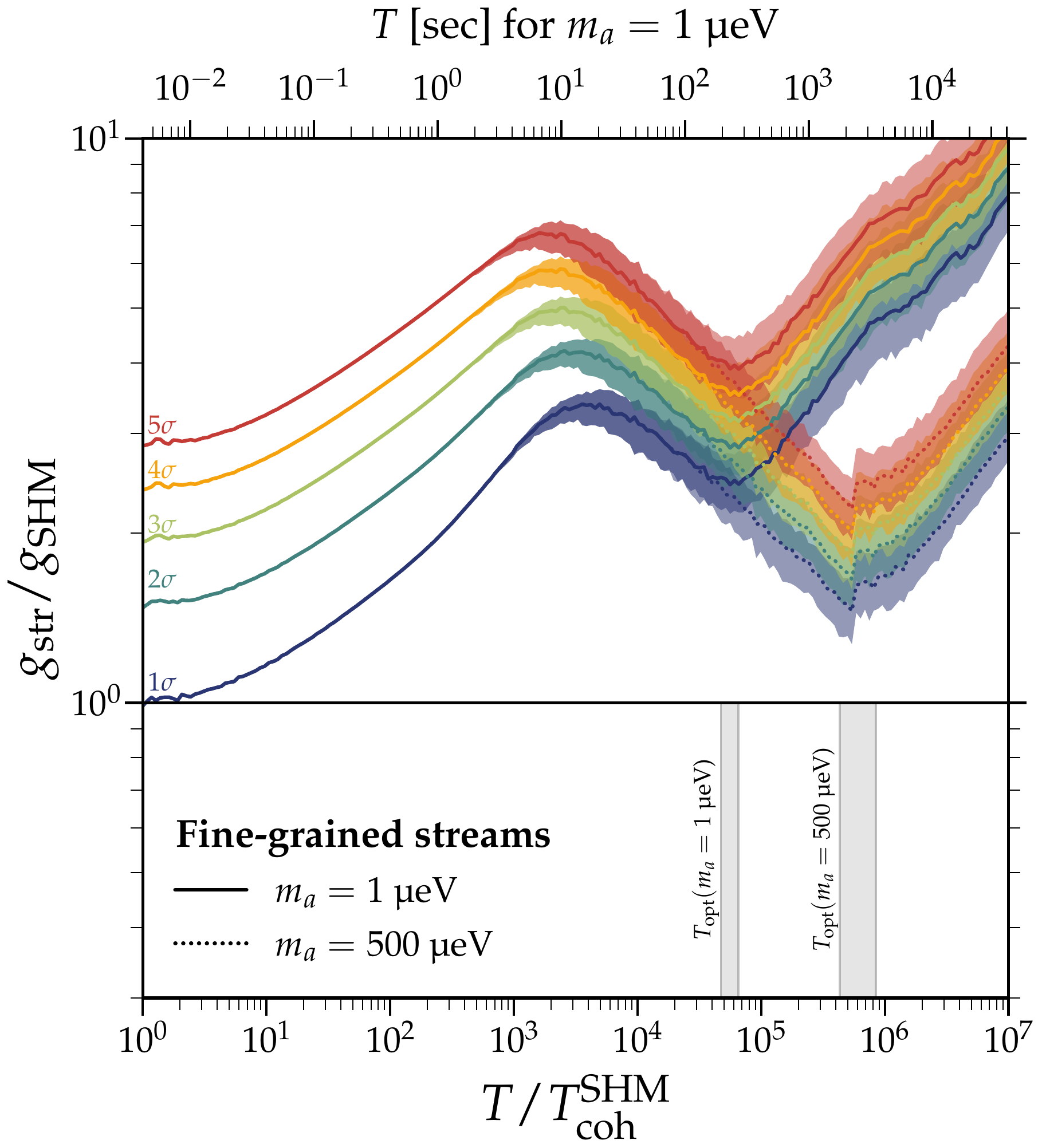}
    \includegraphics[width=0.49\linewidth]{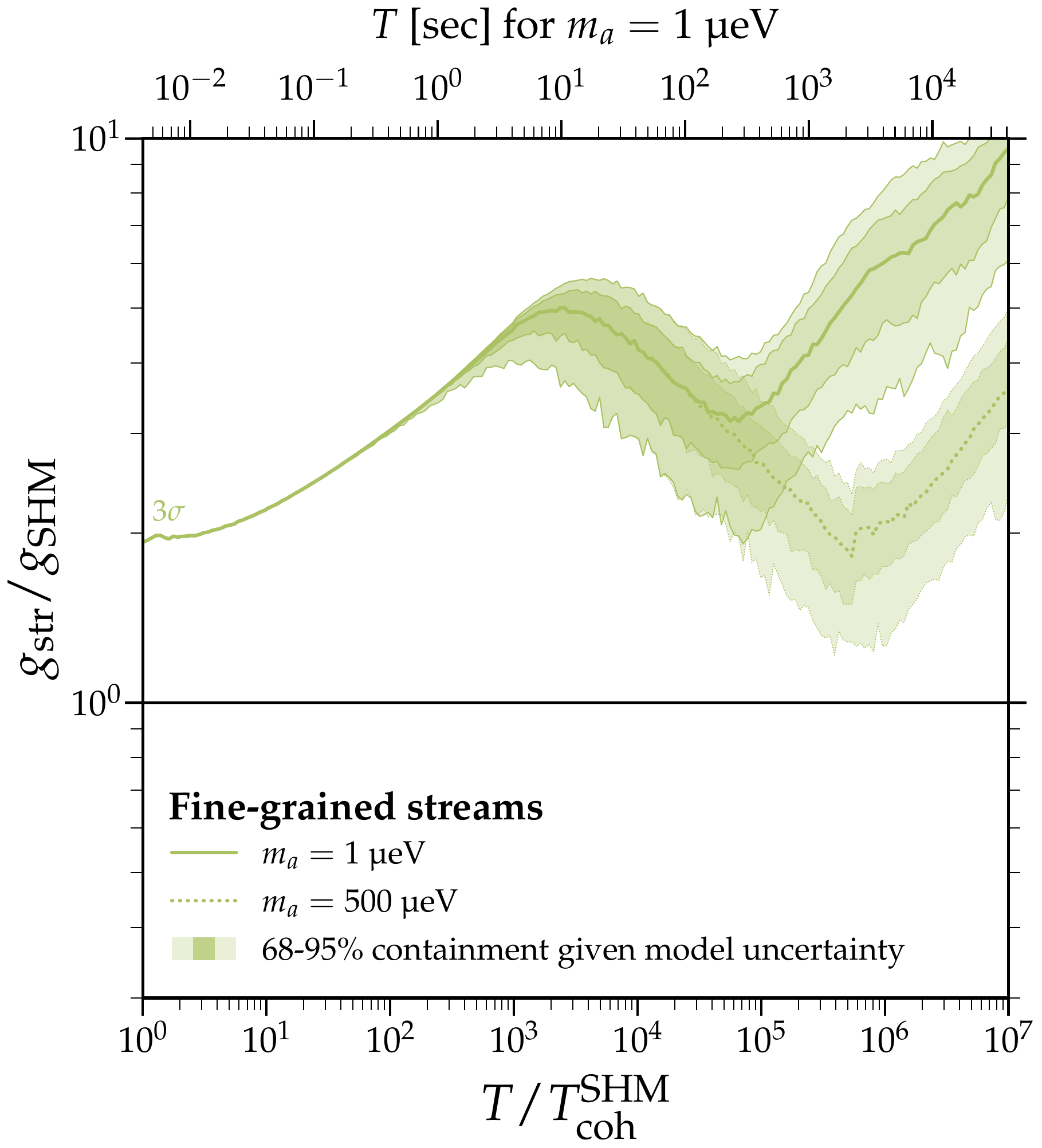}
    \caption{Coupling required for the detection of candidate streams relative to the coupling required to exclude the axion under the SHM.  To construct this plot, we fix the value of $\mathcal{A}/\mathcal{B}$ to be what is required to give a 95\%~CL exclusion of the axion under the SHM assumption for a given value of $T$. We then ask how large the axion-photon coupling needs to be compared to this value for the median experiment to see an $n\sigma$ significant extreme power fluctuation due to a stream---this is then expressed as a ratio $g_{\rm str}/g_{\rm SHM}$. The left and right-hand panels show the same information, but with the left focusing on the different significance levels while leaving the shaded region to contain only 25\% of the model uncertainty for clarity. The right-hand panel selects just the $3\sigma$ case but shows the full 68 and 95\% containment when sampling over all possible minicluster stream models, demonstrating that extended sensitivity ($g_{\rm str}/g_{\rm SHM}<1)$ is difficult to achieve in this case with high confidence, even if the integration time is chosen optimally, i.e. $T = T_{\rm opt}$---see Eq.(\ref{eq:Topt_VWstreams}). The solid and dotted lines shown in both plots correspond to two different axion masses, demonstrating more favourable prospects for heavier axion masses.}
    \label{fig:VWstreams_PmaxProb_Coupling}
\end{figure}

\begin{figure}
    \centering
    \includegraphics[width=0.69\linewidth]{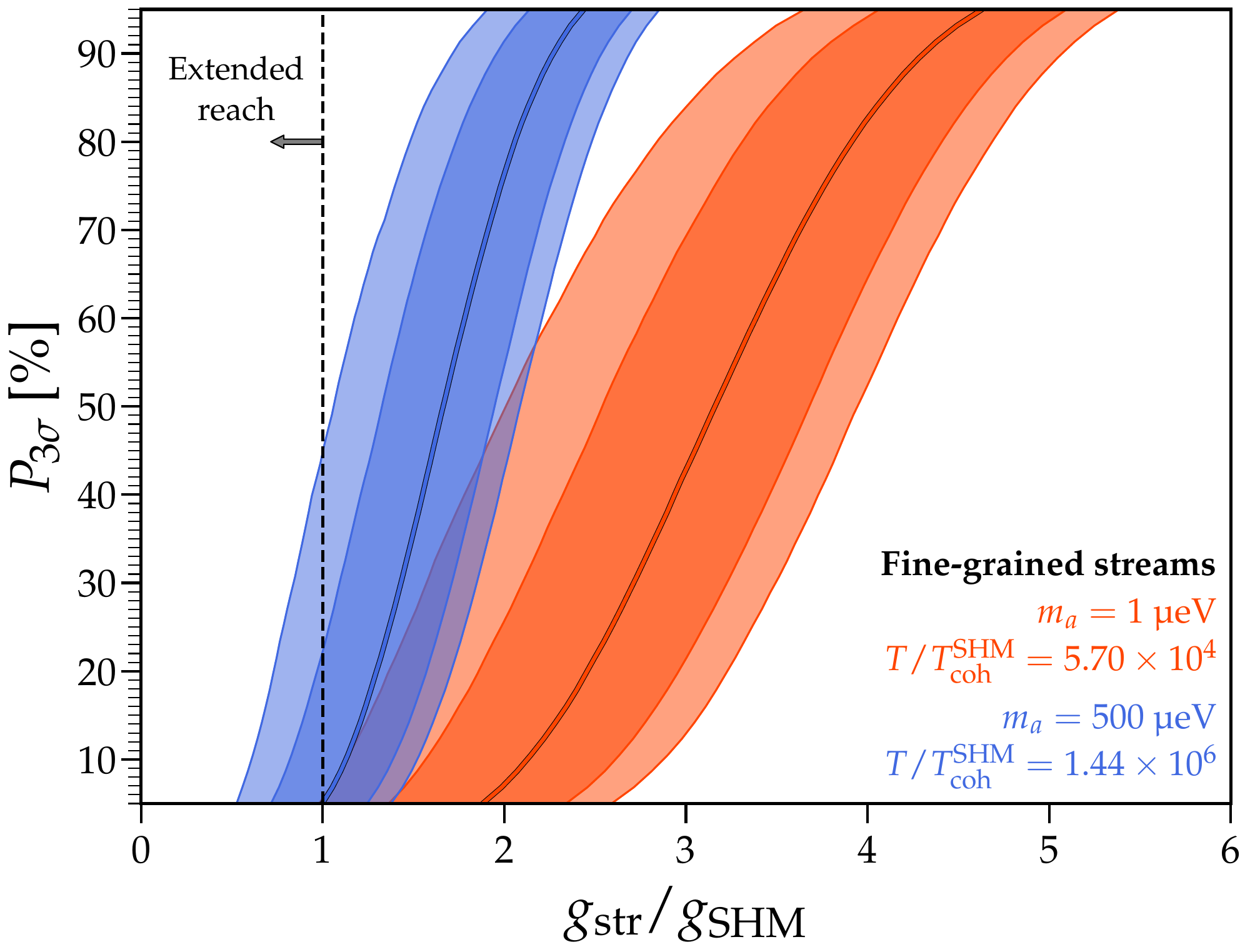}
    \caption{Probability of seeing a $3\sigma$ significant extreme value statistic $P_{\rm max}$ across the axion signal bandwidth as a function of the coupling ratio $g_{\rm str}/g_{\rm SHM}$ as defined in the text. In this case, we are now fixing the integration time to the optimal values expressed in Eq.(\ref{eq:Topt_VWstreams}), which, unlike the minicluster case, is now dependent on the axion mass. The shaded bands contain 68 and 95\% of our models sampled over the parameter priors in Eq.(\ref{eq:minicluster_prior}). As shown in Fig.~\ref{fig:VWstreams_PmaxProb_Coupling}, there is a dependence on the axion mass here, so we show two example masses, chosen to span a larger range than in the previous plot for clarity. All cases where this coupling ratio is less than one constitute a scenario in which a real candidate signal would emerge in a high-resolution analysis, even when a low-resolution analysis of the same data would not discover the axion.}
    \label{fig:VWstreams_ExtendedReach}
\end{figure}

The haloscope sensitivity to our fine-grained stream lineshape is shown in Figs.~\ref{fig:VWstreams_PmaxProb_Coupling} and~\ref{fig:VWstreams_ExtendedReach}. These figures are essentially the same as Figs.~\ref{fig:MiniclusterStreams_PmaxProb_Coupling} and~\ref{fig:MiniclusterStreams_ExtendedReach} that we introduced for the minicluster case. We refer to the explanation at the start of Sec.~\ref{sec:minicluster_sensitivity} for how to interpret these plots, as that explanation applies here too. 

Firstly, Fig.~\ref{fig:VWstreams_PmaxProb_Coupling} shows the ratio between the axion-photon coupling required for the median experiment to see an $n\sigma$ extreme value statistic due to a stream relative to the coupling required to set a 95\% CL exclusion limit on the axion under the SHM. As in the minicluster case, we can identify the ``optimal'' integration time to be the one where a high-resolution analysis has the best chance to reveal a candidate signal due to a prominent stream. The difference in this case compared to the miniclusters is that the optimal value of $T/T^{\rm coh}_{\rm SHM}$ is now dependent on the axion mass. The reason for this is the daily modulation. Because our stream velocity distributions are now delta functions, their widths in the measured lineshape are dictated entirely by the shift in the frequency over the integration time due to $\mathbf{v}_{\rm lab}(t)$. Approximately, $T_{\rm opt}$ is when this frequency shift matches the frequency resolution of the lineshape---the solution to the following equation,
\begin{equation}
    T_{\rm opt} \approx \frac{2\pi}{m_a |\mathbf{v}_{\rm lab} - \mathbf{v}_{\rm str}|\Delta v_{\rm rot}(T_{\rm opt})} \, .
\end{equation}
In contrast to Eq.(\ref{eq:Topt_miniclusters_step}), the dependence on $m_a$ does not drop out when we express $T$ as a multiple of $T/T_{\rm coh}^{\rm SHM}$. This is because there is now a second timescale relevant to the problem. As a result of this dependence, we now observe a noticeable difference between the two different axion mass cases shown in Fig.~\ref{fig:VWstreams_PmaxProb_Coupling} as solid and dotted lines. 

Numerically, we find the optimal time as a function of axion mass by taking an average of all of the stream frequency shifts in the lineshape, weighted by their corresponding densities. We then perform this over a range of axion masses and obtain the following fit,
\begin{equation}\label{eq:Topt_VWstreams}
    T_{\rm opt} \approx 5.7 \times 10^4 \,T_{\rm coh}^{\rm SHM}(m_a) \left( \frac{m_a}{1\,\upmu{\rm eV}}\right)^{0.52} = 230\,{\rm sec} \left( \frac{1\,\upmu{\rm eV}}{m_a}\right)^{0.48}
\end{equation}
We note that there is a scatter in the overall timescale and the scaling index of the axion mass dependence, which is not exactly 0.5. This is due to the fact that $\Delta v_{\rm rot}(T)$ is related to $\cos({2\pi T/{\rm 1 \,day}})$ but has an amplitude and phase that are different for each stream.

Fig.~\ref{fig:VWstreams_PmaxProb_Coupling} also reveals that the sensitivity once the streams are resolved---i.e.~the value of $g_{\rm str}/g_{\rm SHM}$ around $T_{\rm opt}$---is better for higher axion masses, but quite poor for the lowest axion mass example shown here. This can be understood by looking back to Fig.~\ref{fig:EVS_Significance_stream_comparison}, where we notice that the EVS $Z$-score for very narrow streams will continue to increase until the daily modulation effect spreads the stream across multiple bins. When the axion mass is larger, the coherence time is smaller, which means the statistical power of the EVS test is able to build up more as a function of $T$ before the daily modulation effect takes over and suppresses the power of the test. A simpler way to understand it is to consider it as simply a manifestation of the fact that the axion period is longer for lower mass axions---the shift in frequency due to daily modulation will always be larger for lower mass axions than higher mass axions when integrating over the same number of coherence times.

To finish this section, we come to Fig.~\ref{fig:VWstreams_ExtendedReach} where, like in Fig.~\ref{fig:MiniclusterStreams_ExtendedReach}, we fix the integration time to be the optimal value for two axion masses and show how values of $g_{\rm str}/g_{\rm SHM}<1$ can be reached in certain rare cases; albeit only for larger axion masses due to the reasons stated above. So we conclude that for small axion masses, the fine-grained streams are very unlikely to emerge as candidate excess above the noise in a high-resolution search, but for higher axion masses, this becomes an (albeit rare) possibility. We could also relax the significance requirement to say $2\sigma$ (corresponding to a $14\sigma$ local power excess for $T/T_{\rm coh}^{\rm SHM} = 10^4$), which would improve all of these probabilities; however in this case the experiment would likely have to re-scan many more candidate peaks that would turn out to be noise fluctuations. Ultimately, the threshold for the significance level warranting re-scans is something that must be chosen on an experiment-by-experiment basis, so we will end the discussion here.

\section{Conclusion}
With this study, we aimed to motivate the continuation of high-resolution analyses of axion haloscope data by outlining the various classes of ultra-fine-grained velocity substructure that may be present in the dark matter distribution around the Earth. We have developed several physically plausible models that depend only on the cosmology of the axion dark matter. We also explored several statistical tests that could be employed to capitalise on the fact that very narrow features in the velocity distribution can generate large signal excesses in haloscope data, if power spectra are taken at fine enough spectral resolution. Our main takeaway messages and recommendations for experiments can be summarised as follows:

\begin{itemize}
    \item Haloscope experiments should conduct dedicated analyses of their data at as high a spectral resolution as is manageable. As we have shown, these have the possibility to reveal candidate signals as extreme single-frequency-bin excesses that, upon interrogation with more data, would persist and lead to a discovery of an axion. Depending upon the exact nature of the dark matter velocity substructure, we have found that analyses of this type \textit{do} have the possibility to reveal the axion, even if a lower-resolution search using the same data did not reveal a statistically significant signal. 
    \item The prospects for accelerated discovery get stronger for larger axion masses, but, unfortunately, are unlikely for axion masses less than $\upmu$eV, due to the long coherence times, making a lower-resolution search more optimal. For low axion masses, the streams could still emerge, but likely only in a post-discovery setting.
    \item We emphasise that the presence of streams allows for there to be large signal excesses, however these fluctuations are only a statistical possibility, and moreover, we are only able to specify the axion lineshape for a given model in probabilistic terms (the expected distribution of stream densities etc.). We therefore recommend that haloscopes continue to be conservative when setting exclusion limits on the axion by adopting the baseline SHM lineshape assumption, but while leaving open the possibility of there being narrow features which may emerge only in high-resolution analyses of the same data. However, to be clear: the non-detection of a stream in a given narrow window of frequencies cannot be interpreted as the non-detection of an axion signal with a mass in that vicinity. The approach taken by e.g.~Ref.~\cite{ADMX:2024pxg} to constrain the DM density of a stream at a particular frequency while fixing the axion-photon coupling to some value (KSVZ, DFSZ etc.) is therefore appropriate for reporting null results from analyses of this kind.
    \item In a generic CDM cosmology, like in a pre-inflationary axion or ALP model, there are an expected $\sim 10^{14}$ streams overlapping in the solar neighbourhood, which come about as a direct result of the halo formation process, as opposed to any specific dark matter production mechanism. Of these $10^{14}$ streams, only a handful are expected to contribute more than 0.1\% of the local dark matter density (Fig.~\ref{fig:VWstreams_CDF}). For axions produced via misalignment, their velocity dispersions are negligible, so the optimal spectral resolution to capture these features is to match it to the shift in the frequency of the streams due to the Earth's rotation (Fig.~\ref{fig:VWstreams_PmaxProb_Coupling}). For this model, we find the optimal integration time to be: $T_{\rm opt}  = 230\,{\rm sec} \left( \frac{1\,\upmu{\rm eV}}{m_a}\right)^{0.48}$.
    \item In the post-inflationary scenario, the axion distribution forms small-scale bound substructures called miniclusters or minihalos, which are mostly destroyed through tidal interactions, leading to on the order of several hundred to a thousand overlapping streams locally. The most massive handful of these streams will amount to around $0.1$--$1\%$ of the local dark matter density, and they will have velocity dispersions between $10^{-5}$ and $10^{-2}$ km/s (Fig.~\ref{fig:MiniclusterStream_properties}), subject to theoretical uncertainties related to the initial minicluster mass function and the disruption process. These can be captured much more readily in haloscopes operating in the post-inflationary mass window, where we find the optimal integration time to be: $  T_{\rm opt} = 6.56 \, {\rm sec} \, \left(\frac{100\,\upmu{\rm eV}}{m_a}\right)$ (Fig.~\ref{fig:MiniclusterStreams_PmaxProb_Coupling}). This possibility strongly motivates high-resolution analysis of data from experiments targeting this window, such as QUAX, ORGAN, BREAD, BRASS, MADMAX and CADEx.    
\end{itemize}

There are several directions in which the models we have presented here may be improved upon, as well as alternative experimental strategies that could be explored. In this work, we have chosen to neglect the possibility of gravitational focusing~\cite{Griest:1987vc, Sikivie:2002bj, Alenazi:2006wu, Lee:2013wza, Bozorgnia:2014dqa, DelNobile:2015nua, Kim:2021yyo} that can lead to extremely rare but potentially large enhancements of streams that possess a favourable alignment with the Earth-Sun system~\cite{Kryemadhi:2022vuk,Arza:2022dng,DeMiguel:2024cwb}. Including these effects may enhance the prospects further beyond what we have presented here. We have also outlined in Sec.~\ref{sec:miniclusters} several aspects of the minicluster model that require refinements through further simulations---doing so would improve the uncertainty bands presented in Figs.~\ref{fig:MiniclusterStream_properties} and~\ref{fig:MiniclusterStreams_PmaxProb_Coupling}. 

On the experimental side, we have so far neglected the discussion of axion experiments exploiting other couplings than the axion-photon coupling, as well as other forms of wave-like dark matter such as scalars~\cite{Hui:2016ltb,Banerjee:2018xmn,Antypas:2022asj,Cyncynates:2024ufu,Cyncynates:2024bxw} and vectors/dark photons~\cite{Arias:2012az,Graham:2015rva,Caputo:2021eaa,Cyncynates:2023zwj}. In particular, experiments that couple in a way that invokes a dependence on the dark matter \textit{velocity}, rather than just the speed, may be significantly impacted by the models we have developed here. These examples include experiments that couple to the axion field gradient $\nabla a$~\cite{Gramolin:2021mqv, Lisanti:2021vij} (e.g.~through axion-fermion couplings~\cite{JacksonKimball:2017elr, Garcon:2019inh, Lee:2022vvb, Gao:2022nuq}), as well as proposed time-correlated networks of experiments that compare the phases of the axion field oscillations over spatial scales larger than a few axion coherence lengths, as explored in Refs.~\cite{Foster:2020fln}. We hope this work sparks further investigation into this topic so as to uncover more possibilities to reveal the nature of dark matter in a direct detection experiment through its fine-grained substructure.

\section*{Acknowledgements}
CAJO is supported by the Australian Research Council under the grant numbers DE220100225 and CE200100008. GP is supported by the Australian Research Council's Discovery Project (DP240103130) scheme. This article involves work from COST Action COSMIC WISPers CA21106, supported by COST (European Cooperation in Science and Technology).

\appendix

\section{Analytic expressions for the mean axion lineshape}\label{app:lineshape}
\begin{figure}
    \centering
    \includegraphics[height=0.49\linewidth]{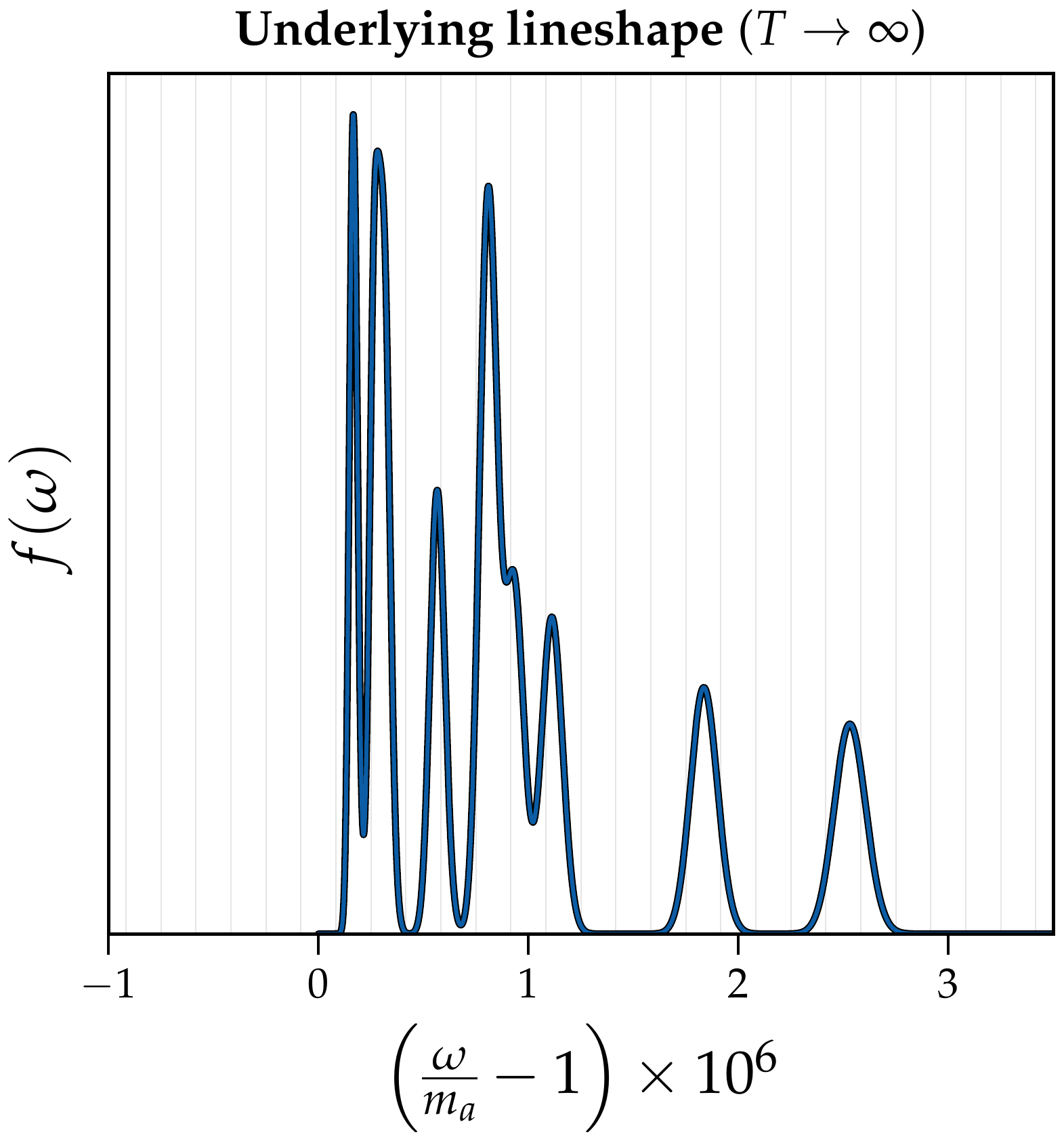}
    \includegraphics[height=0.49\linewidth]{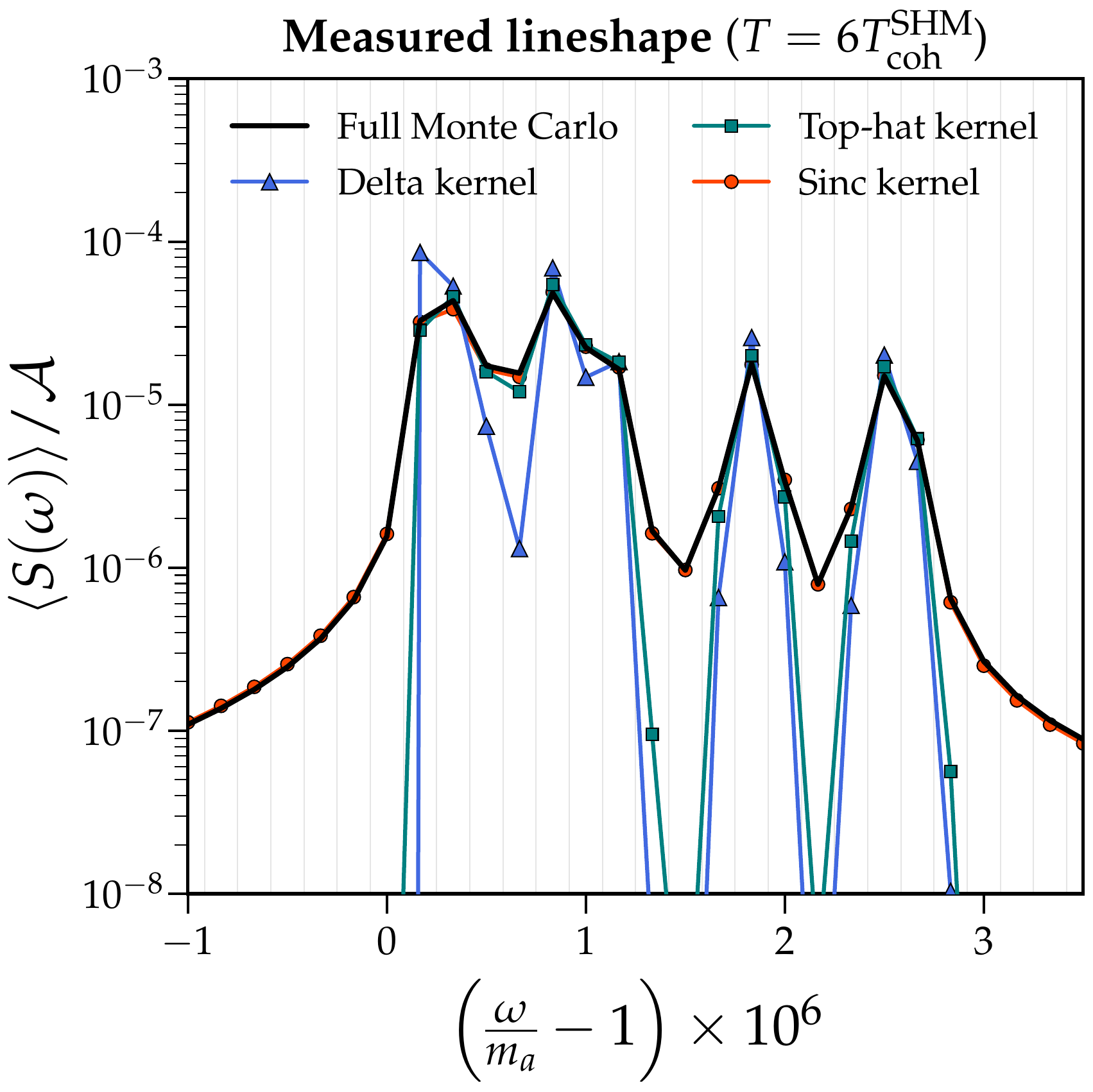}
    \caption{Comparison of various calculations of the mean axion lineshape. The underlying functional form of the lineshape is shown in the left-hand plot, while various calculations of the \textit{measured} lineshape for an integration time of $T = 6T^{\rm SHM}_{\rm coh}$ are shown in the right-hand plot. As a cross-check, we also show the average of the lineshapes from many Monte Carlo realisations of the axion field as a black line---confirming that the Sinc kernel calculation gives the correct shape for all values of $T/T_{\rm coh}^{\rm SHM}$, while the other calculations are approximations that break down at small values of $T/T_{\rm coh}^{\rm SHM}$. Note that this example lineshape is entirely arbitrary and is just chosen to exaggerate the differences in these calculations---all of them will converge in the limit of large $T$, and the differences are much less pronounced for lineshapes that do not have features that are comparable to the frequency bin width. For comparison, the bins are shown with vertical grey lines.}
    \label{fig:LineshapeCalculationComparisons}
\end{figure}

In Section~\ref{sec:lineshape}, we introduced a general expression for the axion lineshape given a finite integration time $T$,
\begin{equation}
    \langle S(\omega) \rangle = \frac{\mathcal{A}T}{2} \int_0^\infty {\rm d}v f(v)\, {\rm sinc}^2\left( \frac{1}{2}(\omega_v - \omega)T\right) \,.
\end{equation}
This expression is equivalent to other expressions presented in the literature previously, but only under certain limits. For example if we take $T\to \infty$---which is to say that $T$ is much longer than any other timescale---then ${\rm sinc}^2(T(\omega_v -\omega)/2) \to 2\pi/T \delta(\omega_v-\omega)$ and so we recover~\cite{Foster:2017hbq},
\begin{equation}
    \langle S(\omega) \rangle = \frac{\mathcal{A} \pi f(v)}{m_a v}\bigg|_{v=\sqrt{2 \omega / m_a-2}} \, .
\end{equation}
Alternatively one can approximate the ${\rm sinc}^2$ kernel with a top-hat,  
\begin{equation}
    {\rm sinc}^2\left( \frac{1}{2}(\omega_v - \omega)T\right) \to \mathcal{T}(\omega_v-\Delta \omega/2,\omega_v+\Delta \omega/2)
\end{equation}
where we define $\mathcal{T}(x_0,x_1)$ to be a top-hat function equal to 1 between $x_0$ and $x_1$ and zero elsewhere. Then the integral becomes
\begin{equation}
\begin{aligned}
    \langle S(\omega) \rangle &= \frac{1}{\Delta \omega} \int_{\omega - \Delta \omega/2}^{\omega+\Delta \omega/2} {\rm d}\omega \frac{\mathcal{A} \pi f(v)}{m_a v}\bigg|_{v=\sqrt{2 \omega / m_a-2}} \, \\
    &= \left\langle \frac{\mathcal{A} \pi f(v)}{m_a v} \right\rangle_{\omega-\Delta \omega/2}^{\omega+\Delta \omega/2} \,\, ,
    \end{aligned}
\end{equation}
which is the mean value of Eq.(\ref{eq:signal0}) over the frequency bin. This form has also appeared occasionally in the literature, e.g.~Refs.~\cite{Brubaker:2017rna, Millar:2022peq}. We compare the accuracy of these expressions for an example signal in Fig.~\ref{fig:LineshapeCalculationComparisons}. For completeness, we compare these against a full Monte-Carlo treatment, which generates the signal by summing a large number of individual axion waves and taking a discrete Fourier transform, as we described in Sec.~\ref{sec:lineshape}. The fact that the black and orange lines are almost identical is simply a confirmation of the mathematical derivation we presented in Sec.~\ref{sec:lineshape}.

The full expression for $S(\omega)$ requires us to convolve a ${\rm sinc}^2$ kernel with the boosted Maxwell-Boltzmann-like form for $f(v)$ presented in Eq.(\ref{eq:gaussian_fv}). Unfortunately, this does not yield an analytic solution. Although the convolution is not costly to evaluate numerically, it becomes inconvenient to do it repeatedly while also scanning over many other parameters, many streams, and for values of $T>10^4 T^{\rm SHM}_{\rm coh}$ where the number of frequency bins making up $S(\omega)$ can become extremely large. We therefore adopt the top-hat kernel, which provides an acceptable approximation that is more accurate than the Dirac kernel for smaller $T$ while admitting an analytic solution. This solution is:
\begin{align}
    \langle S(\omega ; v_c, \sigma)\rangle  \approx & \frac{\mathcal{A}T}{4}\bigg[\sqrt{\frac{2}{\pi}} \frac{\sigma}{v_c} 
\bigg(e^{-\frac{(v_1-v_c)^2}{2 \sigma^2}}-e^{-\frac{(v_2-v_c)^2}{2 \sigma^2}}-e^{-\frac{(v_1+v_c)^2}{2 \sigma^2}}+e^{-\frac{(v_2+v c)^2}{2 \sigma^2}}\bigg) \\ &-
\operatorname{erf}\bigg(\frac{v_1-v_c}{\sqrt{2} \sigma}\bigg)+\operatorname{erf}\bigg(\frac{v_2-v_c}{\sqrt{2} \sigma}\bigg)-\operatorname{erf}\bigg(\frac{v_1+v_c}{\sqrt{2} \sigma}\bigg)+\operatorname{erf}\bigg(\frac{v_2+v_c}{\sqrt{2} \sigma}\bigg)\bigg] \, ,
\end{align}
where $v_{1,2} = \sqrt{2(\omega \mp \Delta \omega/2 - m_a)/m_a}$. Recall that a multistream model is then built out of a weighted sum of these with different parameter values, i.e.
\begin{equation}
    \langle S(\omega)\rangle = \sum_{i=1}^{N_{\rm str}} \frac{\rho_{\rm str}^i}{\rho_{\rm DM}} \left\langle S\left(\omega ; |\mathbf{v}_{\rm lab} - \mathbf{v}^i_{\rm str}|,\sigma_{\rm str}^i\right)\right\rangle \, .
\end{equation}

\bibliography{axion.bib}
\bibliographystyle{bibi}

\end{document}